\PassOptionsToPackage{pdfpagelabels=false}{hyperref}
\documentclass[fleqn,usenatbib]{mnras}
\usepackage{xcolor, graphicx}
\usepackage{amsmath}
\usepackage{amssymb}
\usepackage{soul}
\usepackage{txfonts}
\usepackage{journal_names}
\usepackage{caption}

\newcommand{\Msun}{\rm M_\odot}

\newcommand{\kms}{\mbox{km s$^{-1}$}}
\newcommand{\re}{R_{\rm e}}

\newcommand{\fdm}{f_{\rm DM}}

\newcommand{\rhodm}{\left< \rho_{\rm DM} \right>}
\newcommand{\chalo}{c_{\rm 200}}
\newcommand{\mhalo}{M_{\rm 200}}
\newcommand{\rhalo}{r_{\rm 200}}
\newcommand{\rs}{r_{\rm s}}

\def\Sersic/{{S\'ersic}}

\title[DM fractions and assembly epochs of ETGs]
{The SLUGGS Survey: Dark matter fractions at large radii and assembly epochs of early--type galaxies from globular cluster kinematics}
\author[Alabi~et~al.~ ]
{Adebusola B. Alabi$^{1}$\thanks{Email: aalabi@swin.edu.au} , Duncan A. Forbes$^{1}$, Aaron J. Romanowsky$^{2,3}$, Jean P. Brodie$^{3}$,\\
\\
\normalfont{\LARGE Jay Strader$^{4}$, Joachim Janz$^{1}$, Christopher Usher$^{5}$, Lee R. Spitler$^{6,7,8}$,
Sabine Bellstedt$^{1}$,}\\
\\
\normalfont{\LARGE Anna Ferr{\'e}-Mateu$^{1}$}\\
\\
\\
$^1$ Centre for Astrophysics \& Supercomputing, Swinburne University, Hawthorn VIC 3122, Australia\\
$^2$ Department of Physics and Astronomy, San Jos\'e State University, San Jose, CA 95192, USA\\
$^3$ University of California Observatories, 1156 High Street, Santa Cruz, CA 95064, USA\\
$^4$ Department of Physics and Astronomy, Michigan State University, East Lansing, Michigan 48824, USA\\
$^5$ Astrophysics Research Institute, Liverpool John Moores University, Liverpool L3 5RF, United Kingdom\\
$^6$ Australian Astronomical Observatory, PO Box 915, North Ryde, NSW 1670, Australia\\
$^7$ Department of Physics and Astronomy, Macquarie University, North Ryde NSW 2109, Australia\\
$^8$ Research Centre for Astronomy, Astrophysics \& Astrophotonics, Macquarie University, North Ryde NSW 2109, Australia\\
}

\begin{document}

\date{Accepted today}

\pagerange{\pageref{firstpage}--\pageref{lastpage}} \pubyear{2015}

\maketitle

\label{firstpage}
\begin{abstract}
We use globular cluster kinematics data, primarily from the SLUGGS survey, to measure the dark matter fraction ($\fdm$) and the average dark matter density ($\rhodm$) within the inner 5 effective radii ($\re$) for 32 nearby early--type galaxies (ETGs) with stellar mass log $(M_*/\Msun)$ ranging from $10.1$ to $11.8$. We compare our results with a simple galaxy model based on scaling relations as well as with cosmological hydrodynamical simulations where the dark matter profile has been modified through various physical processes.

We find a high $\fdm$ ($\geq0.6$) within 5~$\re$ in most of our sample, which we interpret as a signature of a late mass assembly history that is largely devoid of gas-rich major mergers. However, around log $(M_*/\Msun) \sim 11$, there is a wide range of $\fdm$ which may be challenging to explain with any single cosmological model. We find tentative evidence that lenticulars (S0s), unlike ellipticals, have mass distributions that are similar to spiral galaxies, with decreasing $\fdm$ within 5~$\re$ as galaxy luminosity increases. However, we do not find any difference between the $\rhodm$ of S0s and ellipticals in our sample, despite the differences in their stellar populations. We have also used $\rhodm$ to infer the epoch of halo assembly ($z{\sim}2-4$). %Our ETGs generally reside in haloes that formed at $z{\sim}3$, with haloes associated with ${\sim}L^*$ ETGs forming earlier ($z{\sim}4$) while those of massive ETGs form later ($z{\sim}2$) than average. 
By comparing the age of their central stars with the inferred epoch of halo formation, we are able to gain more insight into their mass assembly histories. Our results suggest a fundamental difference in the dominant late-phase mass assembly channel between lenticulars and elliptical galaxies.
\end{abstract}


\section{Introduction}
The natural expectation within the hierarchical structure formation paradigm and $\rm \Lambda$CDM cosmology (e.g. \citealt{Peebles_1982}) is that dark matter haloes and their resident galaxies grow in tandem. The growth channels include mergers (major/minor, with/without gas) and gas accretion (smooth or clumpy) (e.g. \citealt{Genel_2010, Rodriguez_2016}), with galaxy assembly showing more complexities due to the baryonic processes involved (e.g. gas dissipation, star formation, feedback processes due to active galactic nuclei or supernova, etc.). These baryonic processes may also alter the dark matter  distribution, especially in the most-central parts, through adiabatic halo contraction \citep[e.g.][]{Blumenthal_1986} or halo expansion through gravitational heating from infalling gas clumps \citep[e.g.][]{Johansson_2009} or outflows linked to feedback events \citep[e.g][]{Maccio_2012}. The implication is that the relative distributions of DM and baryons in present-day galaxies contain clues about when (epoch of formation) and how (nature of the mass assembly) they formed and how they have evolved.

Both halo and galaxy growth have been divided into two phases in the literature (e.g. \citealt{Zhao_2003, Oser_2010, Klypin_2016}). At early times, when the universe was denser, dark matter haloes grew rapidly via frequent mergers and accretion. In parallel, gas-rich, dissipative events were very common, and the stellar cores of present--day galaxies were formed. Galaxies then grew predominantly by rapidly forming stars \textit{in-situ}, such that the dark matter fraction ($\fdm$) at the centres of young galaxies is relatively low and the galaxies themselves are very compact (e.g. \citealt{vanDokkum_2008, Naab_2009, Napolitano_2010, Remus_2016}). Dark matter haloes experienced an increase in both core size and extent (core size and halo extent are usually parametrized by the scale radius ($\rs$) and virial radius ($\rhalo$), respectively). Thus, the halo concentration ($\chalo\equiv\rhalo/\rs$) is kept fairly constant during this phase \citep{Klypin_2016}. The halo concentration is directly linked to the density of the universe at the epoch when the halo formed ($\chalo \propto (1 + z_{\rm form})^{-1}$ e.g. \citealt{Bullock_2001, Wechsler_2002}) and is well described by the $\chalo-\mhalo$ scaling relation (e.g. \citealt{Dutton_2010}), where $\mhalo$ is the virial mass.

At later times ($z\leq2$), in the two-phase paradigm, when gas-rich events become fewer, mass growth (be it baryonic or dark matter) occurs predominantly in the galaxy outskirts. The progenitors of present-day massive early-type galaxies (ETGs) increase rapidly in mass and size through a multitude of minor mergers and/or dry major mergers, thereby increasing significantly their fraction of stars formed \textit{ex-situ} (e.g. \citealt{Naab_2009, Pillepich_2014, Rodriguez_2016, Remus_2016}). This is consistent with the low angular momentum content (e.g. \citealt{Emsellem_2011, Arnold_2014, Raskutti_2014, Moody_2014, Foster_2016}), old central stars (e.g. \citealt{Terlevich_2002, McDermid_2015}) and high $\fdm$ at large radii (e.g. \citealt{Deason_2012, Alabi_2016}) usually reported in studies of massive ETGs. The dark matter haloes also get bigger in size while the core sizes may grow or shrink depending on whether violent relaxation, adiabatic halo contraction or halo expansion events occur (e.g. \citealt{Johansson_2009, Governato_2010, Klypin_2016}). Therefore, the observed dark matter density within the scale radius should reflect the epoch of halo assembly as well as imprints of how baryonic processes have altered the distribution of dark matter within the halo during galaxy evolution.

Dark matter fractions at large radii, e.g. 5 effective radii ($\re$), are becoming increasingly available, especially in ${\sim}L^*$ ETGs, i.e. ETGs with stellar mass ($M_*$) ${\sim}10^{11}\ \Msun$ (e.g. \citealt{Deason_2012, Alabi_2016}). This is due to the use of dynamical mass tracers such as planetary nebulae (PNe) or globular clusters (GCs) which probe further out into the galaxy halo where the light from galaxy stars is faint. At a fiducial 5~$\re$, which is always interior to the halo $\rs$ (it is expected on average that, 5~$\re \sim 0.4 \rs$), dark matter should dominate the mass profiles in ETGs. While most ETGs studied show a high dark matter fraction within 5~$\re$ (on average,  $\fdm\geq0.6$), in agreement with theoretical predictions and an increasing trend with total galaxy stellar mass, some ETGs with $M_*{\sim}10^{11}\ \Msun$ have been found to have surprisingly low dark matter content within 5~$\re$ (e.g. \citealt{Romanowsky_2003, Napolitano_2005, Deason_2012, Alabi_2016}).

Several reasons have been given in the literature for this intriguing tension between observations and simulations including the stellar initial mass function (IMF), the orbital anisotropy of the mass tracers, modification of the dark matter profile during mass assembly (net effect of adiabatic halo contraction and inner dark matter halo expansion), the nature of the DM halo profile i.e. logarithmic or NFW or even a failure of the $\Lambda$~CDM cosmology (e.g. \citealt{Dekel_2005,Thomas_2009,Napolitano_2005, Napolitano_2010, Napolitano_2011, Morganti_2013}). In \citet{Alabi_2016}, we suggested that these galaxies with low dark matter fractions could have different halo structures, i.e., diffuse DM haloes. However, the exact origin of this anomaly is still unclear. Cosmological hydrodynamical simulations of ETGs where baryons have modified the present--day dark matter profiles have also started to report predictions for $\fdm$ at 5~$\re$ (e.g. \citealt{Wu_2014}). More recently, simulations with realistic implementation of active galactic nuclei (AGN) and supernova (SN) feedback recipes, (e.g \citealt{Vogelsberger_2014, Schaye_2015, Remus_2016}), covering the stellar mass range of ETGs in this work, i.e. $10^{10}-10^{12}\ \Msun$ have been released. The time is therefore ripe to systematically compare results from observations with theoretical predictions, and thereby unravel the nature of mass distributions in ETGs within a cosmological context.

Unfortunately, the halo concentration is difficult to directly constrain in ETGs (e.g. \citealt{Napolitano_2005, Napolitano_2009, Samurovic_2014}), due in part to the limited radial extent of kinematics data and the degeneracy between the halo virial mass and concentration. However, it is possible to infer the epoch of halo assembly from the average dark density within the scale radius ($\rhodm \propto (1 + z_{\rm form})^3$, where $\rhodm$ and $z_{\rm form}$ are the average dark matter density and the epoch of halo assembly, respectively). \citet{Thomas_2009} inferred $z_{\rm form}$ for several ETGs in the Coma cluster using $\rhodm$ obtained from stellar kinematics within the inner 2~$\re$ and found that their haloes must have assembled at $z_{\rm form}\approx2-3$.

In this work, we expand our sample of ETGs with homogeneously measured $\fdm$ at 5~$\re$ in \citet{Alabi_2016} from 23 to 32, using GC kinematics data mostly obtained as part of the SLUGGS \footnote{http://sluggs.swin.edu.au} survey \citep{Brodie_2014}. SLUGGS stands for SAGES Legacy Unifying Globulars and  GalaxieS. This brings the number of ${\sim}L^*$ ETGs with total mass measurements within 5~$\re$ up to 16. We also adopt the recently published galaxy sizes, \Sersic/ indices and stellar mass measurements from \citet{Forbes_2016} for our galaxy sample. With this larger sample, homogeneously measured galaxy parameters and the suite of cosmological simulations that are now available, we investigate the cosmological origins of the measured $\fdm$ at large radii in ETGs. We also address the curious cases of ETGs with low dark matter fractions in more detail. We study the structural properties of the dark matter haloes within the inner 5~$\re$ using their average dark matter densities, and infer their halo assembly epochs. Lastly, we use the halo formation epoch, the age of the central stars and the dark matter fraction to probe the nature of mass assembly in ETGs. We pay special attention to the morphology, environment and angular momentum in this exercise.

The paper structure is as follows: In Section \ref{data} we describe the new data we introduce in this work. In Section \ref{methods}, we obtain the dynamical mass estimates and $\fdm$ (for the newly introduced galaxies) and $\rhodm$ (for the combined sample) and compare with cosmological hydrodynamical simulations. We also obtain dark matter halo properties and compare with results from the literature. In Section \ref{discussion}, we discuss the diverse nature of $\fdm$ in ETGs and the nature of their mass assembly. We summarise our results in Section \ref{summary}.

\begin{table*}
{\small \caption{General properties of our galaxies.} \label{tab:gal_prop}}
\begin{tabular}{@{}l l l l l l l l l l l l l l l l} %c r c r r c c c c c c c c c c
\hline
\hline
Galaxy & Dist. & $V_{\rm sys}$ & $\sigma_{\rm kpc}$ & $\epsilon$ & ${\rm env.}$  & ${\rm morph.}$ & Age & $N_{\rm GC}$ & $\re$ & $\log(M_{*})$ & $n$ & $\alpha$ & $\gamma$ & corr & $V_{\rm rot}/\sigma$\\
$\rm [NGC]$ & [Mpc]  & [$\kms$] & [$\kms$]	&	&  & &  [Gyr] &  & [kpc] & $\rm [\Msun]$ & 	&	&	&  & \\
(1) & (2) &	(3) & (4) & (5) & (6) & (7) & (8) & (9) & (10) & (11) & (12) & (13) & (14) & (15) & (16) \\
\hline
720   & 26.9        & 1745 & 227         & 0.49 & F & E      &    $7.8^{b}$  &  69  & 3.80         &   11.27   & 3.8   & 0.106  & 2.71  & 0.92 & $0.42_{-0.17} ^{+0.24}$\\[1.3ex]
821   & 23.4        & 1718 & 193         & 0.35 & F & E      &   11.0  &  69  & 4.90         &   11.00    & 6.0   & 0.230   & 2.88  & 0.98 & $0.40_{-0.18} ^{+0.20}$\\[1.3ex]
1023  & 11.1        &  602 & 183         & 0.63 & G & S0     &   12.3  &  115 & 2.58         &   10.99   & 4.2   & 0.235  & 2.89  & 0.85 & $0.65_{-0.18} ^{+0.21}$\\[1.3ex]
1400  & 26.8        &  558 & 236         & 0.13 & G & E/S0   &   $13.8^{c}$  &  69  & 3.33         &   11.08   & 5.0   & 0.193  & 2.83  & 1.01 & $0.22_{-0.15} ^{+0.20}$\\[1.3ex]
1407  & 26.8        & 1779 & 252         & 0.07 & G & E      &   $12.0^{c}$  &  372 & 12.14        &   11.60    & 4.9   & $-0.046$ & 2.50   & 1.01 & $0.04_{-0.07} ^{+0.08}$\\[1.3ex]
2768  & 21.8        & 1353 & 206         & 0.57 & G & E/S0   &   12.3  &  107 & 6.37         &   11.21   & 3.8   & 0.133  & 2.75  & 0.88 & $0.50_{-0.15} ^{+0.15}$ \\[1.3ex]
$2974^{\dag}$  & 20.9        & 1887 & 231         & 0.36 & F & S0     &    9.3  &   26 & 3.06         &   10.93   & 4.3   & 0.262  & 2.92  & 0.97 & $0.31_{-0.17} ^{+0.12}$\\[1.3ex]
3115  &  9.4        &  663 & 248         & 0.66 & F & S0     &    $9.0^{d}$  &  150 & 1.66         &   10.93   & 4.7   & 0.262  & 2.92  & 0.83 & $0.94_{-0.16} ^{+0.15}$\\[1.3ex]
3377  & 10.9        &  690 & 135         & 0.33 & G & E      &    7.0  &  122 & 2.40         &   10.50    & 5.9   & 0.460   & 3.20  & 0.98 & $0.23_{-0.10} ^{+0.14}$\\[1.3ex]
$3607^{\ddag}$  & 22.2        &  942 & 229         & 0.13 & G & S0     &   10.3  &   36 & 5.19         &   11.39   & 5.3   & 0.051  & 2.63  & 1.01 & $0.18_{-0.15} ^{+0.22}$\\[1.3ex]% confirm the number
3608  & 22.3        & 1226 & 179         & 0.20 & G & E      &    9.9  &   36 & 4.64         &   11.03   & 5.3   & 0.216  & 2.86  & 1.01 & $0.21_{-0.18} ^{+0.26}$\\[1.3ex] % confirm the number
4278  & 15.6        &  620 & 228         & 0.09 & G & E      &   11.8  &  270 & 2.14         &   10.95   & 6.2   & 0.253  & 2.91  & 1.01 & $0.13_{-0.07} ^{+0.08}$\\[1.3ex]
4365  & 23.1        & 1243 & 253         & 0.24 & G & E      &   13.4  &  251 & 8.71         &   11.51   & 4.9   & $-0.005$ & 2.56  & 1.00  & $0.15_{-0.08} ^{+0.10}$\\[1.3ex]
4374  & 18.5        & 1017 & 284         & 0.05 & C & E      &   14.9  &   41 & 12.45        &   11.51   & 8.0   & $-0.005$ & 2.56  & 1.01 & $0.45_{-0.24} ^{+0.25}$\\[1.3ex]
$4459^{\dag}$  & 16.0        & 1192 & 170         & 0.21 & C & S0     &    7.0  &   36 & 3.75         &   10.98   & 5.4   & 0.239  & 2.89  & 1.01 & $0.20_{-0.18} ^{+0.13}$\\[1.3ex]
4473  & 15.2        & 2260 & 189         & 0.43 & C & E      &   13.1  &  106 & 2.23         &   10.96   & 5.0   & 0.248  & 2.91  & 0.95 & $0.23_{-0.11} ^{+0.15}$\\[1.3ex] % confirm the number
$4474^{\dag}$  & 15.5        & 1611 &  88         & 0.42 & C & E/S0   &   10.8  &   23 & 1.50         &   10.23   & 2.8   & 0.584  & 3.37  & 0.95& $2.04_{-1.89} ^{+1.55}$\\[1.3ex]
4486  & 16.7        & 1284 & 307         & 0.16 & C & E      &   17.7  &  702 & 7.01         &   11.62   & 5.1   & $-0.055$ & 2.49  & 1.01 & $0.14_{-0.05} ^{+0.06}$\\[1.3ex]
4494  & 16.6        & 1342 & 157         & 0.14 & G & E      &    8.0  &  107 & 4.23         &   11.02   & 4.5   & 0.221  & 2.87  & 1.01 & $0.51_{-0.14} ^{+0.15}$\\[1.3ex]
4526  & 16.4        &  617 & 233         & 0.76 & C & S0     &   11.0  &  107 & 2.58         &   11.26   & 3.6   & 0.110  & 2.72  & 0.77 & $0.61_{-0.24} ^{+0.23}$\\[1.3ex]
4564  & 15.9        & 1155 & 153         & 0.53 & C & E      &   11.8  &   27 & 1.14         &   10.58   & 3.2   & 0.423  & 3.14  & 0.90 & $1.80_{-0.33} ^{+0.51}$\\[1.3ex]
4649  & 16.5        & 1110 & 308         & 0.16 & C & E/S0   &   17.7  &  431 & 6.34         &   11.60    & 4.6   & $-0.046$ & 2.50  & 1.01 & $0.34_{-0.08} ^{+0.07}$\\[1.3ex]
$4697^{\dag}$  & 12.5        & 1252 & 180         & 0.32 & G & E      &   11.3  &   90 & 5.81         &   11.15   & 5.3   & 0.161  & 2.79  & 0.98  & $0.72_{-1.31} ^{+0.51}$\\[1.3ex]
5846  & 24.2        & 1712 & 231         & 0.08 & G & E/S0   &   17.7  &  190 & 10.54        &   11.46   & 5.2   & 0.018  & 2.59  & 1.01 & $0.08_{-0.07} ^{+0.09}$\\[1.3ex]
$5866^{\ddag}$   & 14.9        &  755 & 163         & 0.58 & G & S0     &    5.9  &   20 & 1.69         &   10.83   & 2.8   & 0.308  & 2.99  & 0.88 & $0.16_{-0.36} ^{+1.06}$\\[1.3ex]
7457  & 12.9        &  844 &  74         & 0.47 & F & S0     &    3.8  &   40 & 2.13         &   10.13   & 2.6   & 0.63   & 3.43  & 0.93 & $1.90_{-0.42} ^{+0.53}$\\[1.3ex]
\hline
1316  & $20.8$  & 1760 & $225$   & 0.28 & C & S0      &   $4.7^{e}$  &  $175^{h}$ & $8.57^{n}$   &   11.55   & $5.0^{n}$   & $-0.023$ & 2.53  & 1.00 & $0.60_{-0.11} ^{+0.10}$ \\[1.3ex]
1399  & $21.2$  & 1425 & $335$   & 0.01 & C & E      &   $11.0^{f}$  &  $514^{i}$ & $15.83^{o}$  &   11.50   & $11.1^{o}$  & 0.002  & 2.57  & 1.00 & $0.09_{-0.05} ^{+0.05}$\\[1.3ex]
4472  & $16.7$  &  981 & $288$   & 0.19 & C & E      &   17.7  &  $263^{j}$ & $15.74^{p}$  &  11.78   & $6.0^{p}$   & $-0.126$ & 2.39  & 1.00 & $0.19_{-0.08} ^{+0.07}$\\[1.3ex]
4594  & 9.77  & 1024 & $231$   & 0.46 & G & Sa$^{\star}$     &   $12.5^{g}$  &  $232^{m}$ & 3.41   &   11.41   & 3.2   & 0.041  & 2.62  & 1.00 & $0.13_{-0.10} ^{+0.08}$\\[1.3ex]
4636  & $14.3$  &  930 & $198$   & 0.23 & C & E      &   13.4  &  $386^{k}$ & $12.71^{p}$  &   11.17   & $5.7^{p}$   & $0.153$ & 2.77  & 1.00 & $0.35_{-0.08} ^{+0.08}$\\[1.3ex]
5128  & $3.8^{a}$   &  547 & $107$   & 0.11 & G & E/S0   &   $--$  &  $549^{l}$ & $2.21^{n}$   &   10.94   & $3.5^{n}$   & 0.258  & 2.92  & 1.00 & $0.17_{-0.07} ^{+0.07}$\\[1.3ex]%
%$\ 224$& $0.79^{b}$ & $-300$ & $157$   & 0.44 & G & Sa     &   $10.0^{i}$  &  $324^{p}$ & $1.59^{t}$   &   10.81   & $2.2^{t}$   & %0.319  & 3.0   & 1.0  & $0.72_{-0.09} ^{+0.11}$\\[1.3ex]
\hline
\end{tabular}
\begin{flushleft}
{\small Columns: (1) galaxy name: $\dag =$ bonus galaxies, $\ddag =$ SLUGGS galaxies not studied in \hyperlink{Alabi+16}{Alabi+16}; (2) distance; (3) systemic velocity; (4) central stellar velocity dispersion within 1~kpc; (5) ellipticity; (6) galaxy environment: F=field, G=group, C=cluster; (7) galaxy morphology, mostly sourced from \citet{Brodie_2014}, otherwise from \citet{Hyperleda}, although NGC~4594 (Sombrero galaxy) is classified as an Sa, we include it in our ETG sample; (8) average luminosity--weighted age of central (1~$\re$) stellar population, mostly from \citet{McDermid_2015} unless otherwise noted and listed below;  (9) number of globular clusters with kinematics data; (10) effective (half--light) radius; (11) stellar mass; (12) \Sersic/ $n$ index; (10)-(12) are mostly from \citet{Forbes_2016} unless otherwise noted and listed below (also see text for more details); (13) the power--law slope of the gravitational potential; (14) the power--law slope of the de--projected globular cluster density profile ; (15) normalising factor to correct for effect of galaxy flattening on dynamical mass estimate and (16) rotation dominance parameter for the globular cluster system [details on the derivation of columns (13)--(16) can be found in \hyperlink{Alabi+16}{Alabi+16}]. Galaxies below the horizontal line are not part of the SLUGGS survey, we have obtained their globular cluster kinematics data from the literature.
References : \begin{small} \textit{a}. \citet{Harris_2010}, \textit{b}. \citet{Rembold_2005}, \textit{c}. \citet{Spolaor_2008}, \textit{d}. \citet{Norris_2006}, \textit{e}. \citet{Koleva_2011}, \textit{f}. \citet{Trager_2000},  \textit{g}. \citet{Sanchez_2006},  \textit{h}. \citet{Richtler_2014}, \textit{i}. \citet{Schuberth_2010}, \textit{j}. \citet{Cote_2003}, \textit{k}. \citet{Schuberth_2012}, \textit{l}. \citet{Woodley_2010},  \textit{m}. \citet{Alves_2010}, \textit{n}. \citet{Sani_2010}, \textit{o}. \citet{Lasker_2014}, \textit{p}. \citet{Kormendy_2009}\end{small}}
\end{flushleft}
\end{table*}

\section{Data}
\label{data}
\subsection{New SLUGGS survey globular cluster kinematics data}
Here we introduce new Keck/DEIMOS globular cluster kinematics data for NGC~2974, NGC~4474, NGC~4459 and NGC~4697. These ETGs have log $(M_*/\Msun) \sim10- 11$. NGC~4697 was already studied in \hypertarget{Alabi+16}{\citet{Alabi_2016}}, hereafter \hyperlink{Alabi+16}{Alabi+16}, but the total number of GCs ($N_{\rm GC}$) with radial velocities was 20. Here, we use an improved dataset for NGC~4697, with $N_{\rm GC}=90$ and radial extent out to 4~$\re$. The remaining newly introduced lower mass ETGs have relatively sparse datasets but always with $N_{\rm GC} > 20$. This limit is set due to the $1/{\sqrt{N_{\rm GC}}}$ nature of the uncertainty on total mass estimate such that below
$N_{\rm GC}=20$, the uncertainty increases rapidly beyond 0.5 dex. We have therefore adopted $N_{\rm GC}=20$ as a sample size limit for our analysis. This is consistent with recent results from \citet{Toloba_2016} and was pointed out earlier in \citet{Strader_2011}.

\subsection{Globular cluster kinematics data from the literature}
We also include six ETGs (NGC~1316, NGC~1399, NGC~4472, NGC~4594/M104, NGC~4636 and NGC~5128) from the literature, with rich GC kinematics datasets ($N_{\rm GC} > 170$) obtained from various telescopes/instruments. For NGC~1316, we use the GC kinematics catalogue published in \citet{Richtler_2014}, obtained from VLT/FORS2. The data for NGC~1399 and NGC~4636 are from \citet{Schuberth_2010} and \citet{Schuberth_2012}, respectively, also obtained from VLT/FORS2. The data for NGC~4472 are from Keck/LRIS \citep{Cote_2003}. The data for NGC~4594 are from Keck/DEIMOS but with a different set-up that did not use the CaT features as we have done in the SLUGGS survey \citep{Alves_2010}. Finally, we also considered NGC~5128 (Centaurus A), using the GC kinematics catalog compiled in \citet{Woodley_2010}. The GC kinematics data for NGC~5128 have been obtained over two decades from an array of telescopes and instruments such as CTIO/SIT, Magellan/LDSS-2, VLT/VIMOS and CTIO/HYDRA.%Finally, we also considered NGC~224/M31 (the \textit{Andromeda} galaxy), using the GC kinematics catalogue compiled in \citet{Caldwell_2016} and we justify the inclusion of this spiral galaxy later in the text.

Unlike Keck/DEIMOS data obtained with the SLUGGS set-up, where the average uncertainty on the kinematics data is ${\sim}15~\kms$, these externally sourced data have an average uncertainty of ${\sim}50~\kms$. Higher uncertainty in the kinematics data tends to wash out subtle but important details in the velocity distributions e.g. kinematics substructures and higher velocity moments (see, for example, \citealt{Amorisco_2012}), needed to accurately determine galaxy dynamical mass. Also, larger uncertainties on the velocities tend to bias mass estimates, especially in galaxies with low velocity dispersions. This bias typically scales with $\Delta V_{i}/\sigma$, where $\Delta V_{i}$ and $\sigma$ are the uncertainties on individual velocity measurements and the central velocity dispersion of the galaxies, respectively.

\subsection{Size and stellar mass measurements}
Systematic deviation of galaxies from the size-stellar mass scaling
relation could create artificial tension between our dark matter fraction measurements and expectations from cosmological simulations. Here, we revisit and update (where necessary) the size and stellar mass measurements used in \hypertarget{Alabi+16}{\citet{Alabi_2016}}, as described below.

The galaxy sizes used in \hypertarget{Alabi+16}{\citet{Alabi_2016}} were taken from \citet{Brodie_2014} which they obtained mostly from the ATLAS$^{\rm 3D}$ survey \citep{Cappellari_2011} based on calibrated near-IR 2MASS and optical RC3 size measurements \citep{de_Vaucouleurs_1991}. Such measurements, however, underestimate galaxy sizes up to a factor of ${\sim}2-3$ for our most massive galaxies. This naturally leads to an underestimation of the measured $\fdm$ within 5~$\re$. Here, we partially correct for this underestimation by adopting the recently published size measurements from \citet{Forbes_2016} obtained using $3.6\ \mu$m \textit{Spitzer} imaging data. This correction only affects the most massive galaxies in our sample, with NGC~4374 being the most severely affected, where we have now revised its $\re$ upwards by a factor of ${\sim}$3 to ${\sim}12$ kpc.

In \hypertarget{Alabi+16}{\citet{Alabi_2016}}, we obtained stellar masses for our galaxy sample from their 2MASS absolute $K$-band magnitude, assuming a stellar-mass-light ratio ${M/L}_K=1$. While this assumption is consistent with a Kroupa/Chabrier stellar initial mass function, it does not account for variations in stellar age or metallicity. The implication of this assumption becomes critical in ETGs whose stellar population is dominated by younger stars, since stellar ${M/L}_K$ for ETGs is known to be age-dependent. For example, at a mean stellar age of ${\sim}6$ Gyr, assuming solar metallicity and Padova isochrones,  \citet{Rock_2016} recently reported a stellar ${M/L}_K{\sim}0.6$. This corresponds to a ${\sim}0.2$~dex decrease in stellar mass and a ${\sim}0.1$ increase in $\fdm$ which may change our earlier conclusions, especially in galaxies with low $\fdm$. One way of addressing this concern would be to apply an age-weighted correction to our previous ${M/L}_K=1$ assumption. \citet{Forbes_2016} applied this correction to their stellar mass estimates (they assumed a Kroupa IMF) from the
$3.6\mu$m \textit{Spitzer} data. We note that there is a one-to-one correspondence with the 2MASS $K$-band stellar mass estimates if they are also corrected for stellar age-variations.

We use the homogeneously measured effective radii and total stellar mass estimates for the 27 galaxies we have in common with \citet{Forbes_2016}. For the remaining galaxies, we obtain their stellar mass estimates from 2MASS absolute $K$-band magnitudes, correcting for sky over-subtraction \citep{Scott_2013} and stellar age-variation \citep{Rock_2016}. Their effective radii are obtained from the literature studies which used $3.6\mu$m \textit{Spitzer} imaging data and a similar effective radius measurement procedure as in \citet{Forbes_2016}. Our final sample of 32 galaxies now spans a log $(M_*/\Msun)$ range of $10.1-11.8$, with the typical uncertainty on $M_*$ and $\re$ being ${\sim}0.1$ dex and ${\sim}0.15$ dex, respectively. Table \ref{tab:gal_prop} contains a summary of the salient properties of the galaxies used in this work.

%\footnotetext[1]{http://leda.univ-lyon1.fr/}

\section{Method and results}
\label{methods}
\subsection{Total mass estimates and dark matter fractions within 5~$\re$}
\label{mass_est}
We use the tracer mass estimator (TME) of \citet{Watkins_2010} to obtain the total mass estimates and subsequently the dark matter fractions for our galaxy sample, following the implementation described in \hyperlink{Alabi+16}{Alabi+16}. We give a brief summary of the implementation below and encourage interested readers to see \hyperlink{Alabi+16}{Alabi+16} for more details.

The TME assumes that the discrete dynamical tracers follow a power-law density distribution when de-projected and a power-law description for the gravitational potential. The pressure-supported mass within a sphere with projected radius $R$, $M_{\rm p}(<R)$, is then
\begin{equation}
\label{mass_est_eqn}
M_{\rm p}(<R) \propto f(\alpha, \beta, \gamma) \times \left\langle V^{2}_{\rm los}R^{\alpha} \right\rangle
\end{equation}
where $\alpha$ and $\gamma$ are the power-law slopes of the gravitational potential and the de-projected GC number density profile, respectively. $\beta$ is the Binney anisotropy parameter \citep{Binney_1987}, assumed to be constant with radius. For each galaxy, we use eqns. 8 and 10 from \hyperlink{Alabi+16}{Alabi+16} to obtain $\alpha$ and $\gamma$, respectively. In \hyperlink{Alabi+16}{Alabi+16}, we obtained $\alpha$ from the logarithmic slope of the circular velocity curves of realistic galaxies (from the cosmological simulation reported in \citealt{Wu_2014}) at 5~$\re$. We obtained $\gamma$ from a compilation of GC densityprofiles from the literature. We found that $\alpha$ and $\gamma$ are well approximated by:
\begin{equation}\label{eq:fit_alpha}
\alpha = (-0.46\pm0.06) \times {\rm log}(M_*/\Msun) + (5.29\pm0.68)
\end{equation}
and
\begin{equation}\label{eq:fit_gamma}
\gamma = ( -0.63 \pm 0.17) \times {\rm log}(M_*/\Msun) + (9.81 \pm 1.94)
\end{equation}
, respectively. The most massive galaxies in our sample have $\alpha\sim0$, i.e. they are nearly isothermal, and the less massive ones are more Keplarian. Also, $\gamma$ defined this way is such that $2 \leq \gamma \leq 4$, with the most massive galaxies well described by shallow GC density profiles.

\begin{table*}
\footnotesize
\captionsetup{width=.96\linewidth}
{\small
\caption{Summary of mass estimates and dark matter fractions assuming different anisotropy. The results shown here have been obtained using the tracer mass estimator and a stellar $M/L$ that accounts for stellar age-variation. $M_{\rm p}$ is the pressure-supported mass and has been corrected for the effect of galaxy flattening. $M_{\rm rot}$ is the rotationally-supported mass.
$M_{\rm tot}$ is the total mass after correcting for galaxy flattening, rotation in the GC system and the presence of kinematics substructures. $f_{\rm DM}$ is the dark matter fraction. We list masses enclosed within spheres of radius 5~$\re$ and $R_{\rm max}$, the maximum galactocentric radius where we have GC kinematics data. Note that the kinematics data for NGC~4374, NGC~4472, NGC~4636 and NGC~4697 do not extend out to 5~$\re$.} \label{tab:mass_summary}}
\begin{tabular}{lcccccccccl}
\hline\hline
Galaxy & $\beta$ & $M_{\rm rot}(<5{\re})$ & $M_{\rm p}(<5{\re})$ & $M_{\rm tot}(<5{\re})$ & $f_{\rm DM} (<5{\re})$ & $R_{\rm max}$ & $M_{\rm rot}(<R_{\rm max})$ & $M_{\rm p}(<R_{\rm max})$ & $M_{\rm tot}(<R_{\rm max})$& $f_{\rm DM}(<R_{\rm max})$ \\
$\rm [NGC]$ &  & $[10^{10} \Msun]$ & $[10^{11} \Msun]$ & $[10^{11} \Msun]$ &  & [$\re$] & $[10^{10} \Msun]$ & $[10^{11} \Msun]$ & $[10^{11} \Msun]$ & 	\\
\hline
  $720 $ & $   0 $ & $1.7 \pm 0.3$ & $2.4  \pm 0.6$ & $2.6  \pm 0.5$ & $0.36 \pm 0.20$ & $22.91$ & $7.6  \pm 1.6$ & $ 11.4 \pm 2.2$ & $12.2 \pm 2.0$ & $0.85 \pm 0.04$ \\
  $    $ & $ 0.5 $ &               & $2.4  \pm 0.6$ & $2.6  \pm 0.5$ & $0.34 \pm 0.21$ & $     $ &                & $ 11.1 \pm 2.2$ & $11.9 \pm 2.0$ & $0.84 \pm 0.04$ \\
  $    $ & $-0.5 $ &               & $2.5  \pm 0.6$ & $2.6  \pm 0.6$ & $0.37 \pm 0.18$ & $     $ &                & $ 11.6 \pm 2.3$ & $12.4 \pm 2.2$ & $0.85 \pm 0.04$ \\
  \hline
  $821 $ & $   0 $ & $2.1 \pm 0.4$ & $4.3  \pm 0.8$ & $4.5  \pm 0.8$ & $0.81 \pm 0.05$ & $8.06 $ & $3.4  \pm 0.6$ & $ 5.6  \pm 1.0$ & $5.9  \pm 1.0$ & $0.85 \pm 0.04$ \\
  $    $ & $ 0.5 $ &               & $4.4  \pm 0.8$ & $4.6  \pm 0.9$ & $0.82 \pm 0.05$ & $     $ &                & $ 5.8  \pm 1.0$ & $6.1  \pm 1.0$ & $0.85 \pm 0.04$ \\
  $    $ & $-0.5 $ &               & $4.2  \pm 0.8$ & $4.4  \pm 0.8$ & $0.81 \pm 0.05$ & $     $ &                & $ 5.5  \pm 1.0$ & $5.8  \pm 1.0$ & $0.85 \pm 0.04$ \\
  \hline
  $1023$ & $   0 $ & $2.4 \pm 0.4$ & $1.4  \pm 0.3$ & $1.6  \pm 0.2$ & $0.47 \pm 0.12$ & $16.15$ & $7.7  \pm 1.4$ & $ 3.2  \pm 0.6$ & $3.9  \pm 0.5$ & $0.76 \pm 0.05$ \\
  $    $ & $ 0.5 $ &               & $1.5  \pm 0.3$ & $1.7  \pm 0.2$ & $0.49 \pm 0.12$ & $     $ &                & $ 3.3  \pm 0.6$ & $4.1  \pm 0.5$ & $0.76 \pm 0.05$ \\
  $    $ & $-0.5 $ &               & $1.4  \pm 0.3$ & $1.6  \pm 0.2$ & $0.46 \pm 0.13$ & $     $ &                & $ 3.1  \pm 0.5$ & $3.9  \pm 0.5$ & $0.75 \pm 0.05$ \\
  \hline
  $1400$ & $   0 $ & $0.3 \pm 0.1$ & $2.2  \pm 0.5$ & $2.2  \pm 0.6$ & $0.54 \pm 0.21$ & $22.55$ & $1.6  \pm 0.6$ & $ 7.1  \pm 1.3$ & $7.2  \pm 1.3$ & $0.84 \pm 0.04$ \\
  $    $ & $ 0.5 $ &               & $2.2  \pm 0.6$ & $2.3  \pm 0.6$ & $0.54 \pm 0.2 $ & $     $ &                & $ 7.2  \pm 1.3$ & $7.4  \pm 1.3$ & $0.84 \pm 0.04$ \\
  $    $ & $-0.5 $ &               & $2.2  \pm 0.5$ & $2.2  \pm 0.5$ & $0.53 \pm 0.17$ & $     $ &                & $ 7.0  \pm 1.3$ & $7.1  \pm 1.2$ & $0.83 \pm 0.04$ \\
  \hline
  $1407$ & $   0 $ & $0.1 \pm 0.1$ & $18.6 \pm 1.5$ & $18.6 \pm 1.5$ & $0.82 \pm 0.04$ & $9.54 $ & $0.2  \pm 0.1$ & $ 36.4 \pm 2.7$ & $36.4 \pm 2.7$ & $0.90 \pm 0.02$ \\
  $    $ & $ 0.5 $ &               & $16.5 \pm 1.4$ & $16.5 \pm 1.4$ & $0.79 \pm 0.04$ & $     $ &                & $ 32.2 \pm 2.4$ & $32.2 \pm 2.4$ & $0.88 \pm 0.02$ \\
  $    $ & $-0.5 $ &               & $19.7 \pm 1.6$ & $19.7 \pm 1.7$ & $0.83 \pm 0.03$ & $     $ &                & $ 38.5 \pm 2.8$ & $38.5 \pm 2.7$ & $0.90 \pm 0.02$ \\
  \hline
  $2768$ & $   0 $ & $4.3 \pm 0.4$ & $6.3  \pm 1.1$ & $6.8  \pm 0.9$ & $0.78 \pm 0.05$ & $11.36$ & $9.8  \pm 0.9$ & $ 12.0 \pm 1.9$ & $13.0 \pm 1.7$ & $0.88 \pm 0.02$ \\
  $    $ & $ 0.5 $ &               & $6.3  \pm 1.0$ & $6.7  \pm 0.9$ & $0.78 \pm 0.05$ & $     $ &                & $ 11.9 \pm 1.9$ & $12.9 \pm 1.7$ & $0.88 \pm 0.03$ \\
  $    $ & $-0.5 $ &               & $6.4  \pm 1.1$ & $6.8  \pm 0.9$ & $0.79 \pm 0.05$ & $     $ &                & $ 12.1 \pm 1.9$ & $13.1 \pm 1.7$ & $0.88 \pm 0.02$ \\
  \hline
  $2974$ & $   0 $ & $0.5 \pm 0.1$ & $0.8  \pm 0.4$ & $0.8  \pm 0.4$ & $0.08 \pm 0.45$ & $10.69$ & $1.0  \pm 0.3$ & $ 4.9  \pm 1.5$ & $5.0  \pm 1.5$ & $0.84 \pm 0.09$ \\
  $    $ & $ 0.5 $ &               & $0.8  \pm 0.4$ & $0.9  \pm 0.4$ & $0.12 \pm 0.44$ & $     $ &                & $ 5.1  \pm 1.6$ & $5.2  \pm 1.6$ & $0.84 \pm 0.11$ \\
  $    $ & $-0.5 $ &               & $0.8  \pm 0.4$ & $0.8  \pm 0.4$ & $0.06 \pm 0.48$ & $     $ &                & $ 4.8  \pm 1.5$ & $4.9  \pm 1.4$ & $0.83 \pm 0.09$ \\
  \hline
  $3115$ & $   0 $ & $3.4 \pm 0.6$ & $1.7  \pm 0.3$ & $2.1  \pm 0.3$ & $0.64 \pm 0.08$ & $17.6 $ & $11.8 \pm 2.1$ & $ 4.4  \pm 0.6$ & $5.6  \pm 0.6$ & $0.85 \pm 0.03$ \\
  $    $ & $ 0.5 $ &               & $1.8  \pm 0.4$ & $2.2  \pm 0.3$ & $0.65 \pm 0.08$ & $     $ &                & $ 4.6  \pm 0.7$ & $5.8  \pm 0.6$ & $0.86 \pm 0.03$ \\
  $    $ & $-0.5 $ &               & $1.7  \pm 0.3$ & $2.0  \pm 0.3$ & $0.63 \pm 0.09$ & $     $ &                & $ 4.3  \pm 0.6$ & $5.5  \pm 0.6$ & $0.85 \pm 0.03$ \\
  \hline
  $3377$ & $   0 $ & $0.1 \pm 0.1$ & $0.7  \pm 0.1$ & $0.7  \pm 0.1$ & $0.62 \pm 0.09$ & $11.37$ & $0.3  \pm 0.1$ & $ 1.3  \pm 0.2$ & $1.3  \pm 0.2$ & $0.77 \pm 0.05$ \\
  $    $ & $ 0.5 $ &               & $0.8  \pm 0.1$ & $0.8  \pm 0.1$ & $0.66 \pm 0.09$ & $     $ &                & $ 1.4  \pm 0.2$ & $1.5  \pm 0.2$ & $0.80 \pm 0.04$ \\
  $    $ & $-0.5 $ &               & $0.6  \pm 0.1$ & $0.7  \pm 0.1$ & $0.60 \pm 0.10$ & $     $ &                & $ 1.2  \pm 0.2$ & $1.2  \pm 0.2$ & $0.76 \pm 0.05$ \\
  \hline
  $3607$ & $   0 $ & $0.4 \pm 0.1$ & $2.4  \pm 0.7$ & $2.5  \pm 0.7$ & $0.16 \pm 0.44$ & $16.76$ & $1.3  \pm 0.5$ & $ 10.2 \pm 2.4$ & $10.3 \pm 2.4$ & $0.77 \pm 0.09$ \\
  $    $ & $ 0.5 $ &               & $2.3  \pm 0.6$ & $2.4  \pm 0.6$ & $0.11 \pm 0.38$ & $     $ &                & $ 9.6  \pm 2.3$ & $9.7  \pm 2.2$ & $0.75 \pm 0.08$ \\
  $    $ & $-0.5 $ &               & $2.5  \pm 0.7$ & $2.6  \pm 0.6$ & $0.18 \pm 0.38$ & $     $ &                & $ 10.4 \pm 2.5$ & $10.5 \pm 2.6$ & $0.77 \pm 0.24$ \\
  \hline
  $3608$ & $   0 $ & $0.5 \pm 0.2$ & $3.9  \pm 1.1$ & $4.0  \pm 1.2$ & $0.77 \pm 0.19$ & $6.82 $ & $0.7  \pm 0.3$ & $ 4.4  \pm 1.2$ & $4.5  \pm 1.2$ & $0.78 \pm 0.12$ \\
  $    $ & $ 0.5 $ &               & $4.1  \pm 1.2$ & $4.1  \pm 1.2$ & $0.78 \pm 0.11$ & $     $ &                & $ 4.5  \pm 1.2$ & $4.6  \pm 1.2$ & $0.79 \pm 0.1 $ \\
  $    $ & $-0.5 $ &               & $3.9  \pm 1.1$ & $3.9  \pm 1.1$ & $0.77 \pm 0.2 $ & $     $ &                & $ 4.3  \pm 1.2$ & $4.4  \pm 1.2$ & $0.78 \pm 0.12$ \\
  \hline
  $4278$ & $   0 $ & $0.1 \pm 0.1$ & $2.6  \pm 0.3$ & $2.7  \pm 0.4$ & $0.72 \pm 0.07$ & $16.81$ & $0.5  \pm 0.2$ & $ 6.5  \pm 0.6$ & $6.6  \pm 0.6$ & $0.87 \pm 0.02$ \\
  $    $ & $ 0.5 $ &               & $2.8  \pm 0.4$ & $2.8  \pm 0.4$ & $0.73 \pm 0.06$ & $     $ &                & $ 6.8  \pm 0.6$ & $6.9  \pm 0.6$ & $0.88 \pm 0.02$ \\
  $    $ & $-0.5 $ &               & $2.6  \pm 0.3$ & $2.6  \pm 0.3$ & $0.72 \pm 0.07$ & $     $ &                & $ 6.3  \pm 0.6$ & $6.4  \pm 0.5$ & $0.87 \pm 0.02$ \\
  \hline
  $4365$ & $   0 $ & $1.5 \pm 0.2$ & $16.5 \pm 1.6$ & $16.6 \pm 1.6$ & $0.83 \pm 0.03$ & $8.79 $ & $2.6  \pm 0.3$ & $ 29.7 \pm 2.6$ & $30.0 \pm 2.6$ & $0.90 \pm 0.02$ \\
  $    $ & $ 0.5 $ &               & $15.1 \pm 1.4$ & $15.2 \pm 1.5$ & $0.82 \pm 0.04$ & $     $ &                & $ 27.1 \pm 2.4$ & $27.4 \pm 2.4$ & $0.89 \pm 0.02$ \\
  $    $ & $-0.5 $ &               & $17.3 \pm 1.6$ & $17.4 \pm 1.6$ & $0.84 \pm 0.03$ & $     $ &                & $ 31.0 \pm 2.8$ & $31.3 \pm 2.7$ & $0.90 \pm 0.02$ \\
  \hline
  $4374$ & $   0 $ & $16.1\pm 1.4$ & $20.9 \pm 4.8$ & $22.5 \pm 4.9$ & $0.89 \pm 0.04$ & $3.51 $ & $-$           &                $-$  &         $-$        &		$-$ 				\\
  $    $ & $ 0.5 $ &               & $19.1 \pm 4.4$ & $20.7 \pm 4.4$ & $0.88 \pm 0.04$ & $     $ &    $-$             &		$-$ 			& 	$-$ 			 & 		$-$ 				\\
  $    $ & $-0.5 $ &               & $21.8 \pm 5.0$ & $23.4 \pm 5.3$ & $0.90 \pm 0.04$ & $     $ &     $-$            &		$-$ 		 &	$-$ 			 &			$-$ 			\\
  \hline
  $4459$ & $   0 $ & $0.2 \pm 0.1$ & $2.1  \pm 0.6$ & $2.2  \pm 0.6$ & $0.62 \pm 0.32$ & $7.27 $ & $0.3  \pm 0.1$ & $ 2.6  \pm 0.7$ & $2.7  \pm 0.7$ & $0.68 \pm 0.14$ \\
  $    $ & $ 0.5 $ &               & $2.2  \pm 0.6$ & $2.2  \pm 0.6$ & $0.64 \pm 0.17$ & $     $ &                & $ 2.7  \pm 0.7$ & $2.8  \pm 0.7$ & $0.69 \pm 0.13$ \\
  $    $ & $-0.5 $ &               & $2.1  \pm 0.6$ & $2.1  \pm 0.6$ & $0.62 \pm 0.39$ & $     $ &                & $ 2.6  \pm 0.7$ & $2.6  \pm 0.7$ & $0.67 \pm 0.13$ \\
  \hline
  $4473$ & $   0 $ & $0.2 \pm 0.1$ & $1.6  \pm 0.3$ & $1.6  \pm 0.3$ & $0.51 \pm 0.15$ & $15.51$ & $0.8  \pm 0.4$ & $ 3.6  \pm 0.5$ & $3.7  \pm 0.5$ & $0.76 \pm 0.05$ \\
  $    $ & $ 0.5 $ &               & $1.7  \pm 0.3$ & $1.7  \pm 0.3$ & $0.53 \pm 0.12$ & $     $ &                & $ 3.7  \pm 0.6$ & $3.8  \pm 0.5$ & $0.77 \pm 0.05$ \\
  $    $ & $-0.5 $ &               & $1.5  \pm 0.3$ & $1.6  \pm 0.3$ & $0.50 \pm 0.14$ & $     $ &                & $ 3.5  \pm 0.5$ & $3.6  \pm 0.5$ & $0.75 \pm 0.05$ \\
  \hline
\end{tabular}
\end{table*}
\addtocounter{table}{-1}
\begin{table*}
\footnotesize
{\small \caption{Continued.}}
\begin{tabular}{lcccccccccl}
\hline\hline
Galaxy & $\beta$ & $M_{\rm rot}(<5{\re})$ & $M_{\rm p}(<5{\re})$ & $M_{\rm tot}(<5{\re})$ & $f_{\rm DM} (<5{\re})$ & $R_{\rm max}$ & $M_{\rm rot}(<R_{\rm max})$ & $M_{\rm p}(<R_{\rm max})$ & $M_{\rm tot}(<R_{\rm max})$& $f_{\rm DM}(<R_{\rm max})$ \\
$\rm [NGC]$ &  & $[10^{10} \Msun]$ & $[10^{11} \Msun]$ & $[10^{11} \Msun]$ &  & [$\re$] & $[10^{10} \Msun]$ & $[10^{11} \Msun]$ & $[10^{11} \Msun]$ & 	\\
  \hline
  $4474$ & $   0 $ & $1.2 \pm 1.2$ & $0.4  \pm 0.2$ & $0.6  \pm 0.2$ & $0.71 \pm 0.42$ & $17.18$ & $4.2  \pm 4.6$ & $ 0.9  \pm 0.3$ & $1.3  \pm 0.6$ & $0.87 \pm 0.38$ \\
  $    $ & $ 0.5 $ &               & $0.5  \pm 0.2$ & $0.6  \pm 0.3$ & $0.75 \pm 0.41$ & $     $ &                & $ 1.0  \pm 0.4$ & $1.4  \pm 0.6$ & $0.88 \pm 0.44$ \\
  $    $ & $-0.5 $ &               & $0.4  \pm 0.2$ & $0.5  \pm 0.2$ & $0.70 \pm 0.45$ & $     $ &                & $ 0.8  \pm 0.3$ & $1.2  \pm 0.6$ & $0.86 \pm 0.44$ \\
  \hline
  $4486$ & $   0 $ & $1.7 \pm 0.2$ & $25.5 \pm 1.8$ & $25.7 \pm 1.9$ & $0.86 \pm 0.03$ & $28.55$ & $9.8  \pm 1.0$ & $ 148.0\pm 8.4$ & $149.0\pm 8.4$ & $0.97 \pm 0.01$ \\
  $    $ & $ 0.5 $ &               & $22.5 \pm 1.6$ & $22.7 \pm 1.6$ & $0.84 \pm 0.03$ & $     $ &                & $ 130.0\pm 7.3$ & $131.0\pm 7.6$ & $0.97 \pm 0.01$ \\
  $    $ & $-0.5 $ &               & $27.1 \pm 1.9$ & $27.3 \pm 1.9$ & $0.87 \pm 0.03$ & $     $ &                & $ 157.0\pm 8.8$ & $158.0\pm 9.2$ & $0.97 \pm 0.01$ \\
  \hline
  $4494$ & $   0 $ & $1.3 \pm 0.1$ & $1.4  \pm 0.2$ & $1.5  \pm 0.2$ & $0.4  \pm 0.14$ & $7.95 $ & $2.0  \pm 0.2$ & $ 1.9  \pm 0.3$ & $2.1  \pm 0.3$ & $0.53 \pm 0.11$ \\
  $    $ & $ 0.5 $ &               & $1.4  \pm 0.2$ & $1.6  \pm 0.2$ & $0.41 \pm 0.13$ & $     $ &                & $ 1.9  \pm 0.3$ & $2.2  \pm 0.3$ & $0.54 \pm 0.10$ \\
  $    $ & $-0.5 $ &               & $1.4  \pm 0.2$ & $1.5  \pm 0.2$ & $0.39 \pm 0.14$ & $     $ &                & $ 1.9  \pm 0.3$ & $2.1  \pm 0.3$ & $0.52 \pm 0.10$ \\
  \hline
  $4526$ & $   0 $ & $2.1 \pm 0.6$ & $2.7  \pm 0.8$ & $2.9  \pm 0.6$ & $0.43 \pm 0.18$ & $16.75$ & $7.0  \pm 2.1$ & $ 6.5  \pm 1.3$ & $7.2  \pm 1.1$ & $0.75 \pm 0.05$ \\
  $    $ & $ 0.5 $ &               & $2.6  \pm 0.8$ & $2.8  \pm 0.6$ & $0.41 \pm 0.20$ & $     $ &                & $ 6.4  \pm 1.3$ & $7.1  \pm 1.0$ & $0.74 \pm 0.06$ \\
  $    $ & $-0.5 $ &               & $2.7  \pm 0.8$ & $2.9  \pm 0.6$ & $0.43 \pm 0.19$ & $     $ &                & $ 6.6  \pm 1.4$ & $7.3  \pm 1.1$ & $0.75 \pm 0.06$ \\
  \hline
  $4564$ & $   0 $ & $1.8 \pm 0.4$ & $0.7  \pm 0.2$ & $0.9  \pm 0.2$ & $0.59 \pm 0.22$ & $11.25$ & $4.0  \pm 0.9$ & $ 0.9  \pm 0.3$ & $1.3  \pm 0.3$ & $0.71 \pm 0.09$ \\
  $    $ & $ 0.5 $ &               & $0.8  \pm 0.3$ & $0.9  \pm 0.3$ & $0.62 \pm 0.15$ & $     $ &                & $ 1.0  \pm 0.3$ & $1.4  \pm 0.3$ & $0.73 \pm 0.09$ \\
  $    $ & $-0.5 $ &               & $0.6  \pm 0.2$ & $0.8  \pm 0.2$ & $0.57 \pm 0.2 $ & $     $ &                & $ 0.9  \pm 0.3$ & $1.3  \pm 0.3$ & $0.70 \pm 0.08$ \\
\hline
  $4649$ & $   0 $ & $4.6 \pm 0.3$ & $12.9 \pm 1.0$ & $13.4 \pm 1.0$ & $0.74 \pm 0.05$ & $20.2 $ & $18.6 \pm 1.3$ & $ 53.9 \pm 3.8$ & $55.8 \pm 3.8$ & $0.93 \pm 0.01$ \\
  $    $ & $ 0.5 $ &               & $11.4 \pm 0.9$ & $11.9 \pm 0.9$ & $0.71 \pm 0.05$ & $     $ &                & $ 47.8 \pm 3.3$ & $49.7 \pm 3.4$ & $0.92 \pm 0.01$ \\
  $    $ & $-0.5 $ &               & $13.7 \pm 1.1$ & $14.2 \pm 1.1$ & $0.75 \pm 0.05$ & $     $ &                & $ 57.0 \pm 4.0$ & $58.9 \pm 3.9$ & $0.93 \pm 0.01$ \\
  \hline
  $4697$ & $   0 $ & $3.1 \pm 0.2$ & $3.6  \pm 0.6$ & $3.9  \pm 0.6$ & $0.7  \pm 0.07$ & $4.0  $ & $-$           &                $-$  &         $-$        &		$-$ 				\\
  $    $ & $ 0.5 $ &               & $3.6  \pm 0.6$ & $3.9  \pm 0.6$ & $0.7  \pm 0.07$ & $     $ & $-$           &                $-$  &         $-$        &		$-$ 				\\
  $    $ & $-0.5 $ &               & $3.5  \pm 0.6$ & $3.9  \pm 0.5$ & $0.7  \pm 0.07$ & $     $ & $-$           &                $-$  &         $-$        &		$-$ 				\\
  \hline
  $5846$ & $   0 $ & $0.4 \pm 0.1$ & $18.4 \pm 1.9$ & $18.4 \pm 1.9$ & $0.87 \pm 0.03$ & $8.99 $ & $0.8  \pm 0.2$ & $ 32.5 \pm 3.3$ & $32.6 \pm 3.3$ & $0.92 \pm 0.02$ \\
  $    $ & $ 0.5 $ &               & $17.0 \pm 1.8$ & $17.0 \pm 1.8$ & $0.86 \pm 0.03$ & $     $ &                & $ 30.1 \pm 3.0$ & $30.2 \pm 3.1$ & $0.91 \pm 0.02$ \\
  $    $ & $-0.5 $ &               & $19.0 \pm 2.0$ & $19.0 \pm 2.0$ & $0.87 \pm 0.03$ & $     $ &                & $ 33.7 \pm 3.4$ & $33.8 \pm 3.4$ & $0.92 \pm 0.02$ \\
\hline
  $5866$ & $   0 $ & $0.1 \pm 0.1$ & $1.0  \pm 0.5$ & $1.0  \pm 0.4$ & $0.34 \pm 0.45$ & $8.85 $ & $0.1  \pm 0.6$ & $ 1.2  \pm 0.5$ & $1.2  \pm 0.4$ & $0.44 \pm 0.39$ \\
  $    $ & $ 0.5 $ &               & $1.0  \pm 0.5$ & $1.0  \pm 0.4$ & $0.38 \pm 0.45$ & $     $ &                & $ 1.3  \pm 0.5$ & $1.3  \pm 0.5$ & $0.48 \pm 0.42$ \\
  $    $ & $-0.5 $ &               & $0.9  \pm 0.5$ & $0.9  \pm 0.4$ & $0.32 \pm 0.44$ & $     $ &                & $ 1.1  \pm 0.5$ & $1.2  \pm 0.4$ & $0.42 \pm 0.38$ \\
\hline
  $7457$ & $   0 $ & $1.8 \pm 0.2$ & $0.9  \pm 0.3$ & $1.1  \pm 0.2$ & $0.88 \pm 0.04$ & $6.61 $ & $2.4  \pm 0.3$ & $ 0.9  \pm 0.3$ & $1.2  \pm 0.2$ & $0.89 \pm 0.04$ \\
  $    $ & $ 0.5 $ &               & $1.1  \pm 0.3$ & $1.2  \pm 0.3$ & $0.90 \pm 0.04$ & $     $ &                & $ 1.1  \pm 0.3$ & $1.3  \pm 0.3$ & $0.90 \pm 0.03$ \\
  $    $ & $-0.5 $ &               & $0.8  \pm 0.2$ & $1.0  \pm 0.2$ & $0.87 \pm 0.04$ & $     $ &                & $ 0.8  \pm 0.2$ & $1.1  \pm 0.2$ & $0.88 \pm 0.04$ \\
\hline
  $1316$ & $   0 $ & $12.0\pm 0.4$ & $14.6 \pm 2.1$ & $15.8 \pm 2.0$ & $0.81 \pm 0.04$ & $9.45 $ & $22.6 \pm 0.8$ & $ 29.3 \pm 3.6$ & $31.6 \pm 3.4$ & $0.89 \pm 0.02$ \\
  $    $ & $ 0.5 $ &               & $13.2 \pm 1.9$ & $14.4 \pm 1.9$ & $0.79 \pm 0.05$ & $     $ &                & $ 26.4 \pm 3.2$ & $28.7 \pm 3.4$ & $0.88 \pm 0.02$ \\
  $    $ & $-0.5 $ &               & $15.4 \pm 2.2$ & $16.6 \pm 2.1$ & $0.82 \pm 0.04$ & $     $ &                & $ 30.8 \pm 3.7$ & $33.1 \pm 3.6$ & $0.90 \pm 0.02$ \\
\hline
  $1399$ & $   0 $ & $0.2 \pm 0.1$ & $40.8 \pm 2.7$ & $40.8 \pm 2.6$ & $0.94 \pm 0.01$ & $6.06 $ & $0.2  \pm 0.1$ & $ 49.4 \pm 3.2$ & $49.4 \pm 3.1$ & $0.95 \pm 0.01$ \\
  $    $ & $ 0.5 $ &               & $37.4 \pm 2.5$ & $37.4 \pm 2.4$ & $0.94 \pm 0.01$ & $     $ &                & $ 45.3 \pm 2.9$ & $45.3 \pm 3.0$ & $0.95 \pm 0.01$ \\
  $    $ & $-0.5 $ &               & $42.5 \pm 2.8$ & $42.5 \pm 2.9$ & $0.94 \pm 0.01$ & $     $ &                & $ 51.5 \pm 3.3$ & $51.5 \pm 3.4$ & $0.95 \pm 0.01$ \\
\hline
  $4472$ & $   0 $ & $0.2 \pm 0.1$ & $29.6 \pm 2.8$ & $29.6 \pm 2.8$ & $0.85 \pm 0.04$ & $3.12 $ & $-$           &                $-$  &         $-$        &		$-$ 				\\
  $    $ & $ 0.5 $ &               & $24.6 \pm 2.3$ & $24.6 \pm 2.4$ & $0.82 \pm 0.04$ & $     $ & $-$           &                $-$  &         $-$        &		$-$ 				\\
  $    $ & $-0.5 $ &               & $32.2 \pm 3.1$ & $32.2 \pm 3.0$ & $0.86 \pm 0.03$ & $     $ & $-$           &                $-$  &         $-$        &		$-$ 				\\
  \hline
  $4594 $ & $   0 $ & $0.3 \pm 0.1$ & $5.6  \pm 0.6$ & $5.7  \pm 0.6$ & $0.58 \pm 0.08$ & $10.0 $ & $0.6  \pm 0.3$ & $ 9.9  \pm 0.9$ & $9.9  \pm 0.9$ & $0.75 \pm 0.05$ \\
  $    $ & $ 0.5 $ &               & $5.3  \pm 0.5$ & $5.3  \pm 0.6$ & $0.56 \pm 0.09$ & $     $ &                & $ 9.3  \pm 0.9$ & $9.4  \pm 0.9$ & $0.73 \pm 0.05$ \\
  $    $ & $-0.5 $ &               & $5.8  \pm 0.6$ & $5.8  \pm 0.6$ & $0.59 \pm 0.08$ & $     $ &               & $ 10.2 \pm 1.0$ & $10.3 \pm 1.0$ & $0.75 \pm 0.05$ \\
\hline
  $4636$ & $   0 $ & $1.0 \pm 0.1$ & $9.5  \pm 0.7$ & $9.6  \pm 0.7$ & $0.88 \pm 0.02$ & $3.48 $ & $-$           &                $-$  &         $-$        &		$-$ 				\\
  $    $ & $ 0.5 $ &               & $9.5  \pm 0.7$ & $9.6  \pm 0.7$ & $0.88 \pm 0.03$ & $     $ & $-$           &                $-$  &         $-$        &		$-$ 				\\
  $    $ & $-0.5 $ &               & $9.5  \pm 0.7$ & $9.6  \pm 0.7$ & $0.88 \pm 0.02$ & $     $ & $-$           &                $-$  &         $-$        &		$-$ 				\\
\hline
  $5128$ & $   0 $ & $0.1 \pm 0.1$ & $1.7  \pm 0.2$ & $1.7  \pm 0.2$ & $0.53 \pm 0.09$ & $21.73$ & $0.6  \pm 0.3$ & $ 7.4  \pm 0.5$ & $7.5  \pm 0.5$ & $0.88 \pm 0.02$ \\
  $    $ & $ 0.5 $ &               & $1.7  \pm 0.2$ & $1.7  \pm 0.2$ & $0.54 \pm 0.09$ & $     $ &                & $ 7.6  \pm 0.5$ & $7.7  \pm 0.5$ & $0.89 \pm 0.02$ \\
  $    $ & $-0.5 $ &               & $1.6  \pm 0.2$ & $1.6  \pm 0.1$ & $0.52 \pm 0.09$ & $     $ &                & $ 7.3  \pm 0.5$ & $7.4  \pm 0.5$ & $0.88 \pm 0.02$ \\
\hline
%  $224  $ & $   0 $ & $2.5 \pm 0.3$ & $1.6  \pm 0.1$ & $1.8  \pm 0.1$ & $0.66 \pm 0.06$ & $16.53$ & $8.3  \pm 0.9$ & $ 3.9  \pm 0.3$ & $4.8  \pm 0.3$ & $0.87 \pm 0.02$ \\
%  $    $ & $ 0.5 $ &               & $1.7  \pm 0.2$ & $1.9  \pm 0.2$ & $0.68 \pm 0.06$ & $     $ &                & $ 4.2  \pm 0.3$ %& $5.0  \pm 0.3$ & $0.87 \pm 0.02$ \\
%  $    $ & $-0.5 $ &               & $1.5  \pm 0.1$ & $1.8  \pm 0.1$ & $0.66 \pm 0.06$ & $     $ &                & $ 3.8  \pm 0.3$ %& $4.7  \pm 0.3$ & $0.86 \pm 0.02$ \\
%\hline
\end{tabular}
\end{table*}

Since the TME assumes that the GC system is pressure-supported, we first subtract the contribution of rotation, $V_{\rm rot}$, from the line-of-sight velocity, $V_{\rm los}$, before evaluating Equation \ref{mass_est_eqn}. $V_{\rm rot}$ is obtained by fitting an inclined-disc model to the GC kinematics data. Our total mass estimate, $M_{\rm tot}$, is then evaluated as the sum of the rotationally-supported mass, $M_{\rm rot}$, and the pressure-supported mass, $M_{\rm p}$. The contribution from rotation to the total mass is usually small,	 ${\sim}6$ per cent. We evaluate $M_{\rm tot}$ assuming that the orbital anisotropy of the GC system is either strongly radial, mildly tangential or isotropic, i.e, $\beta=\pm0.5, 0$, respectively. Since our mass estimates are largely insensitive to the choice of $\beta$ (deviating by $\leq10$ per cent), we adopt $M_{\rm tot}$ obtained when $\beta=0$, i.e. the velocity distribution is isotropic. As a further test, we have also obtained $M_{\rm tot}$ assuming a more extreme velocity anisotropy of $\beta=-1$. This is motivated by recent results from dynamical studies and cosmological simulations where such anisotropies were reported (e.g. \citealt{Pota_2015, Zhang_2015, Rottgers_2014}). Even with such extreme anisotropies, the maximum deviation in total mass is less than $20$ per cent, never producing a shift in dark matter fraction greater than $0.1$ (see Appendix \ref{extreme_ani} for more details). Note that we have also applied small corrections to $M_{\rm p}$ to account for galaxy flattening and projection effects (on average, ${\sim}5$ per cent) and non-equilibrium conditions (${\sim}20$ per cent when kinematics substructures are identified in the GC system).

For the newly introduced galaxies and at the lower $M_*$ end, the average fractional uncertainty on $M_{\rm tot}$ is 0.4 dex. At the high $M_*$ end of our sample, there is no significant difference between the fractional uncertainties on $M_{\rm tot}$ for galaxies with externally sourced kinematics data, compared to galaxies with Keck/DEIMOS data. We are unable to identify any kinematics substructures that may be in the GC systems of these newly introduced galaxies. This is due to the large uncertainties on the individual radial velocities. Since our kinematics data extend well beyond 5~$\re$ for most galaxies in our sample, we also obtain $M_{\rm tot}$ enclosed within the maximum radial extent, $R_{\rm max}$, of our data. We obtain the dark matter fraction, $f_{\rm DM}$, as $1 - M_{*}(<R)/M_{\rm tot}(<R)$ where we assume that all of the baryonic mass within $R$ in our galaxies is stellar in nature. We describe the total stellar mass within 5~$\re$ with de-projected \Sersic/ profiles, using the \Sersic/ indices from Table \ref{tab:gal_prop}. Table \ref{tab:mass_summary} contains a summary of the total masses and dark matter fractions enclosed within 5~$\re$ and $R_{\rm max}$.

\begin{figure*}
    \includegraphics[height=0.3\textheight, width=0.48\textwidth]{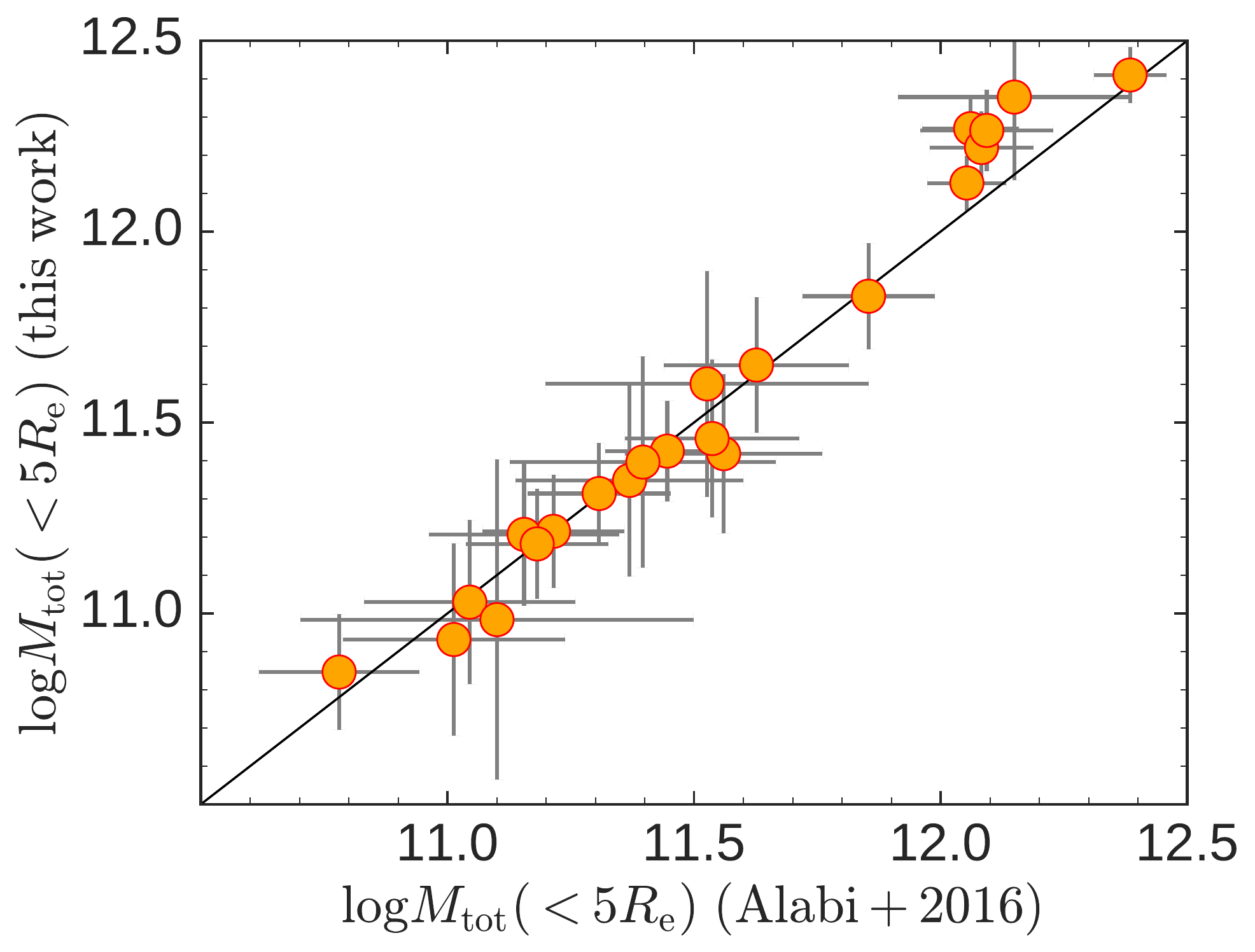}\hspace{0.01\textwidth}%
    \includegraphics[height=0.3\textheight, width=0.48\textwidth]{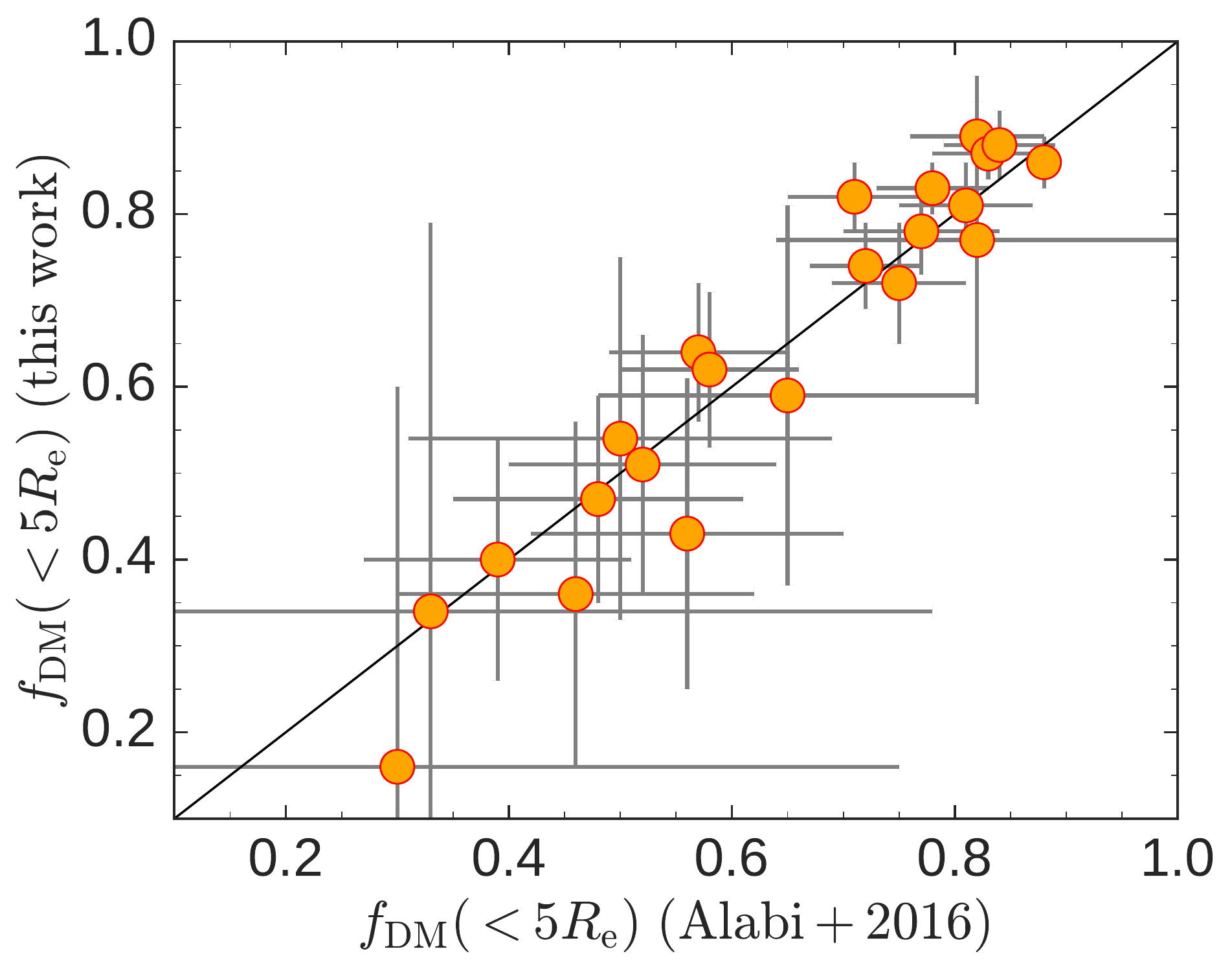}\hspace{0.01\textwidth}
	\caption{\label{fig:cmp_Res} Comparison of total mass estimates (\textit{left panel}) and dark matter fractions (\textit{right panel}) obtained using different galaxy sizes and stellar masses (see text for details). Note that we have excluded NGC~4697 from these plots due to the additional change to its kinematics data.}
\bigskip
    \includegraphics[width=1.0\textwidth]{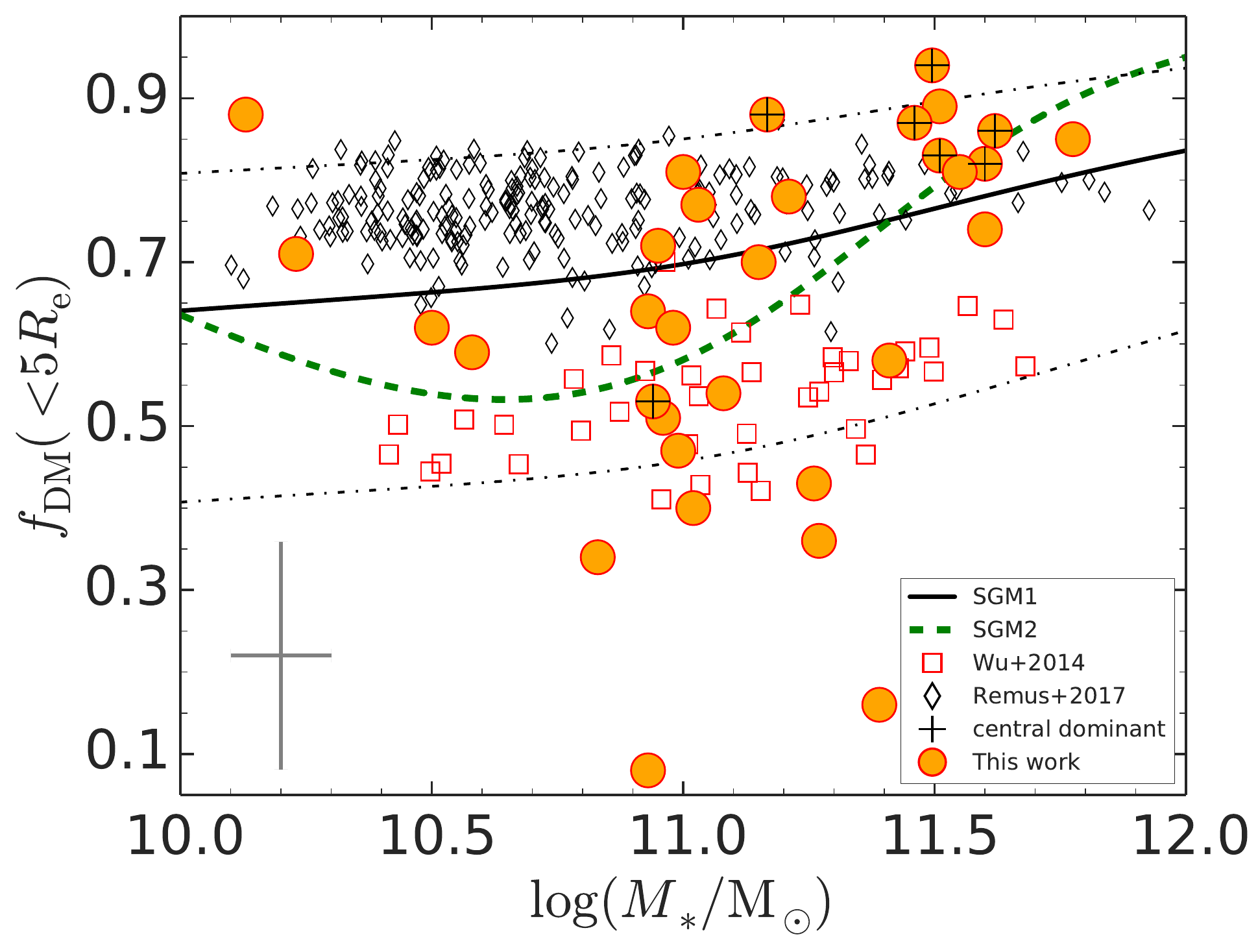}\hspace{0.01\textwidth}\\
	\caption{\label{fig:predict_fdm1} Measured dark matter fraction, $f_{\rm DM}$, within 5 effective radii ($\re$) versus the total stellar mass, $M_*$ assuming $\beta=0$, i.e. isotropic velocity distribution. The solid black line (SGM1) shows the predicted dark matter fraction within 5~$\re$ assuming \textit{Planck} cosmology and Kroupa IMF for a simple galaxy model based on scaling relations for early-type galaxies. The dot-dashed black lines are the $1\sigma$ scatter in the predicted dark matter fractions from the adopted $\re-M_*$ relation. We also show results from the cosmological hydrodynamical simulations reported in \citet{Wu_2014} and \citet{Remus_2016} for comparison. The dashed green line (SGM2) is the predicted dark matter fraction from a simple galaxy model using a fit to galaxy sizes and stellar masses in Table \ref{tab:gal_prop}. The orange-coloured circles and the lower left representative errorbar are for our galaxy sample. Galaxies with log $(M_*/\Msun) \sim 11$ have a larger spread in their measured $f_{\rm DM}$, with a few of them having $f_{\rm DM}$ lower than predicted by any cosmological model. At any stellar mass, central dominant galaxies (marked with crosses) mostly have higher $f_{\rm DM}$.}
\end{figure*}

For the galaxies originally studied in \hyperlink{Alabi+16}{Alabi+16}, we compare results in Figure \ref{fig:cmp_Res} to see how the newly adopted sizes and stellar masses affect both parameters of interest. We remark that while these new galaxy parameters result in changes to the total mass and dark matter fraction estimates within 5~$\re$ on a galaxy by galaxy basis, their overall distributions, which we will present shortly, for our galaxy sample remain unchanged. In particular, at the high $M_*$ end, galaxies are now more massive within the 5~$\re$ aperture and a few ${\sim}L^*$ galaxies also have slightly lowered dark matter fractions compared to \hyperlink{Alabi+16}{Alabi+16}.

Figure \ref{fig:predict_fdm1} shows the $f_{\rm DM}$ versus $M_*$ for our galaxy sample, assuming $\beta=0$ and a Kroupa IMF. For most galaxies in our combined sample, the DM content already dominates the mass budget at 5~$\re$ with the DM domination increasing as we probe beyond 5~$\re$ into the outer haloes. There is a wide diversity in the measured $f_{\rm DM}$ within 5~$\re$, ranging from $0.1-0.9$, generally increasing with galaxy stellar mass, with some ${\rm log~}(M_*/\Msun) \sim 11$ galaxies having very low $f_{\rm DM}$, i.e. $\leq0.4$, less than what a simple galaxy model predicts. This trend persists for a variety of stellar $M/L$ assumptions, assumed slope of the gravitational potential and orbital anisotropies. The large spread in $f_{\rm DM}$ is driven exclusively by ${\rm log~}(M_*/\Msun) \sim 11$ ETGs. The updated list of galaxies with $f_{\rm DM}$ within 5~$\re$ lower than the prediction from our simple galaxy model now consist of NGC~720, NGC~2974, NGC~3607, NGC~4494, NGC~4526 and NGC~5866. A complete inventory of our galaxy sample shows that 2 out of 5 field galaxies, 3 out of 15 group galaxies and 1 out of 12 cluster galaxies have low dark matter fractions. This is the same as 2 out of 16 ellipticals, 4 out of 10 lenticulars and none of the 6 galaxies with ambiguous morphological classification having low dark matter fractions.

The results we have obtained for NGC~2974 may appear to be at odds with that reported in the HI study of \citet{Weijmans_2008} where they obtained a dark matter fraction of 0.55 within 5~$\re$ and a $M_{\rm tot} (< 5~\re) = 2.7 \times 10^{11}\ \Msun$ for their maximal disk model, compared to our values (for the isotropic orbit case) of $0.08 \pm 0.45$ and $M_{\rm tot} (<5~\re) = 0.8 \pm 0.4 \times 10^{11}\ \Msun$. The stellar mass and galaxy size used in the two studies are comparable. If we assume $\alpha = 0$ (i.e. the isothermal case as suggested by their flat rotation curve), we would derive a somewhat higher dark matter fraction of 0.15. The main limitation on our DM fraction measurement for NGC~2974 is likely the limited sample size of 26 GCs; this results in a relatively large error. Despite different datasets and modelling assumptions in both studies, our low DM fraction is consistent within ${\sim}1\sigma$ of the Weijmans et al. value. Therefore, these intriguing results of very low dark matter fractions would need confirmation by future studies to rule out that it is not the poor number statistics that is driving these results, although we also find $f_{\rm DM} {\sim} 0.8$ in NGC~3608 with $M_*{\sim}10^{11}\ \Msun$ and similarly sparse GC kinematics data.

%Key assumptions in the Weijmans et al. modelling include corrections for the inclination and asymmetric drift (the latter of which increases their rotation velocity by ${\sim}120\ \kms$).%

\begin{figure}
    \includegraphics[width=0.48\textwidth]{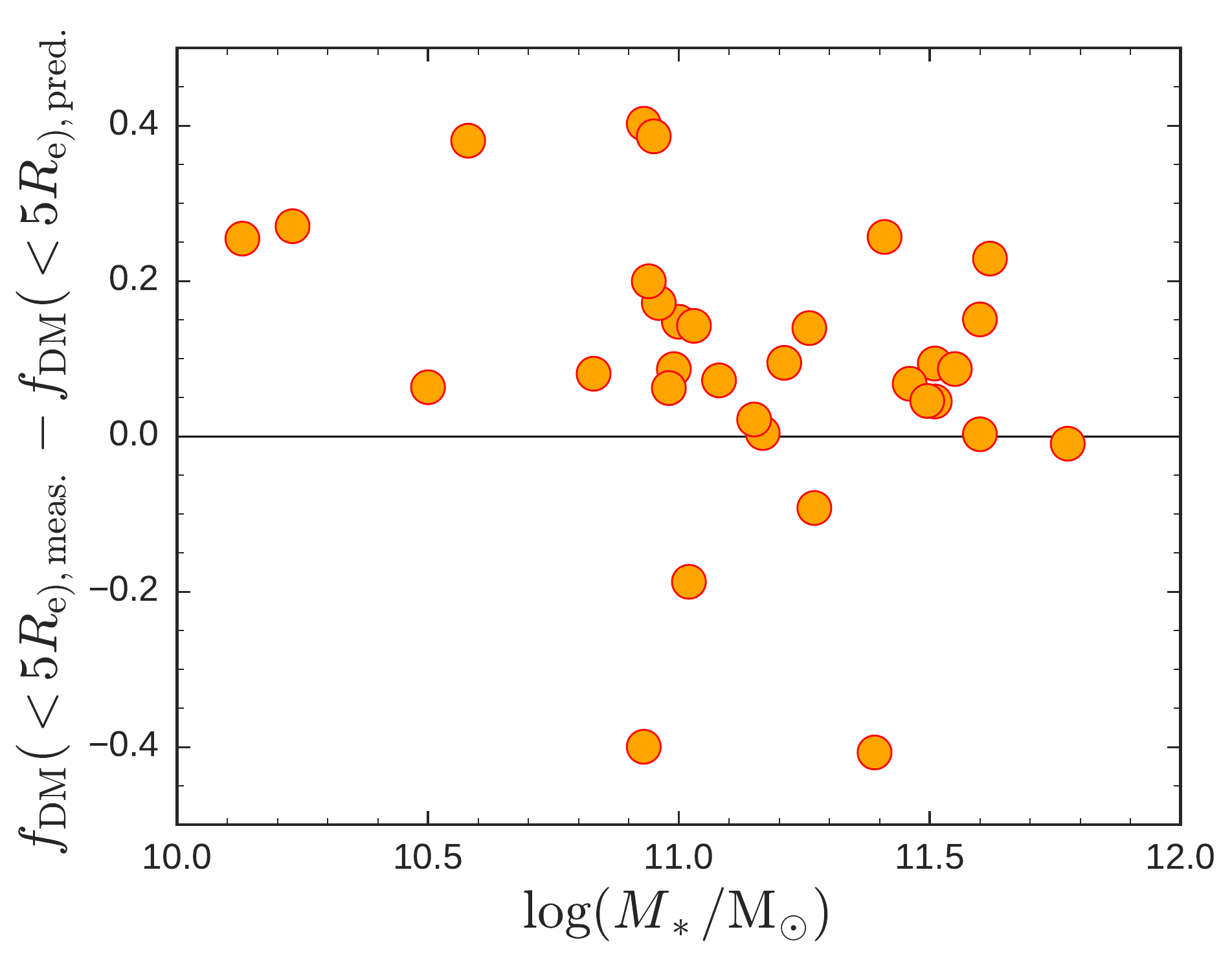}\hspace{0.01\textwidth}\\
	\caption{\label{fig:cmp_residuals} Residuals between observed and predicted dark matter fractions, assuming a \textit{Planck} cosmology and using galaxy sizes, stellar masses and \Sersic/ indices from Table \ref{tab:gal_prop} in the simple galaxy model , i.e. SGM2.}
\end{figure}

To properly understand our observed $f_{\rm DM}$, we compare our results with expectations from simple galaxy models and the cosmological hydrodynamical simulations reported in \hypertarget{Wu+14}{\citet[][\hyperlink{Wu+14}{hereafter Wu+14}]{Wu_2014}} and in \hypertarget{Remus+17}{\citet[][\hyperlink{Remus+17}{hereafter Remus+17}]{Remus_2016}}. The simple galaxy model, labelled SGM1, (details of which are presented in \hyperlink{Alabi+16}{Alabi+16}) does not account directly for processes that are believed to alter the distribution of baryons and non-baryons in present-day ETGs during their evolution. It takes as input the $\re-M_*$, $M_*-\mhalo$ and $\mhalo-\chalo$ scaling relations from the literature, adopts the \textit{Planck} cosmology and predicts the $f_{\rm DM}$ and the $M_{\rm tot}$ within 5~$\re$ for a given $M_*$. On the other hand, the mass distribution is explicitly modified in the cosmological simulation of \hyperlink{Wu+14}{Wu+14} via dissipative and/or non-dissipative processes during galaxy assembly. However, they did not include feedback models from AGN and/or SN winds in their simulations. The immediate effect of this is that their galaxies contain more baryons relative to dark matter when compared to conventional $M_*-\mhalo$ scaling relations for ETGs. If we allow for a factor of $2-3$ excess stellar mass at any defined $\mhalo$ in our SGM1 model, we adequately predict the $\fdm$ reported in \hyperlink{Wu+14}{Wu+14}, as shown in Figure \ref{fig:predict_fdm1}. The predicted $f_{\rm DM}$ from our SGM1 is then reduced by ${\sim}0.1$ at all galaxy stellar mass. The cosmological simulation of \hyperlink{Remus+17}{Remus+17} is an improvement on \hyperlink{Wu+14}{Wu+14} in that they have included a feedback model which accounts for AGN and SN winds effects. However, at low $M_*$, their galaxies are larger than the expectations from conventional $\re-M_*$ scaling relations for ETGs (e.g. \citealt{Lange_2015}). This probably indicates AGN feedback that is too strong in their lower stellar mass regime. Lastly, we construct a variant of our SGM1 where we use the galaxy sizes, stellar masses and \Sersic/ indices listed in Table \ref{tab:gal_prop}, and compare the predicted dark matter fractions with what we have measured within 5~$\re$. We show the comparison in Figure \ref{fig:cmp_residuals} as a function of galaxy stellar mass. We also show this simple galaxy model (labelled SGM2) in Figure \ref{fig:predict_fdm1} where we have used a double-power law fit to the galaxy sizes and stellar masses in Table \ref{tab:gal_prop}, and show that it is consistent (within $1\sigma$) with the dark matter fractions predicted by SGM1. This galaxy model captures the shape of our measured dark matter fractions better than SGM1.

\subsection{Average Dark matter density}
We obtain the average enclosed DM density, $\rhodm$, within a sphere with radius $R$, as in \citet{Thomas_2009} using
\begin{equation}
\label{rhodef}
\rhodm = \frac{M_{\rm DM}(<R)}{(4\pi/3) R^3}
\end{equation}
\begin{figure*}
    \includegraphics[width=1.0\textwidth]{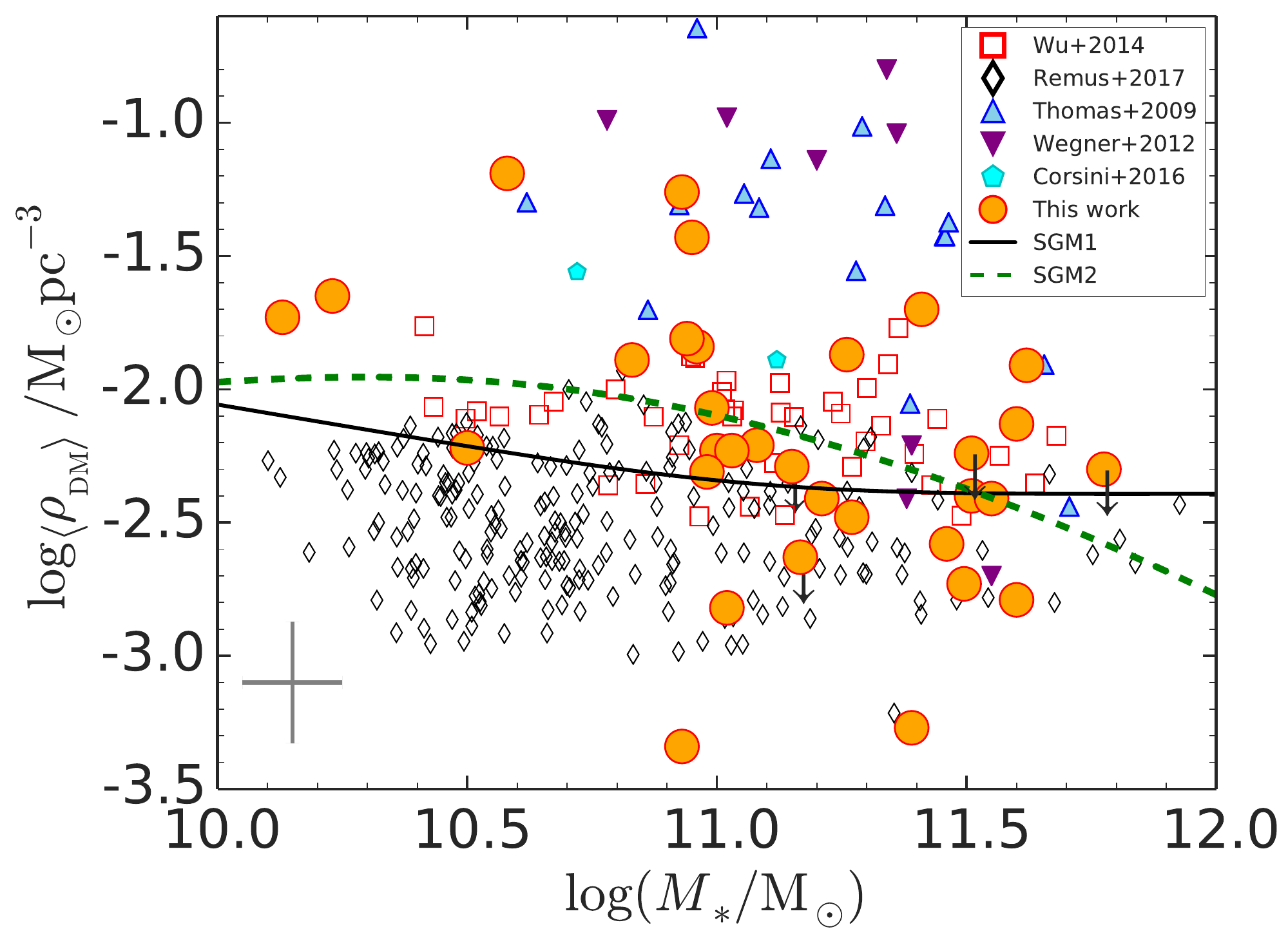}\hspace{0.01\textwidth}\\
	\caption{\label{fig:aveden} Average dark matter density within 5~$\re$, $\rhodm$, versus the stellar mass for our galaxy sample. Results from \citet{Thomas_2009} for ETGs in the Coma cluster, \citet{Wegner_2012} for eight galaxies in the nearby, but poor Abell~262 cluster and \citet{Corsini_2016} for two galaxies in low-density environments,  within 2~$\re$ are also shown. The offset from our data is due to differences in the apertures used. Galaxies with sub-5~$\re$ kinematics data are marked with downward-pointing arrows. The solid and dashed lines are the predicted average dark matter densities within 5~$\re$ from our simple galaxy models, i.e., SGM1 and SGM2, respectively. We have also included results from the cosmological simulations of \citet{Wu_2014} and \citet{Remus_2016}. Average DM density within 5~$\re$ (as well as within any other aperture) decreases mildly with total stellar mass, with a larger scatter around log $(M_*/\Msun){\sim}11$.}
\end{figure*}
where $M_{\rm DM}$ is the enclosed DM mass, evaluated as $M_{\rm tot}(<R)-M_*(<R)$. Again we have followed the approach in \hyperlink{Alabi+16}{Alabi+16}, where we assume that all the baryonic matter within our 5~$\re$ aperture is in the stellar component.

%from \hyperlink{Wu+14}{Wu+14} and \hyperlink{Remus+17}{Remus+17} (also obtained within their 5~$\re$) as a function of galaxy stellar mass and size

Figure \ref{fig:aveden} shows the log~$\rhodm$ within 5~$\re$ for our galaxy sample. We also show similar data from \citet{Thomas_2009}, \citet{Wegner_2012} and \citet{Corsini_2016}, but obtained within 2~$\re$ for several ETGs in the Coma cluster, the nearby Abell~262 cluster and low-density environments, respectively. The offset between these literature results and our measurements is due to the difference between the apertures used. The general trend, regardless of the adopted aperture, is for $\rhodm$ to decrease with $M_*$, with an enhanced scatter around log~$(M_*/\Msun){\sim}11$. A wide range of average DM densities is possible at any stellar mass, in agreement with theory, where galaxies are expected to have diverse mass assembly histories at any stellar mass. We do not see any difference between the mean average densities for the lenticulars or ellipticals in our galaxy sample, neither do we see any significant trend with galaxy environment. The increasing trend earlier observed in $f_{\rm DM}$ as a function of mass is now reversed when $\rhodm$ is compared with $M_*$. This is due to the steep increase of $\re$ with $M_*$, such that in the more massive galaxies, our fiducial radius now encloses more dark matter within a much more increased volume, hence the lowered density. The increased offset at the low stellar mass end between our measurements and the predictions from \hyperlink{Remus+17}{Remus+17} is due to the relatively large galaxies produced in their simulations.

\subsection{Inferring DM halo properties from dynamical measurements}
Next we turn to one of the main questions we wish to address in this work, i.e., given dynamical mass measurements at some fiducial radii (in our case, 5~$\re$), which are smaller than the typical scale radii in DM haloes, can we \textit{reliably} infer the structural properties of these haloes, i.e., $\mhalo$ (virial mass), $\chalo$ (halo concentration) and $z_{\rm form}$ (the halo assembly epoch)?

DM haloes can generally be described by Navarro--Frenk--White (NFW) profiles \citep{Navarro_1996} where the average enclosed DM density, $\rhodm$, can be expressed as
\begin{equation}
\label{rhodef2}
\rhodm(<R) =  \frac{200}{x^3} \frac{{\rm ln}(1+cx)-cx/(cx+1)}{{\rm ln}(1+c)-c/(c+1)} \rho_{\rm crit}
\end{equation}
where $x\equiv R/\rhalo$ and $c\equiv\chalo\equiv\rhalo/r_{\rm s}$ ($\rhalo$ is the virial radius and $\chalo$ is the NFW DM halo concentration), with $\rs$ being the scale radius of the DM halo, corresponding to the radius where the logarithmic slope of the DM density profile is $-2$. The virial over-density, $\Delta_{\rm 200}$, is $200$ times the critical density, $\rho_{\rm crit} = 1.37 \times 10^2\ \Msun\ kpc^{-3}$. Since the RHS of equation \ref{rhodef2} has 2 unknown quantities ($\chalo$ and $\rhalo$), it cannot be solved uniquely without some extra assumptions. We solve Equation \ref{rhodef2} using the mock galaxies in our SGM1 with the following steps:

\begin{itemize}
\item {For mock galaxies with $M_*$ identical to our galaxy sample, obtain galaxy sizes using the $\re-M_*$ relation from \citet{Lange_2015}.}
\item {Use the $\re-n$ relation from \citet{Graham_2013} to obtain \Sersic/ indices (\citealt{Sersic_1968}) for each mock galaxy.}
\item {Calculate the total $M_*$ enclosed within 5~$\re$ with the de--projected \Sersic/ mass profile \citep{Terzic_2005}.}
\item {Calculate $\mhalo$ for each mock galaxy using the $M_* - \mhalo$ relation for ETGs from \citet{Dutton_2010}, which assumes a \textit{Planck} cosmology. Note that $\rhalo$ follows directly from $\mhalo$.}
\item {Calculate the total DM mass, $M_{\rm DM}$ enclosed within 5~$\re$, from the cumulative NFW DM only profile.}
\item {Calculate the ratio of $\mhalo$ to $M_{\rm tot}$ within 5~$\re$ for each mock galaxy.}
\item {Assume that for any $M_*$, given the ratio of $\mhalo$ to $M_{\rm tot}$ within 5$\re$ obtained from our mock galaxies, we can extrapolate our measured $M_{\rm tot}$ within 5~$\re$ to obtain the corresponding $\mhalo$. }
\end{itemize}

As a sanity test, we have compared the ratio of our measured $M_{\rm tot}$ within 5 $\re$ to the extrapolated $\mhalo$ for our galaxy sample with that from \hyperlink{Wu+14}{Wu+14} (see Appendix for the plot). The scaling ratios we have used are consistent with those inferred from their simulations, bearing in mind that they have over-produced stars by factors of $2-3$.
%\textcolor{red}{Should I also discuss model galaxies from the SAGE/Millennium simulations. Note I used $\re-M_*$ relation to get the sizes of their mock galaxies.}
Armed with the \textit{extrapolated} $\mhalo$, which we transform to $\rhalo$ using $\mhalo=4\pi\Delta_{\rm 200}\rho_{\rm crit}{\rhalo}^3/3$, we then numerically solve the non-linear Equation \ref{rhodef2}, and obtain $\chalo$ given our measured $\rhodm$ within 5~$\re$. We have used Monte Carlo methods to propagate uncertainties from our mass measurements and the scaling relations at every stage of this analysis. From our results, $\chalo$ increases steeply as $\rhodm$ increases, such that an order of magnitude increase in $\rhodm$ corresponds to a factor of ${\sim}3$ increase in $\chalo$.

The final step in estimating the epoch of halo assembly is to transform our inferred halo concentrations into halo assembly epochs. Tools that efficiently do this transformation are now readily available. For each estimated $\chalo$, we use the \textsc{COMMAH} package from \citet{Correa_2015}, which is based on NFW DM profiles, to obtain the corresponding halo formation redshift, $z_{\rm form}$, given the halo concentration, $\chalo$, while adopting the \textit{Planck} cosmology. This NFW parametrisation matches well with our earlier preference to describe our DM haloes with NFW profiles, rather than with cored logarithmic DM haloes or other alternate parametrisations and thus enables us to calibrate our dark matter densities directly into halo assembly epochs. In the COMMAH package, $z_{\rm form}$ is the epoch when the virial mass of any progenitor halo is equivalent to the mass within its present day scale radius. Table \ref{tab:halo_prop} contains a summary of the inferred halo properties for our galaxy sample. From Table \ref{tab:halo_prop}, low $\rhodm$ corresponds to more recently formed haloes and vice-versa. Galaxies with $\chalo \leq 2$ (NGC~4494, NGC~2974 and NGC~3607), have invalid $z_{\rm form}$ from the \textsc{COMMAH} package and for these, we fix their halo formation epoch at $z_{\rm form}\sim0.1$. These are all galaxies with low $\fdm$ within 5~$\re$.

\begin{table*}
\begin{minipage}{0.45\textwidth}
\centering
%\tablewidth=\textwidth
{\small \caption{Summary of inferred halo parameters.} \label{tab:halo_prop}}
\begin{tabular}{@{}l l l l l l } %c r c r r c c c c c c c c c c
\hline
Galaxy & log\ $\mhalo$  & log\ $\rhodm$ & $\chalo$ & $z_{\rm form}$ & $z_{\rm stars}$\\
$\rm [NGC]$ & [$\Msun$]  & [$\Msun{pc}^{-3}$]  &  &  & 	 \\
\hline
$720 $ & $12.83  \pm 0.38$ & $-2.48 \pm 0.65$ & $3.0   \pm 2.0 $ & $0.6   \pm 0.8 $ & $1.0 $ \\
$821 $ & $12.99  \pm 0.29$ & $-2.23 \pm 0.22$ & $6.4   \pm 1.5 $ & $2.3   \pm 0.6 $ & $2.4 $ \\
$1023$ & $ 12.56 \pm 0.28$ & $-2.07 \pm 0.37$ & $6.2   \pm 2.2 $ & $2.3   \pm 0.9 $ & $4.1 $ \\
$1400$ & $ 12.71 \pm 0.36$ & $-2.21 \pm 0.49$ & $5.4   \pm 2.6 $ & $1.9   \pm 1.1 $ & $>10 $ \\
$1407$ & $ 13.82 \pm 0.40$ & $-2.79 \pm 0.11$ & $2.3   \pm 0.3 $ & $0.5   \pm 0.3 $ & $3.5 $ \\
$2768$ & $ 13.22 \pm 0.33$ & $-2.41 \pm 0.18$ & $4.7   \pm 0.9 $ & $1.6   \pm 0.5 $ & $4.1 $ \\
$2974$ & $ 12.25 \pm 0.55$ & $-3.34 \pm 6.34$ & $0.01  \pm 6.6 $ & $0.1$ & $1.4 $ \\
$3115$ & $ 12.65 \pm 0.25$ & $-1.26 \pm 0.23$ & $17.9  \pm 3.7 $ & $6.9   \pm 1.4 $ & $1.3 $ \\
$3377$ & $ 12.19 \pm 0.20$ & $-2.22 \pm 0.27$ & $5.7   \pm 1.6 $ & $2.0   \pm 0.7 $ & $0.8 $ \\
$3607$ & $ 12.85 \pm 0.45$ & $-3.27 \pm 1.98$ & $0.2   \pm 2.0 $ & $0.1$ & $1.9 $ \\
$3608$ & $ 12.95 \pm 0.38$ & $-2.23 \pm 0.39$ & $6.2   \pm 2.5 $ & $2.2   \pm 1.0 $ & $1.7 $ \\
$4278$ & $ 12.76 \pm 0.26$ & $-1.43 \pm 0.20$ & $15.7  \pm 2.9 $ & $6.1   \pm 1.1 $ & $3.2 $ \\
$4365$ & $ 13.73 \pm 0.39$ & $-2.40 \pm 0.12$ & $4.4   \pm 0.5 $ & $1.4   \pm 0.3 $ & $10.0 $ \\
$4374$ & $ 13.86 \pm 0.43$ & $-2.24 \pm 0.25$ & $5.6   \pm 1.4 $ & $2.0   \pm 0.7 $ & $10.0 $ \\
$4459$ & $ 12.67 \pm 0.37$ & $-2.31 \pm 0.47$ & $5.1   \pm 2.5 $ & $1.7   \pm 1.0 $ & $0.8 $ \\
$4473$ & $ 12.55 \pm 0.29$ & $-1.84 \pm 0.40$ & $8.5   \pm 3.2 $ & $3.2   \pm 1.3 $ & $7.5 $ \\
$4474$ & $ 12.13 \pm 0.46$ & $-1.65 \pm 0.63$ & $11.6  \pm 7.2 $ & $4.5   \pm 2.5 $ & $2.2 $ \\
$4486$ & $ 13.97 \pm 0.40$ & $-1.91 \pm 0.09$ & $7.8   \pm 0.7 $ & $2.9   \pm 0.3 $ & $>10 $ \\
$4494$ & $ 12.53 \pm 0.28$ & $-2.82 \pm 0.45$ & $1.9   \pm 1.1 $ & $0.1$ & $1.0 $ \\
$4526$ & $ 12.87 \pm 0.38$ & $-1.87 \pm 0.54$ & $7.2   \pm 3.5 $ & $2.7   \pm 1.4 $ & $2.4 $ \\
$4564$ & $ 12.27 \pm 0.29$ & $-1.19 \pm 0.44$ & $18.6  \pm 7.3 $ & $7.2   \pm 2.7 $ & $3.2 $ \\
$4649$ & $ 13.68 \pm 0.40$ & $-2.13 \pm 0.11$ & $5.7   \pm 0.6 $ & $2.0   \pm 0.3 $ & $>10 $ \\
$4697$ & $ 12.96 \pm 0.32$ & $-2.29 \pm 0.23$ & $5.4   \pm 1.3 $ & $1.9   \pm 0.6 $ & $2.6 $ \\
$5846$ & $ 13.75 \pm 0.38$ & $-2.58 \pm 0.13$ & $3.5   \pm 0.5 $ & $0.9   \pm 0.3 $ & $>10 $ \\
$5866$ & $ 12.31 \pm 0.46$ & $-1.89 \pm 1.26$ & $6.9   \pm 6.3 $ & $2.6   \pm 2.0 $ & $0.6 $ \\
$7457$ & $ 12.43 \pm 0.24$ & $-1.73 \pm 0.24$ & $11.4  \pm 2.7 $ & $4.4   \pm 1.1 $ & $0.3 $ \\
\hline
$1316$ & $ 13.73 \pm 0.41$ & $-2.41 \pm 0.17$ & $4.1   \pm 0.7 $ & $1.3   \pm 0.4 $ & $0.4 $ \\
$1399$ & $ 14.11 \pm 0.38$ & $-2.73 \pm 0.07$ & $2.9   \pm 0.2 $ & $0.5   \pm 0.2 $ & $2.4 $ \\
$4472$ & $ 14.11 \pm 0.42$ & $-2.30 \pm 0.12$ & $4.5   \pm 0.5 $ & $1.4   \pm 0.3 $ & $>10 $ \\
$4594$ & $ 12.75 \pm 0.14$ & $-1.70 \pm 0.21$ & $15.3  \pm 3.4 $ & $5.9   \pm 1.3 $ & $4.6 $ \\
$4636$ & $ 13.36 \pm 0.30$ & $-2.63 \pm 0.09$ & $3.7   \pm 0.4 $ & $1.0   \pm 0.2 $ & $>10 $ \\
$5128$ & $ 12.56 \pm 0.24$ & $-1.81 \pm 0.24$ & $8.9   \pm 2.0 $ & $3.4   \pm 0.8 $ & $-$ \\
%$ 224$ & $ 12.24 \pm 0.11$ & $-1.67 \pm 0.14$ & $17.5  \pm 2.7 $ & $6.7   \pm 1.0 $ & $1.7 $ \\
 \hline
\end{tabular}
\begin{flushleft}
{\small Columns: (1) galaxy name; (2) average dark matter density within 5~$\re$; (3) halo mass; (4) halo concentration;
(5) halo assembly epoch (we have set the halo assembly epoch of galaxies with invalid $z_{\rm form}$ to 0.1; these are all galaxies with very low dark matter fractions); (6) Redshift corresponding to the mean luminosity--weighted stellar age from Table \ref{tab:gal_prop}. For galaxies with mean stellar ages comparable to or older than the age of the universe, we have set their corresponding $z_{\rm stars}$ to a lower limit of $10$.}
\end{flushleft}
\end{minipage}\hfill
%
%
%\begin{table}[t]
\begin{minipage}{0.45\textwidth}
\centering
{\small \caption{Comparison of halo properties with literature results.} \label{tab:halo_lit} }
{\renewcommand{\arraystretch}{1.1}
\resizebox{\linewidth}{!}{%
\begin{tabular}{@{}r l l l l l}
\hline
\hline
Galaxy & ${\rm log} \mhalo$ & $\chalo$ & ${\rm log} \mhalo$ & $\chalo$ & $\rm {Notes}$ \\
$\rm [NGC]$ & [$\Msun$] &  & [$\Msun$]& &$\Delta_{\rm x}$,${\rm log} M_{\rm x}$, $c_{\rm x}$\\
\hline
& \multicolumn{2}{c}{this work} & \multicolumn{2}{c}{literature}&\\
\hline
$720 $ & $12.83$ & $3.0 $ & $12.77$ & $14.15$ & $101.6, 12.82, 18.50^a$\\
$821 $ & $12.99$ & $6.4 $ & $17.52$ & $1.73$  & $101  , 17.68,  2.45^b$\\
$821 $ &         &        & $14.31$ & $1.77$  & $101  , 14.46,  2.50^c$\\
$1400$ & $12.71$ & $5.4 $ & $11.92$ & $3.98 $ & $101  , 12.02,  5.38^d$\\
$1407$ & $13.82$ & $2.3 $ & $13.23$ & $7.77 $ & $101  , 13.3 , 10.30^e$\\
       &         &        & $13.78$ & $6.85 $ & $101  , 13.85,  9.10^f$\\
       &         &        & $12.99$ & $5.63 $ & $101  , 13.07,  7.53^d$\\
       &         &        & $13.57$ & $12.11$ & $200  , 13.57, 12.11^g$\\
       &         &        & $13.34$ & $18.59$ & $200  , 13.34, 18.60^h$\\
       &         &        & $12.96$ & $13.68$ & $101.7, 13.02, 17.88^a$\\
$2768$ & $13.22$ & $4.7 $ & $11.69$ & $4.72 $ & $101  , 11.78,  6.35^d$\\
$3115$ & $12.65$ & $17.9$ & $12.08$ & $13.79$ & $101  , 12.14, 18.07^d$\\
$3377$ & $12.19$ & $5.7 $ & $11.28$ & $5.25 $ & $101  , 11.36,  7.03^d$\\
$4278$ & $12.76$ & $15.7$ & $11.71$ & $13.23$ & $101  , 11.77, 17.35^d$\\
$4365$ & $13.73$ & $4.4 $ & $12.53$ & $11.01$ & $101  , 12.59, 14.48^d$\\
$4374$ & $13.86$ & $5.6 $ & $13.14$ & $10.97$ & $178  , 13.15, 11.50^i$\\
       &         &        & $13.32$ & $5.59 $ & $100  , 13.4 ,  7.50^j$\\
$4486$ & $13.97$ & $7.8 $ & $13.93$ & $6.98 $ & $101  , 14.0 ,  9.27^k$\\
       &         &        & $12.61$ & $12.78$ & $101  , 12.67, 16.77^d$\\
       &         &        & $14.67$ & $3.21$  & $101  , 14.78,  4.38^l$\\
       &         &        & $13.90$ & $3.88$  & $101  , 14.0,   5.25^m$\\
       &         &        & $13.95$ & $3.83$  & $101  , 14.05,  5.19^n$\\
$4494$ & $12.53$ & $1.9 $ & $11.97$ & $6.23 $ & $101  , 12.05,  8.30^c$\\
       &         &        & $12.02$ & $4.32 $ & $101  , 12.11,  5.82^d$\\
$4649$ & $13.68$ & $5.7 $ & $13.49$ & $15.94$ & $101.5, 13.54, 20.80^a$\\
       &         &        & $13.47$ & $ 5.94$ &${\Delta}\ {\rm not\ stated}^o$\\
$4697$ & $12.96$ & $5.4 $ & $12.66$ & $4.53 $ & $101  , 12.75,  6.10^c$\\
$5846$ & $13.75$ & $3.5 $ & $13.16$ & $6.00  $ & $101  , 13.24, 8.00^p$\\
       &         &        & $12.76$ & $11.40 $ & $101  , 12.82, 14.99^d$\\
       &         &        & $13.10$ & $8.90  $ & $200  , 13.1 , 8.90^q$\\
$1316$ & $13.73$ & $4.1 $ & $13.37$ & $6.10 $ & ${\Delta}\ {\rm not\ stated}^w$\\
$1399$ & $14.11$ & $2.9 $ & $13.31$ & $13.57$ & $101  , 13.26, 11.00^r$\\
       &         &        & $12.89$ & $10.20 $ & $101  , 12.95, 13.44^s$\\
$4472$ & $14.11$ & $4.5 $ & $13.45$ & $9.93 $ & $101.4, 13.51, 13.07^a$\\
       &         &        & $13.94$ &$\text{--}$&    $ 200, eqn. 16^t$\\
$4636$ & $13.36$ & $3.7 $ & $13.13$ & $7.37 $ & $101  , 13.07,  5.90^u$\\
       &         &        & $12.99$ & $20.09$ & $200  , 12.99, 20.10^v$\\
\hline
\end{tabular}}}
\begin{flushleft}
{\small
Columns: (1) galaxy name; (2) halo mass; (3) halo concentration; (4) \textit{corrected} halo mass from the literature; (5) \textit{corrected} halo concentration from the literature; (6) virial over--density, halo mass and concentration (where necessary, we have obtained corresponding halo mass and concentration at virial over--density of 200)
References : \textit{a}. \citet{Buote_2007}, \textit{b}. \citet{Forestell_2010}, \textit{c}. \citet{Napolitano_2009}, \textit{d}. \citet{Samurovic_2014}, \textit{e}. \citet{Pota_2015}, \textit{f}. \citet{Romanowsky_2009}, \textit{g}. \citet{Su_2014}, \textit{h}. \citet{Zhang_2007}, \textit{i}. \citet{Zhu_2014}, \textit{j}. \citet{Napolitano_2011}, \textit{k}. \citet{Oldham_2016},  \textit{l}. \citet{McLaughlin_1999}, \textit{m}. \citet{Strader_2011}, \textit{n}. \citet{Murphy_2011},
\textit{o}. \citet{Shen_2010}, \textit{p}. \citet{Napolitano_2014}, \textit{q}. \citet{Zhu_2016}, \textit{r}. \citet{Richtler_2014}, \textit{s}. \citet{Schuberth_2010}, \textit{t}. \citet{Samurovic_2016}, \textit{u}. \citet{Cote_2003}, \textit{v}. \citet{Schuberth_2012}, \textit{w}. \citet{Johnson_2009}}\end{flushleft}
\end{minipage}\hfill
\end{table*}

\subsection{Comparison of halo properties with literature studies}
Some of the galaxies in our sample have published $\chalo$ and $\mhalo$ results in the literature from various studies. These studies are based on data from extended PNe and/or GC kinematics (sometimes supplemented with stellar kinematics data) and X-ray studies, with different modelling techniques. We compile these results in Table \ref{tab:halo_lit} and where a virial over-density other than $\Delta_{\rm 200}$ has been used in the literature, we rescale the results using the conversion relations from \citet{Hu_2003}. Note that while our $\mhalo$ and literature $\mhalo$ are generally consistent, our $\chalo$ values are generally lower than literature results and by construction, are more consistent with expectations from the $\mhalo - \chalo$ relations. The implication of adopting the $\chalo$ reported in the literature for our galaxy sample is that on average, their haloes are already in place at $z_{\rm form}{\sim}5$, whereas we find a mean $z_{\rm form}{\sim}3$. This is due to the extra constraint from the stellar mass-halo mass relation that we have placed on $\mhalo$. A similar approach was used in \citet{Auger_2013} where the $\mhalo - \chalo$ relation was used as a conditional prior on $\chalo$. This eased the tension between theoretical expectations and results often reported in the literature \citep[e.g.][]{Napolitano_2011, Samurovic_2014, Samurovic_2016}.

\section{Discussion}
\label{discussion}
Present-day galaxies are expected to have experienced different merger (major and/or minor) and gas accretion (smooth and/or clumpy) histories of DM and baryons. It is also expected that signatures of these varied assembly histories should be reflected in their mass distributions and halo structural properties. In this work, we have obtained dark matter fractions and average dark matter densities within the inner 5~$\re$ for our sample of ETGs. We interpret the diversity we observe in these parameters as a reflection of their different mass assembly histories. We also use these homogeneously obtained mass measurements to infer the assembly epochs of their haloes as well as their structural parameters.

%neither of the cosmological simulations nor the simple galaxy model we have considered can completely account for the range of $\fdm$ we have measured. This highlights how complex galaxy formation is.
\subsection{Origin of the diverse dark matter fractions within 5~$\re$ in ETGs}
\label{origin}
From our results above, it appears that galaxies with $\fdm {\sim} 0.7$, in particular, the central dominant types, are well described by the simulation from \hypertarget{Remus+17}{\citet{Remus_2016}}, while those with $\fdm {\sim} 0.5$ appear to be better described by \hypertarget{Wu+14}{\citet{Wu_2014}}. A few galaxies have inferred $\fdm$ well below the results from both cosmological simulations. Again, we note that in the simulations of \hypertarget{Wu+14}{\citet{Wu_2014}} and \hypertarget{Remus+17}{\citet{Remus_2016}}, the dark matter distributions have been modified from the standard NFW-like profiles through baryon--DM interactions. Also, \hypertarget{Remus+17}{\citet{Remus_2016}} included feedback from AGN and SN.

During the late phase of the mass assembly (i.e., $z\leq2$) in these simulations, growth is dominated by dry mergers (major/minor) and happens predominantly in the outer haloes. Due to this mostly non-dissipative growth in size and mass, our fiducial 5~$\re$ now encloses more dark matter relative to stars compared to their high redshift progenitors. At any given stellar mass, ETGs with higher dark matter fractions have experienced a late phase mass assembly that is increasingly dominated by dry mergers. However, the present-day mass distribution, parametrised by the $\fdm$, also depends on the extent to which the inner dark matter halo has been modified by baryonic processes during the mass build up of the galaxies, as well as the initial conditions set by the density of the universe during the initial halo collapse. We explore this in more details below and note that none of the cosmological simulations (nor any that we are aware of in the literature) produce galaxies with $\fdm$ within 5~$\re$ as low as we have measured in some of our ETGs.

While the simple galaxy model (SGM1) clearly captures the mean dark matter fraction at any stellar mass, its use in properly understanding the origin of the diversity in our measured $\fdm$ is hampered by the large scatter around the mean $\fdm$. This scatter is from the combined uncertainties from the input scaling relations (i.e., $\re-M_*, M_*-\mhalo$ and $\mhalo-\chalo$). There is a ${\sim}0.25$ dex scatter in the $\re-M_*$, a ${\sim}0.2$ dex scatter in the $M_*-\mhalo$ and a ${\sim}0.11$ dex scatter in the $\mhalo-\chalo$ scaling relations, with the scatter in the $\re-M_*$ being the most critical. Deviations of individual galaxies from the $\re-M_*$ scaling relation produce an asymmetrical bias towards lowered $\fdm$ (see Figure \ref{fig:cmp_residuals}, where we measure higher $\fdm$ than predicted for most of our galaxies), resulting in the wide range of plausible dark matter fractions. However, these deviations do not explain why the galaxies with low dark matter fractions preferentially have ${\rm log}(M_*/\Msun){\sim}11$. SGM2, the simple galaxy model we constructed based on a fit to the sizes and stellar masses in Table \ref{tab:gal_prop}, could potentially help shed more light on this since it captures better the general trend in our measured $\fdm$. However, to properly address this issue, one would need to study more galaxies with ${\rm log}(M_*/\Msun)\leq10.5$ to rule out any artificial bias from our sample selection. 

The low $\fdm$ ETGs, all with ${\rm log}(M_*/\Msun){\sim}11$, have very low average DM densities within 5~$\re$, and from our preceding analysis, they have unrealistic halo assembly epochs, appearing to be incompatible with the \textit{Planck} cosmology. However, they are \textit{normal} ETGs in that they have $\re$ and $M_*$ that are compatible with the $\re-M_*$ galaxy scaling relation. It is remarkable that ${\rm log}(M_*/\Msun){\sim}11$ corresponds to the sharp upturn in the $\re-M_*$ scaling relation and the knee in the $M_* -\mhalo$ scaling relation. At this stellar mass (also at all redshifts), galaxies are most efficient at converting baryons into stars e.g. \citet{Rodriguez_2016}, such that a low dark matter fraction should then be a natural expectation. Above ${\rm log}(M_*/\Msun){\sim}11$, galaxy haloes are too massive for gas to cool and form stars while below this mass they are not massive enough to hold on to their gas. This makes ${\rm log}(M_*/\Msun){\sim}11$ ETGs interesting as one should be able to observe the effects of extended star formation history on galaxy evolution through their mass distributions.
%of both adiabatic halo contraction and halo expansion which is related to

From our SGM1, we find that haloes of ${\rm log}(M_*/\Msun){\sim}11$ ETGs have significantly lower $\rs / \re$ (ratio of dark matter halo scale radius to galaxy size) compared to ETGs at other stellar masses (see Appendix \ref{fig:appendix_b}). A simple experiment with a mock ${\rm log}(M_*/\Msun){\sim}11$ galaxy, where we increase $\rs$ by a factor of $3$ to reflect a more diffuse dark matter halo, sufficiently reduces the $f_{\rm DM}$ by a factor of $2$, i.e. from ${\sim}0.6$ to ${\sim}0.3$. This implies that any mechanism that can produce \textit{normal} galaxies in diffuse dark matter haloes should be able to explain the low dark matter fractions we have observed in these galaxies. The physical processes to achieve this include halo expansion through dynamical friction from infalling stellar clumps \citep{Johansson_2009} or feedback-induced dark matter outflows \citep{Governato_2010}. The modified dark matter profile would then be non-NFW and as such our analysis here which assumes a NFW-like profile would be inadequate.
%This argument becomes more appealing if one considers the result from a photometric study of NGC~4494 by \citet{Foster_2011} where an intermediate--color GC sub--population was identified. This could be the GC counterpart of the infalling stellar system that produced the diffuse halo we now observe in this galaxy.

The low dark matter fractions could also be due to the preferential tidal stripping of dark matter haloes relative to their stars \citep[e.g.][]{Smith_2016} from gravitational interactions with their neighbours. If this were to be the case, then one would expect to find signatures of depletion in the GC population, especially in the galaxy outskirts, since they are more radially extended than the starlight. However, we find evidence to the contrary from their GC subpopulations \citep[e.g.][]{Forbes_2016}, where for example, the low dark matter fraction galaxy NGC~4494 still retains a high fraction of blue globular clusters relative to its entire globular cluster system in the galaxy outskirts (the blue globular clusters usually dominate the globular cluster system in galaxy outer haloes).

Alternatively, their low dark matter fractions could mean that their dark matter haloes are poorly described by NFW DM profiles. This suggests the need for an alternate halo description, e.g. using logarithmic dark matter haloes (\citealt{Thomas_2009, Morganti_2013}, \hyperlink{Alabi+16}{Alabi+16}). Interestingly, logarithmic DM haloes are characterised by shallow central dark matter densities with a maximal stellar contribution (e.g. \citealt{Gentile_2004, Napolitano_2011}), reinforcing our earlier inference. The presence of self-interacting dark matter in haloes can also lower the central dark matter densities (e.g. \citealt{Rocha_2013, diCintio_2017}) by making the core radius larger, but it would be challenging to separate its effects from those purely driven by feedback outflows, especially in ${\rm log}(M_*/\Msun){\sim}11$ galaxies.%While GC systems on highly radial orbits, or a stellar population that is dominated by bottom--heavy stars, could also conspire to produce the low $\fdm$ we have measured in these ETGs, we have already shown in \hyperlink{Alabi+16}{Alabi+16} that the $\fdm$ for these galaxies are low independent of the adopted velocity anisotropy or stellar $M/L$ assumptions. The argument we have made above also suggests that the high $\fdm$ ($>0.6$) we have measured in some ${\rm log}(M_*/\Msun){\sim}11$ ETGs appear to be compatible with a galaxy evolution scenario where the $\fdm$ could have been enhanced through some DM halo contraction.
\begin{figure*}
    \includegraphics[width=1.0\textwidth]{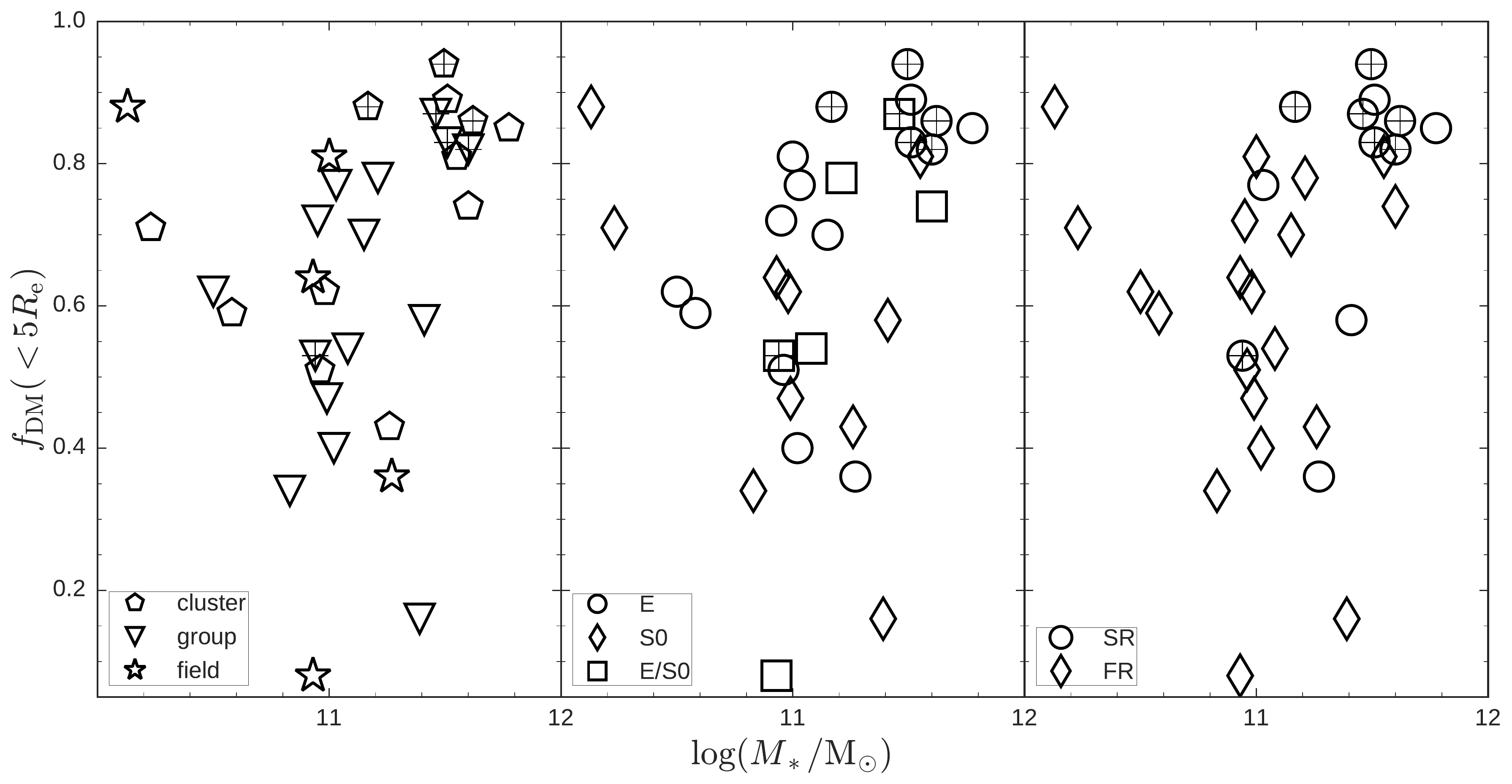}\hspace{0.01\textwidth}%
	\caption{\label{fig:correlation} Dark matter fraction versus total stellar mass, highlighting different galaxy environments in the \textit{left panel}, galaxy morphologies in the \textit{middle panel} and galaxy central kinematics in the \textit{right panel} (SR = slow rotators; FR = fast rotators). Centrally dominant galaxies have been marked with black crosses and they mostly have high dark matter fractions. As a function of morphology, dark matter fraction appears to increase with galaxy stellar mass in ellipticals while lenticulars show a noisy, somewhat decreasing trend with mass and a lower median $\fdm$ compared to elliptical galaxies. The decreasing trend with stellar mass in lenticulars from the \textit{middle panel} is not seen in the fast rotators in the \textit{right panel}.}
\end{figure*}
%
%\begin{figure}
%    \includegraphics[width=0.48\textwidth, height=0.30\textheight]{plots/fig_4b.pdf}\hspace{0.01\textwidth}%0
%	\caption{\label{fig:ATLAS3D_fdm} Dark matter fraction within 1~$\re$ versus total circular velocity for ETGs from ATLAS$^{\rm 3D}$ \citep{Cappellari_2013b}. We only show galaxies with good quality data flag (column $8\ {(\rm Qty > 0)}$ from their table 1). We used the Hubble type parameter from table 1 in \citet{Cappellari_2011} to classify the galaxies as ellipticals ($T\leq-3.5$) or lenticulars ($-3.5 < T \leq-0.5$). The lines are as shown in the plot legend. Mass distribution within the inner 1~$\re$ differs with galaxy morphology, following the same trend we observe at larger radii.}
%\end{figure}
%
%
\begin{figure*}
    \includegraphics[width=1.0\textwidth]{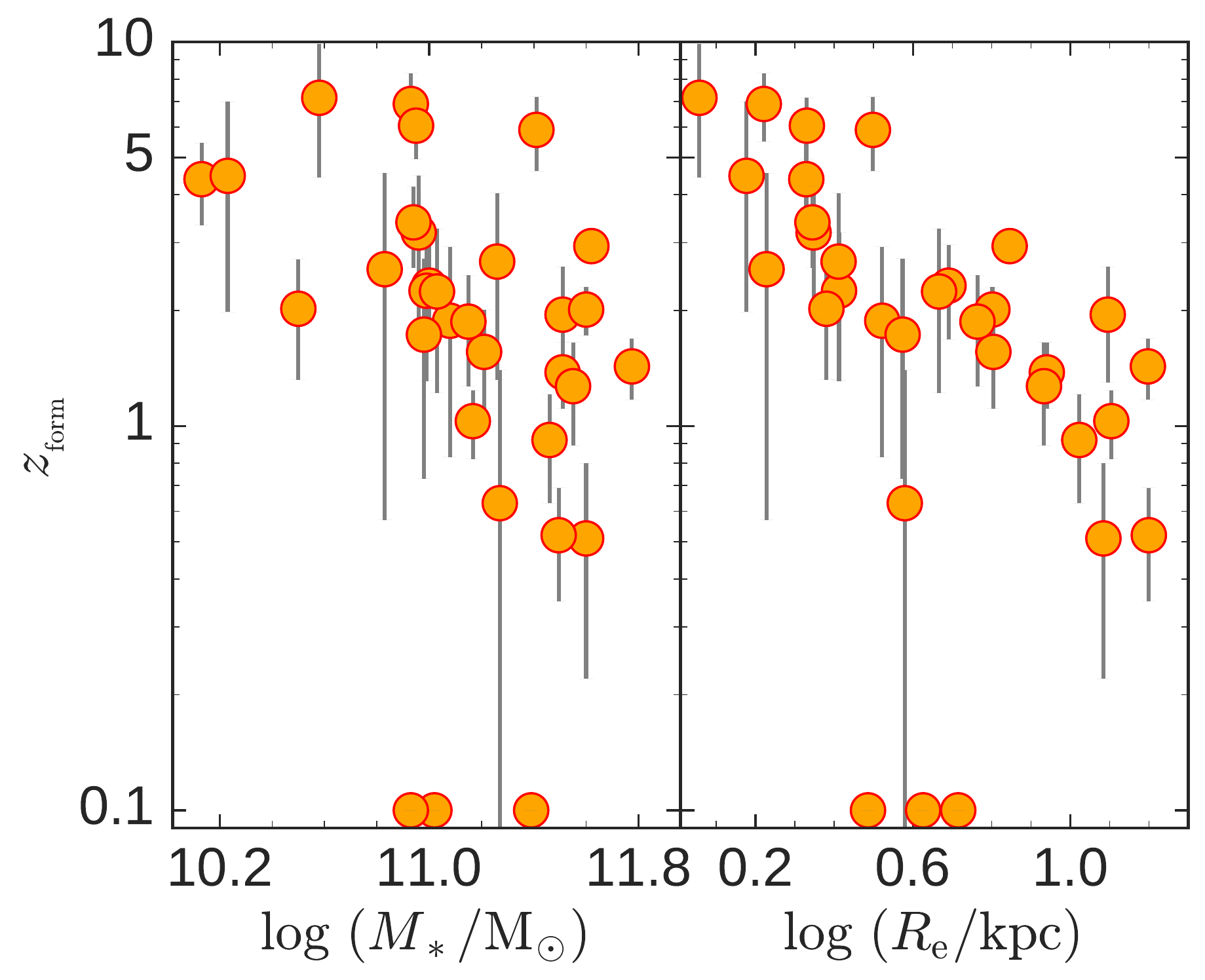}\hspace{0.01\textwidth}\\
	\caption{\label{fig:halo_form} Halo assembly epoch as a function of total stellar mass (\textit{left panel}) and galaxy size (\textit{right panel}). The three galaxies with unrealistic halo assembly epochs are shown at $z_{\rm form}\sim0.1$. Lower mass and/or smaller galaxies reside in haloes which assembled earlier than their more massive and larger counterparts, in agreement with the hierarchical structure growth. However, at any stellar mass or galaxy size, there exists a large spread in the inferred halo formation epoch. Also note the correlation between the average dark matter density from Figure \ref{fig:aveden} and the halo assembly epoch, such that galaxies with high dark matter densities assemble their haloes early.}
\end{figure*}
\begin{figure*}
    \includegraphics[width=1.0\textwidth]{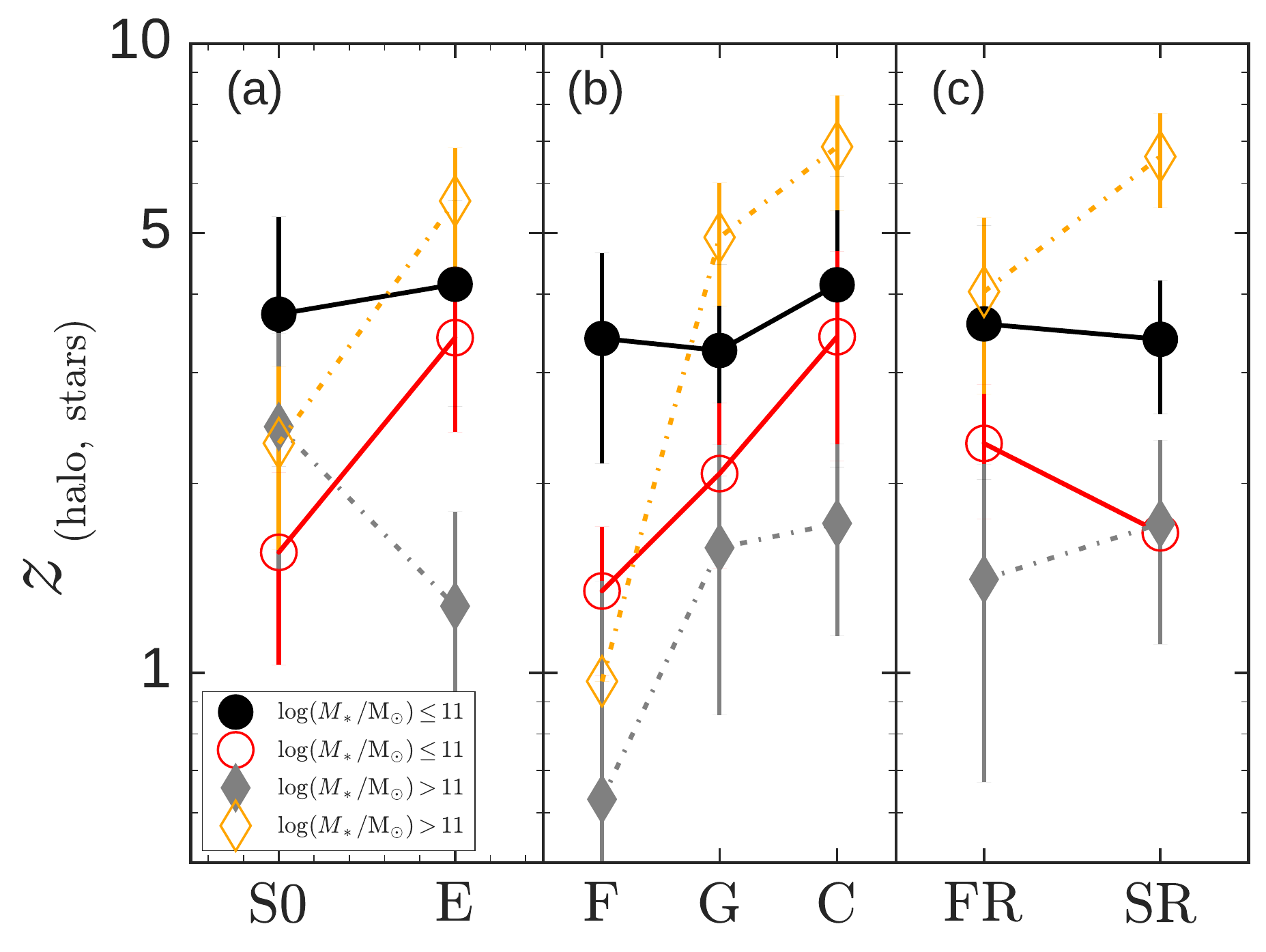}\hspace{0.01\textwidth}\\
	\caption{\label{fig:halo_gal} Summary plot showing the mean halo assembly epoch for our galaxy sample, in low (${\rm log}(M_*/\Msun)\leq11$, circles and solid lines) and high (${\rm log}(M_*/\Msun)>11$, diamonds and dash-dot lines) stellar mass bins. Filled symbols correspond to halo assembly epochs while open symbols show the mean formation epoch that corresponds the luminosity--weighted ages of the central stars in our sample. Panel \textit{a} shows the mean assembly epoch according to galaxy morphology (E=elliptical, S0=lenticular). Note that we have excluded all galaxies with ambiguous morphological classification from this analysis. Panel \textit{b} shows mean assembly epoch as a function of galaxy environment (F=field, G=group, C=cluster) and panel \textit{c} shows the  mean assembly epoch as a function of central galaxy kinematics (FR=fast central rotator, SR=slow central rotator). The errorbars are the standard deviations about the mean. While there are large spreads about the mean, massive ellipticals have haloes which assemble relatively late compared to lenticulars or low mass  ellipticals. We also find that massive ellipticals in the field have haloes which assemble very late in agreement with predictions from the semi-analytic galaxy formation models of \citet{deLucia_2007}.}
\end{figure*}
%
%\begin{figure*}
%    \includegraphics[width=1.0\textwidth]{plots/fig_3.pdf}\hspace{0.01\textwidth}\\
%	\caption{\label{fig:halo_stars} Halo assembly epoch versus the formation epoch of the central stellar population (obtained from \textit{luminosity-weighted} ages) in our galaxy sample, highlighting in panels \textit{a.}, total stellar mass, \textit{b.}, galaxy morphology, \textit{c.}, galaxy rotation classification and \textit{d.}, galaxy environment. We also show in panel \textit{d.}, ETGs from the Coma cluster study of \citet{Thomas_2009}. In all the panels, the solid one--to--one line is to guide the eye and the arrows indicate where stellar ages are lower limits. Lower--mass ETGs reside in haloes that formed early and generally host younger stars in their centres. On the contrary, the most massive galaxies have haloes that assembled late, while their central stars are typically old. The panels also show that S0 galaxies (in our galaxy sample they are always fast rotators) generally form their central stars after their haloes are already assembled, unlike ellipticals which generally form their stars before assembling their haloes. NGC~2974, NGC~3607 and NGC~4494 (DM poor galaxies within 5~$\re$) do not appear in the panels due to their unrealistic $z_{\rm form}$, suggesting that NFW profiles may be inadequate to describe their DM haloes.}
%\end{figure*}
%%

\subsection{Dark matter fractions and correlation with galaxy properties}
We revisit the issue of correlation between the dark matter fractions within 5~$\re$ for our enlarged sample and some of their galaxy properties. We briefly summarise the interesting trends below and show the trends in Figure \ref{fig:correlation}. The Spearman rank correlations between the dark matter fractions and galaxy properties are all statistically insignificant and generally weak, mainly due to the large scatter due to the ${\sim}10^{11}\ \Msun$ galaxies.

First, trends as a function of environment are generally weak. The only stand-out trend we find as a function of galaxy environment is that central dominant ETGs mostly have high dark matter fractions with low dark matter fraction ETGs preferentially residing in less-dense environments. Second, we find that S0s are observed to have lower median dark matter fractions compared to ellipticals, and probably show a hint of an opposite trend in their dark matter fractions with stellar mass compared to ellipticals. This is similar to results reported for spirals in the literature \citep{Persic_1993, Dutton_2011, Courteau_2015}, where the most massive spirals have the lowest central dark matter fractions. This trend was also tentatively identified in the S0s studied in \citet{Tortora_2009}, although, due to their spherical modelling technique, they claimed that the trend may not be real. Interestingly, this trend is lost when our sample is classified according to their central kinematics i.e. fast or slow rotators (using results from the ATLAS$^{\rm 3D}$ \citealt{Cappellari_2013b}).

If this dichotomy in the large scale mass distribution between S0s and ellipticals is real, it implies that S0s are akin to spirals, more than ellipticals. Disk-dominated galaxies would then have a global mass distribution where the dark matter fraction decreases with stellar mass, at least within 1~$\re$ \citep[e.g.][]{Courteau_2015}, and out to large radii. This agrees with results from \citet{Cappellari_2015} where they found that central fast-rotators and disky lenticular galaxies have similar mass distributions out to 4~$\re$. A larger and more complete sample of ETGs probed to large radii would be needed to confirm if indeed this dichotomy is real or not, as well as predictions of the large scale mass distributions from cosmological simulations that produce S0s and ellipticals.

\subsection{When did the haloes of ETGs form?}
We summarise the inferred halo assembly epochs for our galaxy sample in Figure \ref{fig:halo_form}, showing how they vary with galaxy total stellar mass and size. Lower-mass galaxies are associated with haloes that assembled earlier; $z_{\rm form}{\sim}4$, while the more-massive galaxies have haloes that assembled later, at $z_{\rm form}{\sim}2$. Likewise, more compact galaxies are associated with haloes that assembled earlier, and vice-versa. These results are consistent with hierarchical growth of structures such that smaller objects virialise early in gas-rich events, with today's massive galaxies undergoing a more extended halo build-up. The Spearman rank correlation between $z_{\rm form}$ and stellar mass is ${\sim}-0.5$ and only marginally significant, however, if we remove the galaxies with very low dark matter fractions, i.e. NGC~2974, NGC~3607 and NGC~4494, the correlation becomes statistically significant i.e. p-val $< 0.005$. The correlation between $z_{\rm form}$ and size, on the other hand, is stronger (${\sim}-0.7$) and statistically significant (p-val $< 0.005$) regardless of whether we exclude the three galaxies with low dark matter fractions or not. 
%Some of the  galaxies with $z_{\rm form}\leq1$ are thought to have experienced recent gas--rich major merger events, e.g. NGC~1316 \citep{Schweizer_1980, Goudfrooij_2001} and NGC~5866 \citep{Jiang_2009, Duc_2015, Querejeta_2015}, or recent \textit{dry} major merger events, e.g. NGC~720 \citep{Rembold_2005}. Our results here, using GC kinematics data, independently support their late halo assembly.

In Figure \ref{fig:halo_gal}, we compare $z_{\rm form}$ with that of the central stellar populations (obtained from the \textit{luminosity-weighted} ages within their central 1~$\re$, \citealt{McDermid_2015}) for our galaxies. We adopt the cosmological parameters of a flat universe from the \citet{Planck_Collaboration_2013}, i.e. $H_0=67.8\ \kms\ {\rm Mpc}^{-1}, \Omega_M=0.307$ and use the \textsc{astropy.cosmology} package to convert the ages to formation epochs. This exercise enables us to infer the nature of the late mass assembly in our sample of ETGs, i.e. dissipational or non--dissipational, assuming that late gas-rich merger events are always accompanied by central star formation. For some of our galaxies, the stellar age from the literature is comparable to, or more than, the age of the universe (${\sim}13.8$ Gyr). In such cases, we adopt a fixed upper limit of ${\geq}13.3$ Gyr ($z_{\rm stars}\geq10$).

Due to the strong correlation of $z_{\rm form}$ with stellar mass, we make our comparisons after binning our galaxies by their stellar masses. Bearing in mind our modest sample size, we consider two stellar mass bins, i.e. ${\rm log}(M_*/\Msun)\leq11$ and ${\rm log}(M_*/\Msun)>11$ and make the comparison with respect to galaxy morphology, environment and central kinematics. From Figure \ref{fig:halo_gal}, massive ellipticals have haloes that assembled late, i.e. $z_{\rm form}<2$, compared to massive lenticulars, whose haloes assembled earlier ($z_{\rm form}{\sim}4$). The halo assembly epoch of low mass ellipticals is not significantly different from that of lenticulars, in that their haloes also assembled early ($z_{\rm form}{\sim}3$). The late halo assembly in the field for the most massive galaxies is mainly driven by the galaxies with very low dark matter fractions. This is in line with results from semi-analytic models (e.g. \citealt{deLucia_2007}) of galaxy formation where galaxies in low-density environments are expected to be associated with haloes which assembled later than those in cluster environments (see also \citealt{Corsini_2016}, where they arrived at a similar conclusion based on their low-density environment dynamical study). However, we only find this agreement in our most massive field galaxies, i.e. ${\rm log}(M_*/\Msun)>11$. If on the other hand, we disregard the stellar mass binning, halo assembly epoch then has no correlation with galaxy environment (see Appendix \ref{fig:appendix_c} for a version of Figure \ref{fig:halo_gal} but without stellar mass binning). Haloes associated with more massive slow or fast rotating galaxies also have late assembly epochs compared to their low mass counterparts. There is however a strong trend in the central stellar age as a function of galaxy morphology, environment and central kinematics in both stellar mass bins, where the central stars are usually in place at earlier times relative to the halo in bulge-dominated systems. The only exception to this is in low mass, slow rotators which form their central stars relatively late, i.e. $z_{\rm form}{\sim}2$. Our results therefore suggests a dichotomy between the late mass evolution of the bulge-dominated ellipticals and the discy-lenticulars.

This dichotomy, together with our earlier results, where at any stellar mass, more massive and centrally dominant ETGs have higher dark matter fractions and lower average dark matter densities within 5~$\re$, form a consistent picture when considered in the context of the two-phase galaxy formation paradigm for massive ETGs \citep[e.g.][]{Naab_2009, Oser_2010,Forbes_2016}. Dry mergers, after the early dissipational phase, increase the inner dark matter fractions in massive ETGs since they do not bring a significant amount baryons to the galaxy centres but rather lead to a net outward transfer of angular momentum via dynamical friction. They also reduce the inner average dark matter densities since they make the galaxies larger. These findings therefore rule out wet-major merger or gas-rich accretion events as the predominant channel for the late mass build-up of massive ETGs.

%In Figure \ref{fig:halo_stars}, panel \textit{b}, most S0 galaxies, independent of their $M_*$, have experienced some significant, late \textit{in--situ} central star formation, usually after their haloes were assembled. For example, S0s NGC~7457 and NGC~1316, with ${\rm log}(M_*/\Msun)=$ 10.3 and 11.8, respectively, have both experienced late star formation ($z{\sim}0.5$) despite having their haloes in place at $z{\sim}4$ and $0.5$, respectively. If we assume that $z_{\rm form}$ is set by the time a galaxy experienced its most recent major merger, one can safely rule out a late major gas--rich merger in the isolated galaxy NGC~7457, given its high $\fdm{\sim}0.8$. On the other hand, NGC~1316 has experienced a recent gas--rich event \citep[e.g.][${\sim}$3 Gyr]{Goudfrooij_2001}, significant enough to alter its DM distribution (note its relatively low $\fdm$ for its stellar mass in Figure \ref{fig:predict_fdm1}) and initiate central star formation, i.e. a gas--rich major merger. These results highlight the diverse channels through which present--day S0s have evolved and further affirms our earlier inference that a high $\fdm$ is incompatible with late gas--rich major mergers.

\section{Conclusions}
\label{summary}
We have measured the dark matter fraction and average dark matter density within 5~$\re$ in a sample of 32 early-type galaxies (ETGs) using their globular cluster kinematics. We compared our dark matter fractions with predictions from cosmological simulations. We also used our measured dynamical parameters to infer the epochs of assembly of our ETG haloes, assuming  dark matter haloes are well described by Navarro--Frenk--White profiles.
We briefly summarise our results here:
\begin{itemize}
\item Early-type galaxies have a wide range of dark matter fractions within 5~$\re$, ranging from $0.1$ to $0.9$, typically increasing  with galaxy stellar mass, and largely independent of the galaxy's environment. We find that a high dark matter fraction is consistent with a late ($z\leq2$) mass assembly that is dominated by dissipationless mergers.
\item We find that ETGs with low dark matter fractions within 5~$\re$ are typically those with log $(M_*/\Msun){\sim}11$ and diffuse dark matter haloes. We associate their low dark matter fractions with a mass assembly likely dominated by halo expansion.
\item By comparing our results with predictions from a suite of cosmological simulations, we are able to show that modifications of the mass distribution due to physical processes during mass assembly are important in understanding the distribution of dark matter fraction in present-day ETGs.
\item Early-type galaxies, reside in haloes that assembled, on average, $z_{\rm form}{\sim}2-3$. The ${\sim}L^{*}$ ETGs have haloes that assembled earlier ($z{\sim}4$) than their more massive counterparts which assembled later ($z{\sim}2$). We find that massive galaxies, i.e. log $(M_*/\Msun){>}11$, in the field environment have haloes that form late, in agreement with predictions from semi-analytic galaxy formation models.
\item S0s and ellipticals reside in dark matter haloes with similar structural properties and assembly epochs. However, we find hints that there may be a dichotomy in their mass distributions at large radii, with S0s showing signs of a decreasing dark matter fraction with increasing galaxy stellar mass, unlike ellipticals. We attribute this to a fundamental difference in their dominant late--phase mass assembly channel.
\end{itemize}

\section*{Acknowledgements}
We wish to thank Rhea Remus and Alan Duffy for interesting conversations and their assistance with simulation results and software packages. We thank the anonymous referee for comments and suggestions that have been helpful.

The data presented herein were obtained at the W.M. Keck Observatory, which is operated as a scientific partnership among the California
Institute of Technology, the University of California and the National Aeronautics and Space Administration.
The Observatory was made possible by the generous financial support of the W.M. Keck Foundation. The authors wish to
recognize and acknowledge the very significant cultural role and reverence that the summit of Maunakea has always had within
the indigenous Hawaiian community. The analysis pipeline used to reduce the DEIMOS data was developed at UC Berkeley with support from NSF grant AST--0071048. JB acknowledges support from NSF grant AST--1616598. AJR was supported by NSF grant AST--1616710. DAF thanks the ARC for financial support via DP 130100388. JS acknowledges support from the Packard Foundation.
We acknowledge the use of the HYPERLEDA database (http://leda.univ-lyon1.fr), \textsc{astropy} \citep{Astropy_2013} and TOPCAT \citep{Taylor_2005}.

\bibliographystyle{mnras}
\bibliography{halov3}

\begin{thebibliography}{}
\makeatletter
\relax
\def\mn@urlcharsother{\let\do\@makeother \do\$\do\&\do\#\do\^\do\_\do\%\do\~}
\def\mn@doi{\begingroup\mn@urlcharsother \@ifnextchar [ {\mn@doi@}
  {\mn@doi@[]}}
\def\mn@doi@[#1]#2{\def\@tempa{#1}\ifx\@tempa\@empty \href
  {http://dx.doi.org/#2} {doi:#2}\else \href {http://dx.doi.org/#2} {#1}\fi
  \endgroup}
\def\mn@eprint#1#2{\mn@eprint@#1:#2::\@nil}
\def\mn@eprint@arXiv#1{\href {http://arxiv.org/abs/#1} {{\tt arXiv:#1}}}
\def\mn@eprint@dblp#1{\href {http://dblp.uni-trier.de/rec/bibtex/#1.xml}
  {dblp:#1}}
\def\mn@eprint@#1:#2:#3:#4\@nil{\def\@tempa {#1}\def\@tempb {#2}\def\@tempc
  {#3}\ifx \@tempc \@empty \let \@tempc \@tempb \let \@tempb \@tempa \fi \ifx
  \@tempb \@empty \def\@tempb {arXiv}\fi \@ifundefined
  {mn@eprint@\@tempb}{\@tempb:\@tempc}{\expandafter \expandafter \csname
  mn@eprint@\@tempb\endcsname \expandafter{\@tempc}}}

\bibitem[\protect\citeauthoryear{{Alabi} et~al.,}{{Alabi}
  et~al.}{2016}]{Alabi_2016}
{Alabi} A.~B.,  et~al., 2016, \mn@doi [\mnras] {10.1093/mnras/stw1213}, \href
  {http://adsabs.harvard.edu/abs/2016MNRAS.460.3838A} {460, 3838}

\bibitem[\protect\citeauthoryear{{Alves-Brito}, {Hau}, {Forbes}, {Spitler},
  {Strader}, {Brodie}  \& {Rhode}}{{Alves-Brito} et~al.}{2011}]{Alves_2010}
{Alves-Brito} A.,  {Hau} G.~K.~T.,  {Forbes} D.~A.,  {Spitler} L.~R.,
  {Strader} J.,  {Brodie} J.~P.,   {Rhode} K.~L.,  2011, \mn@doi [\mnras]
  {10.1111/j.1365-2966.2011.19368.x}, \href
  {http://adsabs.harvard.edu/abs/2011MNRAS.417.1823A} {417, 1823}

\bibitem[\protect\citeauthoryear{{Amorisco} \& {Evans}}{{Amorisco} \&
  {Evans}}{2012}]{Amorisco_2012}
{Amorisco} N.~C.,  {Evans} N.~W.,  2012, \mn@doi [\mnras]
  {10.1111/j.1365-2966.2012.21307.x}, \href
  {http://adsabs.harvard.edu/abs/2012MNRAS.424.1899A} {424, 1899}

\bibitem[\protect\citeauthoryear{{Arnold} et~al.,}{{Arnold}
  et~al.}{2014}]{Arnold_2014}
{Arnold} J.~A.,  et~al., 2014, \mn@doi [\apj] {10.1088/0004-637X/791/2/80},
  \href {http://adsabs.harvard.edu/abs/2014ApJ...791...80A} {791, 80}

\bibitem[\protect\citeauthoryear{{Astropy Collaboration} et~al.,}{{Astropy
  Collaboration} et~al.}{2013}]{Astropy_2013}
{Astropy Collaboration} et~al., 2013, \mn@doi [\aap]
  {10.1051/0004-6361/201322068}, \href
  {http://adsabs.harvard.edu/abs/2013A%26A...558A..33A} {558, A33}

\bibitem[\protect\citeauthoryear{{Auger}, {Budzynski}, {Belokurov}, {Koposov}
  \& {McCarthy}}{{Auger} et~al.}{2013}]{Auger_2013}
{Auger} M.~W.,  {Budzynski} J.~M.,  {Belokurov} V.,  {Koposov} S.~E.,
  {McCarthy} I.~G.,  2013, \mn@doi [\mnras] {10.1093/mnras/stt1585}, \href
  {http://adsabs.harvard.edu/abs/2013MNRAS.436..503A} {436, 503}

\bibitem[\protect\citeauthoryear{{Binney} \& {Tremaine}}{{Binney} \&
  {Tremaine}}{1987}]{Binney_1987}
{Binney} J.,  {Tremaine} S.,  1987, {Galactic dynamics}

\bibitem[\protect\citeauthoryear{{Blumenthal}, {Faber}, {Flores}  \&
  {Primack}}{{Blumenthal} et~al.}{1986}]{Blumenthal_1986}
{Blumenthal} G.~R.,  {Faber} S.~M.,  {Flores} R.,   {Primack} J.~R.,  1986,
  \mn@doi [\apj] {10.1086/163867}, \href
  {http://adsabs.harvard.edu/abs/1986ApJ...301...27B} {301, 27}

\bibitem[\protect\citeauthoryear{{Brodie} et~al.,}{{Brodie}
  et~al.}{2014}]{Brodie_2014}
{Brodie} J.~P.,  et~al., 2014, \mn@doi [\apj] {10.1088/0004-637X/796/1/52},
  \href {http://adsabs.harvard.edu/abs/2014ApJ...796...52B} {796, 52}

\bibitem[\protect\citeauthoryear{{Bryan}, {Kay}, {Duffy}, {Schaye}, {Dalla
  Vecchia}  \& {Booth}}{{Bryan} et~al.}{2013}]{Bryan_2013}
{Bryan} S.~E.,  {Kay} S.~T.,  {Duffy} A.~R.,  {Schaye} J.,  {Dalla Vecchia} C.,
    {Booth} C.~M.,  2013, \mn@doi [\mnras] {10.1093/mnras/sts587}, \href
  {http://adsabs.harvard.edu/abs/2013MNRAS.429.3316B} {429, 3316}

\bibitem[\protect\citeauthoryear{{Bullock}, {Kolatt}, {Sigad}, {Somerville},
  {Kravtsov}, {Klypin}, {Primack}  \& {Dekel}}{{Bullock}
  et~al.}{2001}]{Bullock_2001}
{Bullock} J.~S.,  {Kolatt} T.~S.,  {Sigad} Y.,  {Somerville} R.~S.,  {Kravtsov}
  A.~V.,  {Klypin} A.~A.,  {Primack} J.~R.,   {Dekel} A.,  2001, \mn@doi
  [\mnras] {10.1046/j.1365-8711.2001.04068.x}, \href
  {http://adsabs.harvard.edu/abs/2001MNRAS.321..559B} {321, 559}

\bibitem[\protect\citeauthoryear{{Buote}, {Gastaldello}, {Humphrey},
  {Zappacosta}, {Bullock}, {Brighenti}  \& {Mathews}}{{Buote}
  et~al.}{2007}]{Buote_2007}
{Buote} D.~A.,  {Gastaldello} F.,  {Humphrey} P.~J.,  {Zappacosta} L.,
  {Bullock} J.~S.,  {Brighenti} F.,   {Mathews} W.~G.,  2007, \mn@doi [\apj]
  {10.1086/518684}, \href {http://adsabs.harvard.edu/abs/2007ApJ...664..123B}
  {664, 123}

\bibitem[\protect\citeauthoryear{{Cappellari} et~al.,}{{Cappellari}
  et~al.}{2011}]{Cappellari_2011}
{Cappellari} M.,  et~al., 2011, \mn@doi [\mnras]
  {10.1111/j.1365-2966.2010.18174.x}, \href
  {http://adsabs.harvard.edu/abs/2011MNRAS.413..813C} {413, 813}

\bibitem[\protect\citeauthoryear{{Cappellari} et~al.,}{{Cappellari}
  et~al.}{2013}]{Cappellari_2013b}
{Cappellari} M.,  et~al., 2013, \mn@doi [\mnras] {10.1093/mnras/stt562}, \href
  {http://adsabs.harvard.edu/abs/2013MNRAS.432.1709C} {432, 1709}

\bibitem[\protect\citeauthoryear{{Cappellari} et~al.,}{{Cappellari}
  et~al.}{2015}]{Cappellari_2015}
{Cappellari} M.,  et~al., 2015, \mn@doi [\apjl] {10.1088/2041-8205/804/1/L21},
  \href {http://adsabs.harvard.edu/abs/2015ApJ...804L..21C} {804, L21}

\bibitem[\protect\citeauthoryear{{Correa}, {Wyithe}, {Schaye}  \&
  {Duffy}}{{Correa} et~al.}{2015}]{Correa_2015}
{Correa} C.~A.,  {Wyithe} J.~S.~B.,  {Schaye} J.,   {Duffy} A.~R.,  2015,
  \mn@doi [\mnras] {10.1093/mnras/stv697}, \href
  {http://adsabs.harvard.edu/abs/2015MNRAS.450.1521C} {450, 1521}

\bibitem[\protect\citeauthoryear{{Corsini}, {Wegner}, {Thomas}, {Saglia}  \&
  {Bender}}{{Corsini} et~al.}{2017}]{Corsini_2016}
{Corsini} E.~M.,  {Wegner} G.~A.,  {Thomas} J.,  {Saglia} R.~P.,   {Bender} R.,
   2017, \mn@doi [\mnras] {10.1093/mnras/stw2935}, \href
  {http://adsabs.harvard.edu/abs/2017MNRAS.466..974C} {466, 974}

\bibitem[\protect\citeauthoryear{{C{\^o}t{\'e}}, {McLaughlin}, {Cohen}  \&
  {Blakeslee}}{{C{\^o}t{\'e}} et~al.}{2003}]{Cote_2003}
{C{\^o}t{\'e}} P.,  {McLaughlin} D.~E.,  {Cohen} J.~G.,   {Blakeslee} J.~P.,
  2003, \mn@doi [\apj] {10.1086/375488}, \href
  {http://adsabs.harvard.edu/abs/2003ApJ...591..850C} {591, 850}

\bibitem[\protect\citeauthoryear{{Courteau} \& {Dutton}}{{Courteau} \&
  {Dutton}}{2015}]{Courteau_2015}
{Courteau} S.,  {Dutton} A.~A.,  2015, \mn@doi [\apjl]
  {10.1088/2041-8205/801/2/L20}, \href
  {http://adsabs.harvard.edu/abs/2015ApJ...801L..20C} {801, L20}

\bibitem[\protect\citeauthoryear{{De Lucia} \& {Blaizot}}{{De Lucia} \&
  {Blaizot}}{2007}]{deLucia_2007}
{De Lucia} G.,  {Blaizot} J.,  2007, \mn@doi [\mnras]
  {10.1111/j.1365-2966.2006.11287.x}, \href
  {http://adsabs.harvard.edu/abs/2007MNRAS.375....2D} {375, 2}

\bibitem[\protect\citeauthoryear{{Deason}, {Belokurov}, {Evans}  \&
  {McCarthy}}{{Deason} et~al.}{2012}]{Deason_2012}
{Deason} A.~J.,  {Belokurov} V.,  {Evans} N.~W.,   {McCarthy} I.~G.,  2012,
  \mn@doi [\apj] {10.1088/0004-637X/748/1/2}, \href
  {http://adsabs.harvard.edu/abs/2012ApJ...748....2D} {748, 2}

\bibitem[\protect\citeauthoryear{{Dekel}, {Stoehr}, {Mamon}, {Cox}, {Novak}  \&
  {Primack}}{{Dekel} et~al.}{2005}]{Dekel_2005}
{Dekel} A.,  {Stoehr} F.,  {Mamon} G.~A.,  {Cox} T.~J.,  {Novak} G.~S.,
  {Primack} J.~R.,  2005, \mn@doi [\nat] {10.1038/nature03970}, \href
  {http://adsabs.harvard.edu/abs/2005Natur.437..707D} {437, 707}

\bibitem[\protect\citeauthoryear{{Di Cintio}, {Tremmel}, {Governato},
  {Pontzen}, {Zavala}, {Bastidas Fry}, {Brooks}  \& {Vogelsberger}}{{Di Cintio}
  et~al.}{2017}]{diCintio_2017}
{Di Cintio} A.,  {Tremmel} M.,  {Governato} F.,  {Pontzen} A.,  {Zavala} J.,
  {Bastidas Fry} A.,  {Brooks} A.,   {Vogelsberger} M.,  2017, preprint, \href
  {http://adsabs.harvard.edu/abs/2017arXiv170104410D} {} (\mn@eprint {arXiv}
  {1701.04410})

\bibitem[\protect\citeauthoryear{{Dutton}, {Conroy}, {van den Bosch}, {Prada}
  \& {More}}{{Dutton} et~al.}{2010}]{Dutton_2010}
{Dutton} A.~A.,  {Conroy} C.,  {van den Bosch} F.~C.,  {Prada} F.,   {More} S.,
   2010, \mn@doi [\mnras] {10.1111/j.1365-2966.2010.16911.x}, \href
  {http://adsabs.harvard.edu/abs/2010MNRAS.407....2D} {407, 2}

\bibitem[\protect\citeauthoryear{{Dutton} et~al.,}{{Dutton}
  et~al.}{2011}]{Dutton_2011}
{Dutton} A.~A.,  et~al., 2011, \mn@doi [\mnras]
  {10.1111/j.1365-2966.2011.19038.x}, \href
  {http://adsabs.harvard.edu/abs/2011MNRAS.416..322D} {416, 322}

\bibitem[\protect\citeauthoryear{{Emsellem} et~al.,}{{Emsellem}
  et~al.}{2011}]{Emsellem_2011}
{Emsellem} E.,  et~al., 2011, \mn@doi [\mnras]
  {10.1111/j.1365-2966.2011.18496.x}, \href
  {http://adsabs.harvard.edu/abs/2011MNRAS.414..888E} {414, 888}

\bibitem[\protect\citeauthoryear{{Forbes}, {Romanowsky}, {Pastorello},
  {Foster}, {Brodie}, {Strader}, {Usher}  \& {Pota}}{{Forbes}
  et~al.}{2016}]{Forbes_2016}
{Forbes} D.~A.,  {Romanowsky} A.~J.,  {Pastorello} N.,  {Foster} C.,  {Brodie}
  J.~P.,  {Strader} J.,  {Usher} C.,   {Pota} V.,  2016, \mn@doi [\mnras]
  {10.1093/mnras/stv3021}, \href
  {http://adsabs.harvard.edu/abs/2016MNRAS.457.1242F} {457, 1242}

\bibitem[\protect\citeauthoryear{{Forestell} \& {Gebhardt}}{{Forestell} \&
  {Gebhardt}}{2010}]{Forestell_2010}
{Forestell} A.~D.,  {Gebhardt} K.,  2010, \mn@doi [\apj]
  {10.1088/0004-637X/716/1/370}, \href
  {http://adsabs.harvard.edu/abs/2010ApJ...716..370F} {716, 370}

\bibitem[\protect\citeauthoryear{{Foster} et~al.,}{{Foster}
  et~al.}{2016}]{Foster_2016}
{Foster} C.,  et~al., 2016, \mn@doi [\mnras] {10.1093/mnras/stv2947}, \href
  {http://adsabs.harvard.edu/abs/2016MNRAS.457..147F} {457, 147}

\bibitem[\protect\citeauthoryear{{Genel}, {Bouch{\'e}}, {Naab}, {Sternberg}  \&
  {Genzel}}{{Genel} et~al.}{2010}]{Genel_2010}
{Genel} S.,  {Bouch{\'e}} N.,  {Naab} T.,  {Sternberg} A.,   {Genzel} R.,
  2010, \mn@doi [\apj] {10.1088/0004-637X/719/1/229}, \href
  {http://adsabs.harvard.edu/abs/2010ApJ...719..229G} {719, 229}

\bibitem[\protect\citeauthoryear{{Gentile}, {Salucci}, {Klein}, {Vergani}  \&
  {Kalberla}}{{Gentile} et~al.}{2004}]{Gentile_2004}
{Gentile} G.,  {Salucci} P.,  {Klein} U.,  {Vergani} D.,   {Kalberla} P.,
  2004, \mn@doi [\mnras] {10.1111/j.1365-2966.2004.07836.x}, \href
  {http://adsabs.harvard.edu/abs/2004MNRAS.351..903G} {351, 903}

\bibitem[\protect\citeauthoryear{{Governato} et~al.,}{{Governato}
  et~al.}{2010}]{Governato_2010}
{Governato} F.,  et~al., 2010, \mn@doi [\nat] {10.1038/nature08640}, \href
  {http://adsabs.harvard.edu/abs/2010Natur.463..203G} {463, 203}

\bibitem[\protect\citeauthoryear{{Graham}}{{Graham}}{2013}]{Graham_2013}
{Graham} A.~W.,  2013, {Elliptical and Disk Galaxy Structure and Modern Scaling
  Laws}.
p.~91, \mn@doi{10.1007/978-94-007-5609-0_2}

\bibitem[\protect\citeauthoryear{{Harris}, {Rejkuba}  \& {Harris}}{{Harris}
  et~al.}{2010}]{Harris_2010}
{Harris} G.~L.~H.,  {Rejkuba} M.,   {Harris} W.~E.,  2010, \mn@doi [\pasa]
  {10.1071/AS09061}, \href {http://adsabs.harvard.edu/abs/2010PASA...27..457H}
  {27, 457}

\bibitem[\protect\citeauthoryear{{Hu} \& {Kravtsov}}{{Hu} \&
  {Kravtsov}}{2003}]{Hu_2003}
{Hu} W.,  {Kravtsov} A.~V.,  2003, \mn@doi [\apj] {10.1086/345846}, \href
  {http://adsabs.harvard.edu/abs/2003ApJ...584..702H} {584, 702}

\bibitem[\protect\citeauthoryear{{Johansson}, {Naab}  \&
  {Ostriker}}{{Johansson} et~al.}{2009}]{Johansson_2009}
{Johansson} P.~H.,  {Naab} T.,   {Ostriker} J.~P.,  2009, \mn@doi [\apjl]
  {10.1088/0004-637X/697/1/L38}, \href
  {http://adsabs.harvard.edu/abs/2009ApJ...697L..38J} {697, L38}

\bibitem[\protect\citeauthoryear{{Johnson}, {Chakrabarty}, {O'Sullivan}  \&
  {Raychaudhury}}{{Johnson} et~al.}{2009}]{Johnson_2009}
{Johnson} R.,  {Chakrabarty} D.,  {O'Sullivan} E.,   {Raychaudhury} S.,  2009,
  \mn@doi [\apj] {10.1088/0004-637X/706/2/980}, \href
  {http://adsabs.harvard.edu/abs/2009ApJ...706..980J} {706, 980}

\bibitem[\protect\citeauthoryear{{Klypin}, {Yepes}, {Gottl{\"o}ber}, {Prada}
  \& {He{\ss}}}{{Klypin} et~al.}{2016}]{Klypin_2016}
{Klypin} A.,  {Yepes} G.,  {Gottl{\"o}ber} S.,  {Prada} F.,   {He{\ss}} S.,
  2016, \mn@doi [\mnras] {10.1093/mnras/stw248}, \href
  {http://adsabs.harvard.edu/abs/2016MNRAS.457.4340K} {457, 4340}

\bibitem[\protect\citeauthoryear{{Koleva}, {Prugniel}, {de Rijcke}  \&
  {Zeilinger}}{{Koleva} et~al.}{2011}]{Koleva_2011}
{Koleva} M.,  {Prugniel} P.,  {de Rijcke} S.,   {Zeilinger} W.~W.,  2011,
  \mn@doi [\mnras] {10.1111/j.1365-2966.2011.19057.x}, \href
  {http://adsabs.harvard.edu/abs/2011MNRAS.417.1643K} {417, 1643}

\bibitem[\protect\citeauthoryear{{Kormendy}, {Fisher}, {Cornell}  \&
  {Bender}}{{Kormendy} et~al.}{2009}]{Kormendy_2009}
{Kormendy} J.,  {Fisher} D.~B.,  {Cornell} M.~E.,   {Bender} R.,  2009, \mn@doi
  [\apjs] {10.1088/0067-0049/182/1/216}, \href
  {http://adsabs.harvard.edu/abs/2009ApJS..182..216K} {182, 216}

\bibitem[\protect\citeauthoryear{{Lange} et~al.,}{{Lange}
  et~al.}{2015}]{Lange_2015}
{Lange} R.,  et~al., 2015, \mn@doi [\mnras] {10.1093/mnras/stu2467}, \href
  {http://adsabs.harvard.edu/abs/2015MNRAS.447.2603L} {447, 2603}

\bibitem[\protect\citeauthoryear{{L{\"a}sker}, {Ferrarese}  \& {van de
  Ven}}{{L{\"a}sker} et~al.}{2014}]{Lasker_2014}
{L{\"a}sker} R.,  {Ferrarese} L.,   {van de Ven} G.,  2014, \mn@doi [\apj]
  {10.1088/0004-637X/780/1/69}, \href
  {http://adsabs.harvard.edu/abs/2014ApJ...780...69L} {780, 69}

\bibitem[\protect\citeauthoryear{{Macci{\`o}}, {Stinson}, {Brook}, {Wadsley},
  {Couchman}, {Shen}, {Gibson}  \& {Quinn}}{{Macci{\`o}}
  et~al.}{2012}]{Maccio_2012}
{Macci{\`o}} A.~V.,  {Stinson} G.,  {Brook} C.~B.,  {Wadsley} J.,  {Couchman}
  H.~M.~P.,  {Shen} S.,  {Gibson} B.~K.,   {Quinn} T.,  2012, \mn@doi [\apjl]
  {10.1088/2041-8205/744/1/L9}, \href
  {http://adsabs.harvard.edu/abs/2012ApJ...744L...9M} {744, L9}

\bibitem[\protect\citeauthoryear{{Makarov}, {Prugniel}, {Terekhova}, {Courtois}
   \& {Vauglin}}{{Makarov} et~al.}{2014}]{Hyperleda}
{Makarov} D.,  {Prugniel} P.,  {Terekhova} N.,  {Courtois} H.,   {Vauglin} I.,
  2014, \mn@doi [\aap] {10.1051/0004-6361/201423496}, \href
  {http://adsabs.harvard.edu/abs/2014A%26A...570A..13M} {570, A13}

\bibitem[\protect\citeauthoryear{{McDermid} et~al.,}{{McDermid}
  et~al.}{2015}]{McDermid_2015}
{McDermid} R.~M.,  et~al., 2015, \mn@doi [\mnras] {10.1093/mnras/stv105}, \href
  {http://adsabs.harvard.edu/abs/2015MNRAS.448.3484M} {448, 3484}

\bibitem[\protect\citeauthoryear{{McLaughlin}}{{McLaughlin}}{1999}]{McLaughlin_1999}
{McLaughlin} D.~E.,  1999, \mn@doi [\apjl] {10.1086/311860}, \href
  {http://adsabs.harvard.edu/abs/1999ApJ...512L...9M} {512, L9}

\bibitem[\protect\citeauthoryear{{Moody}, {Romanowsky}, {Cox}, {Novak}  \&
  {Primack}}{{Moody} et~al.}{2014}]{Moody_2014}
{Moody} C.~E.,  {Romanowsky} A.~J.,  {Cox} T.~J.,  {Novak} G.~S.,   {Primack}
  J.~R.,  2014, \mn@doi [\mnras] {10.1093/mnras/stu1444}, \href
  {http://adsabs.harvard.edu/abs/2014MNRAS.444.1475M} {444, 1475}

\bibitem[\protect\citeauthoryear{{Morganti}, {Gerhard}, {Coccato},
  {Martinez-Valpuesta}  \& {Arnaboldi}}{{Morganti}
  et~al.}{2013}]{Morganti_2013}
{Morganti} L.,  {Gerhard} O.,  {Coccato} L.,  {Martinez-Valpuesta} I.,
  {Arnaboldi} M.,  2013, \mn@doi [\mnras] {10.1093/mnras/stt442}, \href
  {http://adsabs.harvard.edu/abs/2013MNRAS.431.3570M} {431, 3570}

\bibitem[\protect\citeauthoryear{{Murphy}, {Gebhardt}  \& {Adams}}{{Murphy}
  et~al.}{2011}]{Murphy_2011}
{Murphy} J.~D.,  {Gebhardt} K.,   {Adams} J.~J.,  2011, \mn@doi [\apj]
  {10.1088/0004-637X/729/2/129}, \href
  {http://adsabs.harvard.edu/abs/2011ApJ...729..129M} {729, 129}

\bibitem[\protect\citeauthoryear{{Naab}, {Johansson}  \& {Ostriker}}{{Naab}
  et~al.}{2009}]{Naab_2009}
{Naab} T.,  {Johansson} P.~H.,   {Ostriker} J.~P.,  2009, \mn@doi [\apjl]
  {10.1088/0004-637X/699/2/L178}, \href
  {http://adsabs.harvard.edu/abs/2009ApJ...699L.178N} {699, L178}

\bibitem[\protect\citeauthoryear{{Napolitano} et~al.,}{{Napolitano}
  et~al.}{2005}]{Napolitano_2005}
{Napolitano} N.~R.,  et~al., 2005, \mn@doi [\mnras]
  {10.1111/j.1365-2966.2005.08683.x}, \href
  {http://adsabs.harvard.edu/abs/2005MNRAS.357..691N} {357, 691}

\bibitem[\protect\citeauthoryear{{Napolitano} et~al.,}{{Napolitano}
  et~al.}{2009}]{Napolitano_2009}
{Napolitano} N.~R.,  et~al., 2009, \mn@doi [\mnras]
  {10.1111/j.1365-2966.2008.14053.x}, \href
  {http://adsabs.harvard.edu/abs/2009MNRAS.393..329N} {393, 329}

\bibitem[\protect\citeauthoryear{{Napolitano}, {Romanowsky}  \&
  {Tortora}}{{Napolitano} et~al.}{2010}]{Napolitano_2010}
{Napolitano} N.~R.,  {Romanowsky} A.~J.,   {Tortora} C.,  2010, \mn@doi
  [\mnras] {10.1111/j.1365-2966.2010.16710.x}, \href
  {http://adsabs.harvard.edu/abs/2010MNRAS.405.2351N} {405, 2351}

\bibitem[\protect\citeauthoryear{{Napolitano} et~al.,}{{Napolitano}
  et~al.}{2011}]{Napolitano_2011}
{Napolitano} N.~R.,  et~al., 2011, \mn@doi [\mnras]
  {10.1111/j.1365-2966.2010.17833.x}, \href
  {http://adsabs.harvard.edu/abs/2011MNRAS.411.2035N} {411, 2035}

\bibitem[\protect\citeauthoryear{{Napolitano}, {Pota}, {Romanowsky}, {Forbes},
  {Brodie}  \& {Foster}}{{Napolitano} et~al.}{2014}]{Napolitano_2014}
{Napolitano} N.~R.,  {Pota} V.,  {Romanowsky} A.~J.,  {Forbes} D.~A.,  {Brodie}
  J.~P.,   {Foster} C.,  2014, \mn@doi [\mnras] {10.1093/mnras/stt2484}, \href
  {http://adsabs.harvard.edu/abs/2014MNRAS.439..659N} {439, 659}

\bibitem[\protect\citeauthoryear{{Navarro}, {Frenk}  \& {White}}{{Navarro}
  et~al.}{1996}]{Navarro_1996}
{Navarro} J.~F.,  {Frenk} C.~S.,   {White} S.~D.~M.,  1996, \mn@doi [\apj]
  {10.1086/177173}, \href {http://adsabs.harvard.edu/abs/1996ApJ...462..563N}
  {462, 563}

\bibitem[\protect\citeauthoryear{{Norris}, {Sharples}  \&
  {Kuntschner}}{{Norris} et~al.}{2006}]{Norris_2006}
{Norris} M.~A.,  {Sharples} R.~M.,   {Kuntschner} H.,  2006, \mn@doi [\mnras]
  {10.1111/j.1365-2966.2005.09992.x}, \href
  {http://adsabs.harvard.edu/abs/2006MNRAS.367..815N} {367, 815}

\bibitem[\protect\citeauthoryear{{Oldham} \& {Auger}}{{Oldham} \&
  {Auger}}{2016}]{Oldham_2016}
{Oldham} L.~J.,  {Auger} M.~W.,  2016, \mn@doi [\mnras]
  {10.1093/mnras/stv2982}, \href
  {http://adsabs.harvard.edu/abs/2016MNRAS.457..421O} {457, 421}

\bibitem[\protect\citeauthoryear{{Oser}, {Ostriker}, {Naab}, {Johansson}  \&
  {Burkert}}{{Oser} et~al.}{2010}]{Oser_2010}
{Oser} L.,  {Ostriker} J.~P.,  {Naab} T.,  {Johansson} P.~H.,   {Burkert} A.,
  2010, \mn@doi [\apj] {10.1088/0004-637X/725/2/2312}, \href
  {http://adsabs.harvard.edu/abs/2010ApJ...725.2312O} {725, 2312}

\bibitem[\protect\citeauthoryear{{Peebles}}{{Peebles}}{1982}]{Peebles_1982}
{Peebles} P.~J.~E.,  1982, \mn@doi [\apjl] {10.1086/183911}, \href
  {http://adsabs.harvard.edu/abs/1982ApJ...263L...1P} {263, L1}

\bibitem[\protect\citeauthoryear{{Persic}, {Salucci}  \& {Ashman}}{{Persic}
  et~al.}{1993}]{Persic_1993}
{Persic} M.,  {Salucci} P.,   {Ashman} K.~M.,  1993, \aap, \href
  {http://adsabs.harvard.edu/abs/1993A%26A...279..343P} {279, 343}

\bibitem[\protect\citeauthoryear{{Pillepich} et~al.,}{{Pillepich}
  et~al.}{2014}]{Pillepich_2014}
{Pillepich} A.,  et~al., 2014, \mn@doi [\mnras] {10.1093/mnras/stu1408}, \href
  {http://adsabs.harvard.edu/abs/2014MNRAS.444..237P} {444, 237}

\bibitem[\protect\citeauthoryear{{Planck Collaboration} et~al.,}{{Planck
  Collaboration} et~al.}{2014}]{Planck_Collaboration_2013}
{Planck Collaboration} et~al., 2014, \mn@doi [\aap]
  {10.1051/0004-6361/201321591}, \href
  {http://adsabs.harvard.edu/abs/2014A%26A...571A..16P} {571, A16}

\bibitem[\protect\citeauthoryear{{Pota} et~al.,}{{Pota}
  et~al.}{2013}]{Pota_2013}
{Pota} V.,  et~al., 2013, \mn@doi [\mnras] {10.1093/mnras/sts029}, \href
  {http://adsabs.harvard.edu/abs/2013MNRAS.428..389P} {428, 389}

\bibitem[\protect\citeauthoryear{{Pota} et~al.,}{{Pota}
  et~al.}{2015}]{Pota_2015}
{Pota} V.,  et~al., 2015, \mn@doi [\mnras] {10.1093/mnras/stv831}, \href
  {http://adsabs.harvard.edu/abs/2015MNRAS.450.3345P} {450, 3345}

\bibitem[\protect\citeauthoryear{{Raskutti}, {Greene}  \& {Murphy}}{{Raskutti}
  et~al.}{2014}]{Raskutti_2014}
{Raskutti} S.,  {Greene} J.~E.,   {Murphy} J.~D.,  2014, \mn@doi [\apj]
  {10.1088/0004-637X/786/1/23}, \href
  {http://adsabs.harvard.edu/abs/2014ApJ...786...23R} {786, 23}

\bibitem[\protect\citeauthoryear{{Rembold}, {Pastoriza}  \&
  {Bruzual}}{{Rembold} et~al.}{2005}]{Rembold_2005}
{Rembold} S.~B.,  {Pastoriza} M.~G.,   {Bruzual} G.,  2005, \mn@doi [\aap]
  {10.1051/0004-6361:20042464}, \href
  {http://adsabs.harvard.edu/abs/2005A%26A...436...57R} {436, 57}

\bibitem[\protect\citeauthoryear{{Remus}, {Dolag}, {Naab}, {Burkert},
  {Hirschmann}, {Hoffmann}  \& {Johansson}}{{Remus} et~al.}{2017}]{Remus_2016}
{Remus} R.-S.,  {Dolag} K.,  {Naab} T.,  {Burkert} A.,  {Hirschmann} M.,
  {Hoffmann} T.~L.,   {Johansson} P.~H.,  2017, \mn@doi [\mnras]
  {10.1093/mnras/stw2594}, \href
  {http://adsabs.harvard.edu/abs/2017MNRAS.464.3742R} {464, 3742}

\bibitem[\protect\citeauthoryear{{Richtler}, {Hilker}, {Kumar}, {Bassino},
  {G{\'o}mez}  \& {Dirsch}}{{Richtler} et~al.}{2014}]{Richtler_2014}
{Richtler} T.,  {Hilker} M.,  {Kumar} B.,  {Bassino} L.~P.,  {G{\'o}mez} M.,
  {Dirsch} B.,  2014, \mn@doi [\aap] {10.1051/0004-6361/201423525}, \href
  {http://adsabs.harvard.edu/abs/2014A%26A...569A..41R} {569, A41}

\bibitem[\protect\citeauthoryear{{Rocha}, {Peter}, {Bullock}, {Kaplinghat},
  {Garrison-Kimmel}, {O{\~n}orbe}  \& {Moustakas}}{{Rocha}
  et~al.}{2013}]{Rocha_2013}
{Rocha} M.,  {Peter} A.~H.~G.,  {Bullock} J.~S.,  {Kaplinghat} M.,
  {Garrison-Kimmel} S.,  {O{\~n}orbe} J.,   {Moustakas} L.~A.,  2013, \mn@doi
  [\mnras] {10.1093/mnras/sts514}, \href
  {http://adsabs.harvard.edu/abs/2013MNRAS.430...81R} {430, 81}

\bibitem[\protect\citeauthoryear{{R{\"o}ck}, {Vazdekis}, {Ricciardelli},
  {Peletier}, {Knapen}  \& {Falc{\'o}n-Barroso}}{{R{\"o}ck}
  et~al.}{2016}]{Rock_2016}
{R{\"o}ck} B.,  {Vazdekis} A.,  {Ricciardelli} E.,  {Peletier} R.~F.,  {Knapen}
  J.~H.,   {Falc{\'o}n-Barroso} J.,  2016, \mn@doi [\aap]
  {10.1051/0004-6361/201527570}, \href
  {http://adsabs.harvard.edu/abs/2016A%26A...589A..73R} {589, A73}

\bibitem[\protect\citeauthoryear{{Rodriguez-Gomez} et~al.,}{{Rodriguez-Gomez}
  et~al.}{2016}]{Rodriguez_2016}
{Rodriguez-Gomez} V.,  et~al., 2016, \mn@doi [\mnras] {10.1093/mnras/stw456},
  \href {http://adsabs.harvard.edu/abs/2016MNRAS.458.2371R} {458, 2371}

\bibitem[\protect\citeauthoryear{{Romanowsky}, {Douglas}, {Arnaboldi},
  {Kuijken}, {Merrifield}, {Napolitano}, {Capaccioli}  \&
  {Freeman}}{{Romanowsky} et~al.}{2003}]{Romanowsky_2003}
{Romanowsky} A.~J.,  {Douglas} N.~G.,  {Arnaboldi} M.,  {Kuijken} K.,
  {Merrifield} M.~R.,  {Napolitano} N.~R.,  {Capaccioli} M.,   {Freeman} K.~C.,
   2003, \mn@doi [Science] {10.1126/science.1087441}, \href
  {http://adsabs.harvard.edu/abs/2003Sci...301.1696R} {301, 1696}

\bibitem[\protect\citeauthoryear{{Romanowsky}, {Strader}, {Spitler}, {Johnson},
  {Brodie}, {Forbes}  \& {Ponman}}{{Romanowsky} et~al.}{2009}]{Romanowsky_2009}
{Romanowsky} A.~J.,  {Strader} J.,  {Spitler} L.~R.,  {Johnson} R.,  {Brodie}
  J.~P.,  {Forbes} D.~A.,   {Ponman} T.,  2009, \mn@doi [\aj]
  {10.1088/0004-6256/137/6/4956}, \href
  {http://adsabs.harvard.edu/abs/2009AJ....137.4956R} {137, 4956}

\bibitem[\protect\citeauthoryear{{R{\"o}ttgers}, {Naab}  \&
  {Oser}}{{R{\"o}ttgers} et~al.}{2014}]{Rottgers_2014}
{R{\"o}ttgers} B.,  {Naab} T.,   {Oser} L.,  2014, \mn@doi [\mnras]
  {10.1093/mnras/stu1762}, \href
  {http://adsabs.harvard.edu/abs/2014MNRAS.445.1065R} {445, 1065}

\bibitem[\protect\citeauthoryear{{Samurovi{\'c}}}{{Samurovi{\'c}}}{2014}]{Samurovic_2014}
{Samurovi{\'c}} S.,  2014, \mn@doi [\aap] {10.1051/0004-6361/201321459}, \href
  {http://adsabs.harvard.edu/abs/2014A%26A...570A.132S} {570, A132}

\bibitem[\protect\citeauthoryear{{Samurovi{\'c}}}{{Samurovi{\'c}}}{2016}]{Samurovic_2016}
{Samurovi{\'c}} S.,  2016, \mn@doi [\apss] {10.1007/s10509-016-2789-x}, \href
  {http://adsabs.harvard.edu/abs/2016Ap%26SS.361..199S} {361, \#199}

\bibitem[\protect\citeauthoryear{{S{\'a}nchez-Bl{\'a}zquez}, {Gorgas},
  {Cardiel}  \& {Gonz{\'a}lez}}{{S{\'a}nchez-Bl{\'a}zquez}
  et~al.}{2006}]{Sanchez_2006}
{S{\'a}nchez-Bl{\'a}zquez} P.,  {Gorgas} J.,  {Cardiel} N.,   {Gonz{\'a}lez}
  J.~J.,  2006, \mn@doi [\aap] {10.1051/0004-6361:20064845}, \href
  {http://adsabs.harvard.edu/abs/2006A%26A...457..809S} {457, 809}

\bibitem[\protect\citeauthoryear{{Sani}, {Marconi}, {Hunt}  \&
  {Risaliti}}{{Sani} et~al.}{2011}]{Sani_2010}
{Sani} E.,  {Marconi} A.,  {Hunt} L.~K.,   {Risaliti} G.,  2011, \mn@doi
  [\mnras] {10.1111/j.1365-2966.2011.18229.x}, \href
  {http://adsabs.harvard.edu/abs/2011MNRAS.413.1479S} {413, 1479}

\bibitem[\protect\citeauthoryear{{Schaye} et~al.,}{{Schaye}
  et~al.}{2015}]{Schaye_2015}
{Schaye} J.,  et~al., 2015, \mn@doi [\mnras] {10.1093/mnras/stu2058}, \href
  {http://adsabs.harvard.edu/abs/2015MNRAS.446..521S} {446, 521}

\bibitem[\protect\citeauthoryear{{Schuberth}, {Richtler}, {Hilker}, {Dirsch},
  {Bassino}, {Romanowsky}  \& {Infante}}{{Schuberth}
  et~al.}{2010}]{Schuberth_2010}
{Schuberth} Y.,  {Richtler} T.,  {Hilker} M.,  {Dirsch} B.,  {Bassino} L.~P.,
  {Romanowsky} A.~J.,   {Infante} L.,  2010, \mn@doi [\aap]
  {10.1051/0004-6361/200912482}, \href
  {http://adsabs.harvard.edu/abs/2010A%26A...513A..52S} {513, A52}

\bibitem[\protect\citeauthoryear{{Schuberth}, {Richtler}, {Hilker}, {Salinas},
  {Dirsch}  \& {Larsen}}{{Schuberth} et~al.}{2012}]{Schuberth_2012}
{Schuberth} Y.,  {Richtler} T.,  {Hilker} M.,  {Salinas} R.,  {Dirsch} B.,
  {Larsen} S.~S.,  2012, \mn@doi [\aap] {10.1051/0004-6361/201015038}, \href
  {http://adsabs.harvard.edu/abs/2012A%26A...544A.115S} {544, A115}

\bibitem[\protect\citeauthoryear{{Scott}, {Graham}  \& {Schombert}}{{Scott}
  et~al.}{2013}]{Scott_2013}
{Scott} N.,  {Graham} A.~W.,   {Schombert} J.,  2013, \mn@doi [\apj]
  {10.1088/0004-637X/768/1/76}, \href
  {http://adsabs.harvard.edu/abs/2013ApJ...768...76S} {768, 76}

\bibitem[\protect\citeauthoryear{{S\'ersic}}{{S\'ersic}}{1968}]{Sersic_1968}
{S\'ersic} J.~L.,  1968, {Atlas de galaxias australes}

\bibitem[\protect\citeauthoryear{{Shen} \& {Gebhardt}}{{Shen} \&
  {Gebhardt}}{2010}]{Shen_2010}
{Shen} J.,  {Gebhardt} K.,  2010, \mn@doi [\apj] {10.1088/0004-637X/711/1/484},
  \href {http://adsabs.harvard.edu/abs/2010ApJ...711..484S} {711, 484}

\bibitem[\protect\citeauthoryear{{Smith}, {Choi}, {Lee}, {Rhee},
  {Sanchez-Janssen}  \& {Yi}}{{Smith} et~al.}{2016}]{Smith_2016}
{Smith} R.,  {Choi} H.,  {Lee} J.,  {Rhee} J.,  {Sanchez-Janssen} R.,   {Yi}
  S.~K.,  2016, \mn@doi [\apj] {10.3847/1538-4357/833/1/109}, \href
  {http://adsabs.harvard.edu/abs/2016ApJ...833..109S} {833, 109}

\bibitem[\protect\citeauthoryear{{Spolaor}, {Forbes}, {Proctor}, {Hau}  \&
  {Brough}}{{Spolaor} et~al.}{2008}]{Spolaor_2008}
{Spolaor} M.,  {Forbes} D.~A.,  {Proctor} R.~N.,  {Hau} G.~K.~T.,   {Brough}
  S.,  2008, \mn@doi [\mnras] {10.1111/j.1365-2966.2008.12892.x}, \href
  {http://adsabs.harvard.edu/abs/2008MNRAS.385..675S} {385, 675}

\bibitem[\protect\citeauthoryear{{Strader} et~al.,}{{Strader}
  et~al.}{2011}]{Strader_2011}
{Strader} J.,  et~al., 2011, \mn@doi [\apjs] {10.1088/0067-0049/197/2/33},
  \href {http://adsabs.harvard.edu/abs/2011ApJS..197...33S} {197, 33}

\bibitem[\protect\citeauthoryear{{Su}, {Gu}, {White}  \& {Irwin}}{{Su}
  et~al.}{2014}]{Su_2014}
{Su} Y.,  {Gu} L.,  {White} III R.~E.,   {Irwin} J.,  2014, \mn@doi [\apj]
  {10.1088/0004-637X/786/2/152}, \href
  {http://adsabs.harvard.edu/abs/2014ApJ...786..152S} {786, 152}

\bibitem[\protect\citeauthoryear{{Taylor}}{{Taylor}}{2005}]{Taylor_2005}
{Taylor} M.~B.,  2005, in {Shopbell} P.,  {Britton} M.,   {Ebert} R.,  eds,
  Astronomical Society of the Pacific Conference Series Vol. 347, Astronomical
  Data Analysis Software and Systems XIV. p.~29

\bibitem[\protect\citeauthoryear{{Terlevich} \& {Forbes}}{{Terlevich} \&
  {Forbes}}{2002}]{Terlevich_2002}
{Terlevich} A.~I.,  {Forbes} D.~A.,  2002, \mn@doi [\mnras]
  {10.1046/j.1365-8711.2002.05073.x}, \href
  {http://adsabs.harvard.edu/abs/2002MNRAS.330..547T} {330, 547}

\bibitem[\protect\citeauthoryear{{Terzi{\'c}} \& {Graham}}{{Terzi{\'c}} \&
  {Graham}}{2005}]{Terzic_2005}
{Terzi{\'c}} B.,  {Graham} A.~W.,  2005, \mn@doi [\mnras]
  {10.1111/j.1365-2966.2005.09269.x}, \href
  {http://adsabs.harvard.edu/abs/2005MNRAS.362..197T} {362, 197}

\bibitem[\protect\citeauthoryear{{Thomas}, {Saglia}, {Bender}, {Thomas},
  {Gebhardt}, {Magorrian}, {Corsini}  \& {Wegner}}{{Thomas}
  et~al.}{2009}]{Thomas_2009}
{Thomas} J.,  {Saglia} R.~P.,  {Bender} R.,  {Thomas} D.,  {Gebhardt} K.,
  {Magorrian} J.,  {Corsini} E.~M.,   {Wegner} G.,  2009, \mn@doi [\apj]
  {10.1088/0004-637X/691/1/770}, \href
  {http://adsabs.harvard.edu/abs/2009ApJ...691..770T} {691, 770}

\bibitem[\protect\citeauthoryear{{Toloba} et~al.,}{{Toloba}
  et~al.}{2016}]{Toloba_2016}
{Toloba} E.,  et~al., 2016, \mn@doi [\apj] {10.3847/0004-637X/822/1/51}, \href
  {http://adsabs.harvard.edu/abs/2016ApJ...822...51T} {822, 51}

\bibitem[\protect\citeauthoryear{{Tortora}, {Napolitano}, {Romanowsky},
  {Capaccioli}  \& {Covone}}{{Tortora} et~al.}{2009}]{Tortora_2009}
{Tortora} C.,  {Napolitano} N.~R.,  {Romanowsky} A.~J.,  {Capaccioli} M.,
  {Covone} G.,  2009, \mn@doi [\mnras] {10.1111/j.1365-2966.2009.14789.x},
  \href {http://adsabs.harvard.edu/abs/2009MNRAS.396.1132T} {396, 1132}

\bibitem[\protect\citeauthoryear{{Trager}, {Faber}, {Worthey}  \&
  {Gonz{\'a}lez}}{{Trager} et~al.}{2000}]{Trager_2000}
{Trager} S.~C.,  {Faber} S.~M.,  {Worthey} G.,   {Gonz{\'a}lez} J.~J.,  2000,
  \mn@doi [\aj] {10.1086/301299}, \href
  {http://adsabs.harvard.edu/abs/2000AJ....119.1645T} {119, 1645}

\bibitem[\protect\citeauthoryear{{Vogelsberger} et~al.,}{{Vogelsberger}
  et~al.}{2014}]{Vogelsberger_2014}
{Vogelsberger} M.,  et~al., 2014, \mn@doi [\mnras] {10.1093/mnras/stu1536},
  \href {http://adsabs.harvard.edu/abs/2014MNRAS.444.1518V} {444, 1518}

\bibitem[\protect\citeauthoryear{{Watkins}, {Evans}  \& {An}}{{Watkins}
  et~al.}{2010}]{Watkins_2010}
{Watkins} L.~L.,  {Evans} N.~W.,   {An} J.~H.,  2010, \mn@doi [\mnras]
  {10.1111/j.1365-2966.2010.16708.x}, \href
  {http://adsabs.harvard.edu/abs/2010MNRAS.406..264W} {406, 264}

\bibitem[\protect\citeauthoryear{{Wechsler}, {Bullock}, {Primack}, {Kravtsov}
  \& {Dekel}}{{Wechsler} et~al.}{2002}]{Wechsler_2002}
{Wechsler} R.~H.,  {Bullock} J.~S.,  {Primack} J.~R.,  {Kravtsov} A.~V.,
  {Dekel} A.,  2002, \mn@doi [\apj] {10.1086/338765}, \href
  {http://adsabs.harvard.edu/abs/2002ApJ...568...52W} {568, 52}

\bibitem[\protect\citeauthoryear{{Wegner}, {Corsini}, {Thomas}, {Saglia},
  {Bender}  \& {Pu}}{{Wegner} et~al.}{2012}]{Wegner_2012}
{Wegner} G.~A.,  {Corsini} E.~M.,  {Thomas} J.,  {Saglia} R.~P.,  {Bender} R.,
   {Pu} S.~B.,  2012, \mn@doi [\aj] {10.1088/0004-6256/144/3/78}, \href
  {http://adsabs.harvard.edu/abs/2012AJ....144...78W} {144, 78}

\bibitem[\protect\citeauthoryear{{Weijmans}, {Krajnovi{\'c}}, {van de Ven},
  {Oosterloo}, {Morganti}  \& {de Zeeuw}}{{Weijmans}
  et~al.}{2008}]{Weijmans_2008}
{Weijmans} A.-M.,  {Krajnovi{\'c}} D.,  {van de Ven} G.,  {Oosterloo} T.~A.,
  {Morganti} R.,   {de Zeeuw} P.~T.,  2008, \mn@doi [\mnras]
  {10.1111/j.1365-2966.2007.12680.x}, \href
  {http://adsabs.harvard.edu/abs/2008MNRAS.383.1343W} {383, 1343}

\bibitem[\protect\citeauthoryear{{Woodley}, {G{\'o}mez}, {Harris}, {Geisler}
  \& {Harris}}{{Woodley} et~al.}{2010}]{Woodley_2010}
{Woodley} K.~A.,  {G{\'o}mez} M.,  {Harris} W.~E.,  {Geisler} D.,   {Harris}
  G.~L.~H.,  2010, \mn@doi [\aj] {10.1088/0004-6256/139/5/1871}, \href
  {http://adsabs.harvard.edu/abs/2010AJ....139.1871W} {139, 1871}

\bibitem[\protect\citeauthoryear{{Wu}, {Gerhard}, {Naab}, {Oser},
  {Martinez-Valpuesta}, {Hilz}, {Churazov}  \& {Lyskova}}{{Wu}
  et~al.}{2014}]{Wu_2014}
{Wu} X.,  {Gerhard} O.,  {Naab} T.,  {Oser} L.,  {Martinez-Valpuesta} I.,
  {Hilz} M.,  {Churazov} E.,   {Lyskova} N.,  2014, \mn@doi [\mnras]
  {10.1093/mnras/stt2415}, \href
  {http://adsabs.harvard.edu/abs/2014MNRAS.438.2701W} {438, 2701}

\bibitem[\protect\citeauthoryear{{Zhang}, {Xu}, {Wang}, {An}, {Xu}  \&
  {Wu}}{{Zhang} et~al.}{2007}]{Zhang_2007}
{Zhang} Z.,  {Xu} H.,  {Wang} Y.,  {An} T.,  {Xu} Y.,   {Wu} X.-P.,  2007,
  \mn@doi [\apj] {10.1086/510281}, \href
  {http://adsabs.harvard.edu/abs/2007ApJ...656..805Z} {656, 805}

\bibitem[\protect\citeauthoryear{{Zhang} et~al.,}{{Zhang}
  et~al.}{2015}]{Zhang_2015}
{Zhang} H.-X.,  et~al., 2015, \mn@doi [\apj] {10.1088/0004-637X/802/1/30},
  \href {http://adsabs.harvard.edu/abs/2015ApJ...802...30Z} {802, 30}

\bibitem[\protect\citeauthoryear{{Zhao}, {Mo}, {Jing}  \& {B{\"o}rner}}{{Zhao}
  et~al.}{2003}]{Zhao_2003}
{Zhao} D.~H.,  {Mo} H.~J.,  {Jing} Y.~P.,   {B{\"o}rner} G.,  2003, \mn@doi
  [\mnras] {10.1046/j.1365-8711.2003.06135.x}, \href
  {http://adsabs.harvard.edu/abs/2003MNRAS.339...12Z} {339, 12}

\bibitem[\protect\citeauthoryear{{Zhu} et~al.,}{{Zhu} et~al.}{2014}]{Zhu_2014}
{Zhu} L.,  et~al., 2014, \mn@doi [\apj] {10.1088/0004-637X/792/1/59}, \href
  {http://adsabs.harvard.edu/abs/2014ApJ...792...59Z} {792, 59}

\bibitem[\protect\citeauthoryear{{Zhu} et~al.,}{{Zhu} et~al.}{2016}]{Zhu_2016}
{Zhu} L.,  et~al., 2016, \mn@doi [\mnras] {10.1093/mnras/stw1931}, \href
  {http://adsabs.harvard.edu/abs/2016MNRAS.462.4001Z} {462, 4001}

\bibitem[\protect\citeauthoryear{{de Vaucouleurs}, {de Vaucouleurs}, {Corwin},
  {Buta}, {Paturel}  \& {Fouqu{\'e}}}{{de Vaucouleurs}
  et~al.}{1991}]{de_Vaucouleurs_1991}
{de Vaucouleurs} G.,  {de Vaucouleurs} A.,  {Corwin} Jr. H.~G.,  {Buta} R.~J.,
  {Paturel} G.,   {Fouqu{\'e}} P.,  1991, {Third Reference Catalogue of Bright
  Galaxies. Volume I: Explanations and references. Volume II: Data for galaxies
  between 0$^{h}$ and 12$^{h}$. Volume III: Data for galaxies between 12$^{h}$
  and 24$^{h}$.}

\bibitem[\protect\citeauthoryear{{van Dokkum} et~al.,}{{van Dokkum}
  et~al.}{2008}]{vanDokkum_2008}
{van Dokkum} P.~G.,  et~al., 2008, \mn@doi [\apjl] {10.1086/587874}, \href
  {http://adsabs.harvard.edu/abs/2008ApJ...677L...5V} {677, L5}

\makeatother
\end{thebibliography}

\appendix
\newpage

\subsection{Extreme velocity anisotropy, total mass estimates and dark matter fractions}
\label{extreme_ani}
We investigate if the results we have obtained assuming mild velocity anisotropies are robust against extreme anisotropies, by adopting a more extreme velocity anisotropy, i.e. $\beta=-1.0$. This is motivated by results from some dynamical studies (e.g. \citealt{Pota_2015, Zhang_2015}) and cosmological simulations (e.g. \citealt{Rottgers_2014}) where such anisotropies were obtained. There are also indications from cosmological simulations (\citealt{Bryan_2013, Rottgers_2014, Wu_2014}) that the stellar velocity anisotropy out to 5~$\re$ correlates with the fraction of stars that formed in-situ. The nature of the reported correlation is such that in galaxies with low in-situ stellar fractions i.e. galaxies where the late mass assembly is dominated by dry mergers, mostly the slow rotators, the anisotropy is mildly radial ($\beta > 0.2-0.4$). Galaxies with high in-situ stellar fractions, on the other hand, show strongly tangential up to isotropic anisotropies, i.e. $-1.0 \leq \beta < 0.2$. This anisotropy range is similar to what we now explore, however, GCs may not have the same velocity anisotropy as stars. From the few massive galaxies with published GC anisotropy profiles, it is difficult to pick out a clear pattern (e.g \citealt{Pota_2013, Zhu_2016}); see also \citealt{Pota_2015, Zhang_2015}, where GCs are reported to be have strongly tangential anisotropies at large radii. Moreso, in the lower $M_*$ galaxies, GC kinematics data are mostly too sparse to extract any anisotropy information, although from the PNe data associated with these galaxies, the trend is one where the PNe are isotropic near the galaxy centre and radially biased around 5~$\re$. Again, PNe and GCs may not have similar anisotropies.

\begin{figure}
    \includegraphics[width=0.48\textwidth]{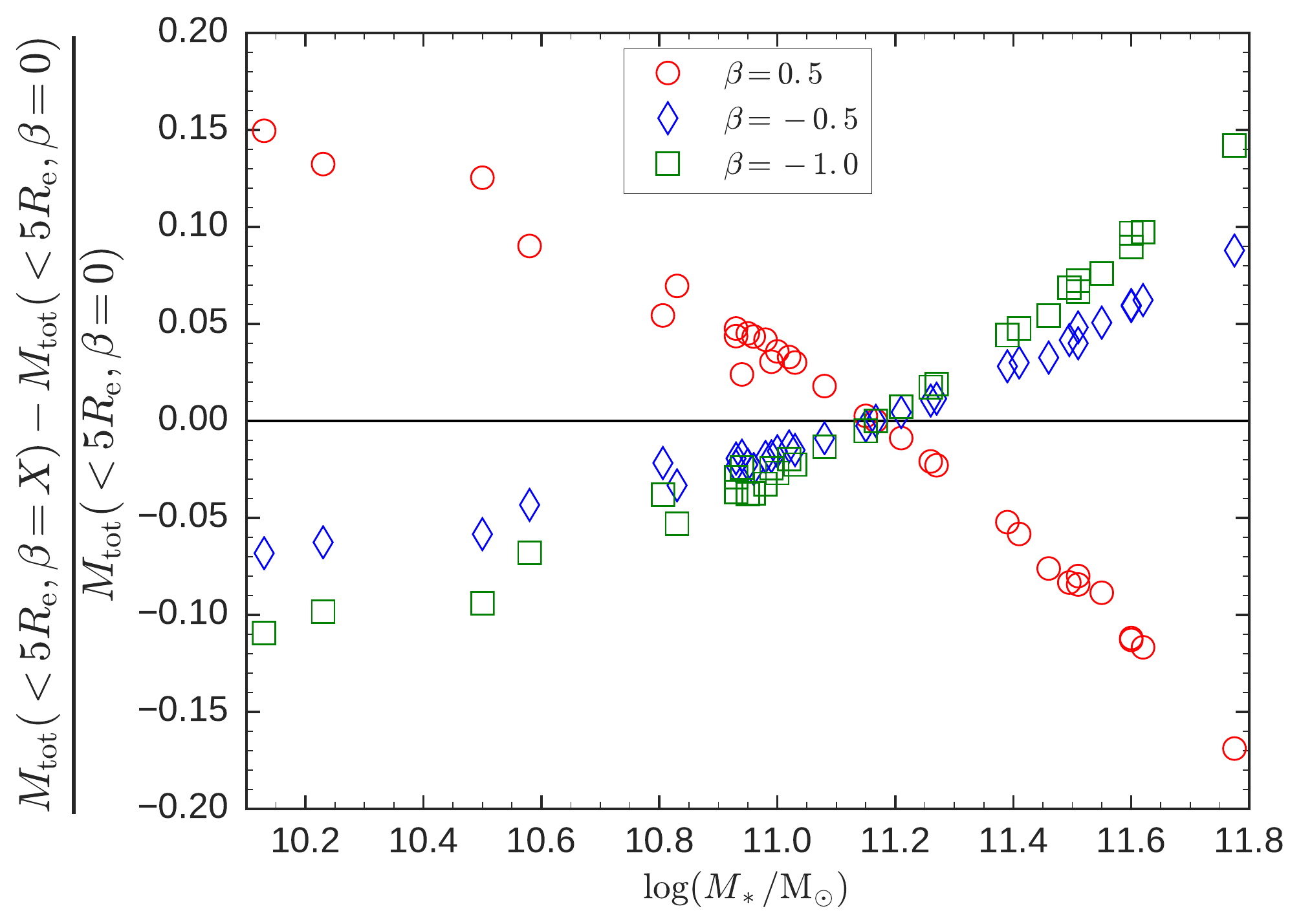}\hspace{0.01\textwidth}\\
    \includegraphics[width=0.48\textwidth]{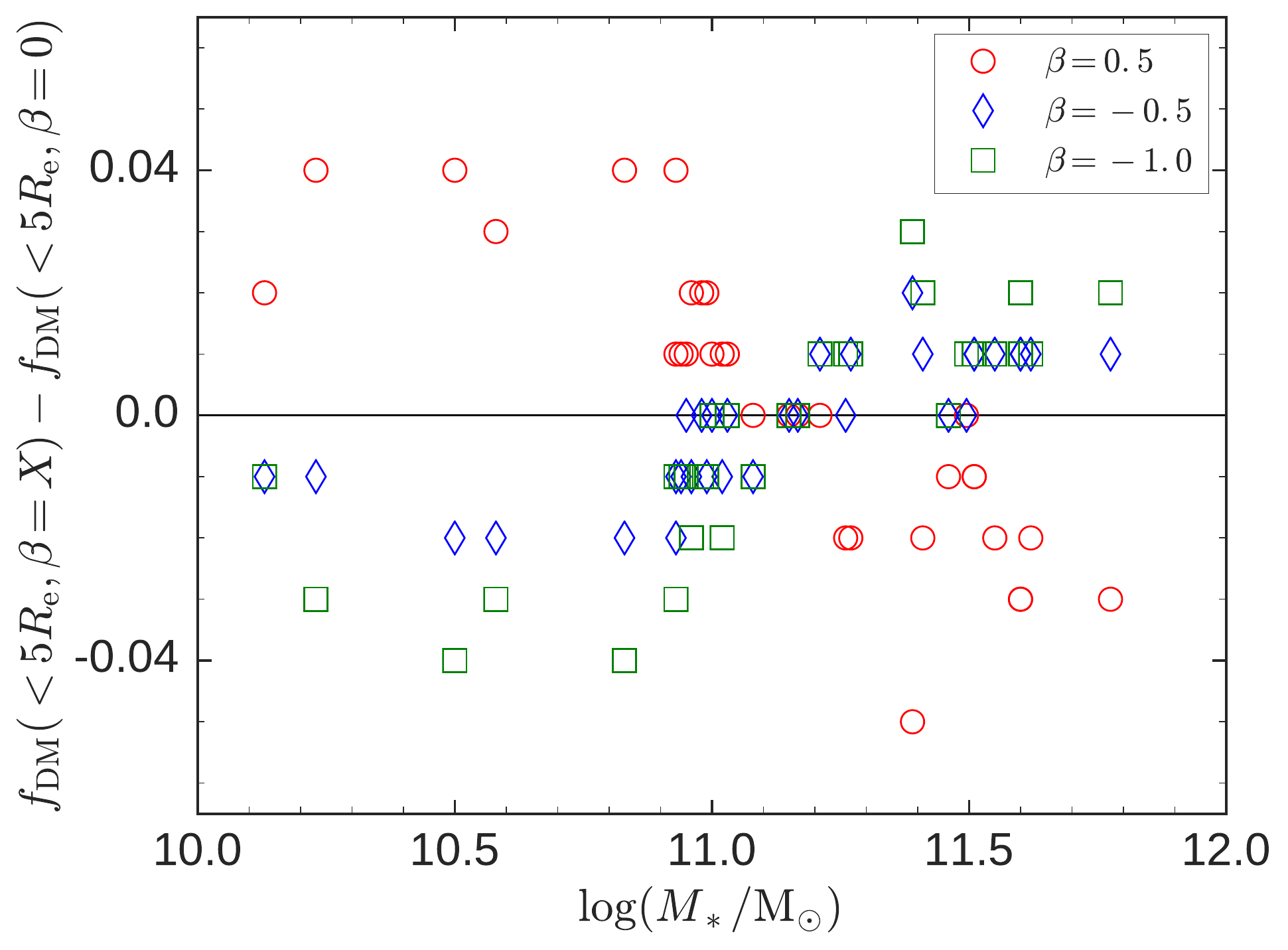}\hspace{0.01\textwidth}\\
	\caption{\label{fig:appendix_a} \textit{Top panel}: Fractional change in total mass within 5~$\re$ for different velocity anisotropy assumptions and, \textit{bbottom panel}: Corresponding change in dark matter fraction within 5~$\re$ for different velocity anisotropy assumptions.}
\end{figure}

The top panel in Figure \ref{fig:appendix_a} shows the fractional changes in $M_{\rm tot}$ as a function of $M_*$ while the bottom panel shows the corresponding changes in $\fdm$ versus $M_*$, for different anisotropy assumptions. Note that assuming a more strongly tangential anisotropy results in an increase in $M_{\rm tot}$ and $\fdm$ only in the most mass massive galaxies and the opposite effects in the lower $M_*$ galaxies in our sample. If we assume a correlation between $\beta$ and $M_*$ that maximises $M_{\rm tot}$ within 5~$\re$, the fractional change in $M_{\rm tot}$ is $<0.2$ dex, and this results in a $< 0.1$ change in $\fdm$. Around ${\rm log~}(M_*/\Msun) \sim 11$, where we measure our the lowest $\fdm$, we now observe the least change in $M_{\rm tot}$. While it is interesting to understand the nature and systematics of GC velocity anisotropy, our analysis suggest that its effect on the total mass estimates and dark matter fractions within large radii is minimal.

\subsection{Variation of $\rs/\re$ with stellar mass}
\label{rs_re}
This plot is in reference to how stellar mass varies with the ratio of the scale radius of the dark matter halo and galaxy size from our simple galaxy model (SGM1). The minimum in $\rs/\re$ observed at ${\rm log~}(M_*/\Msun) \sim 11$ implies that at this stellar mass,  galaxies with may already be structurally different.

\begin{figure}
    \includegraphics[width=0.48\textwidth]{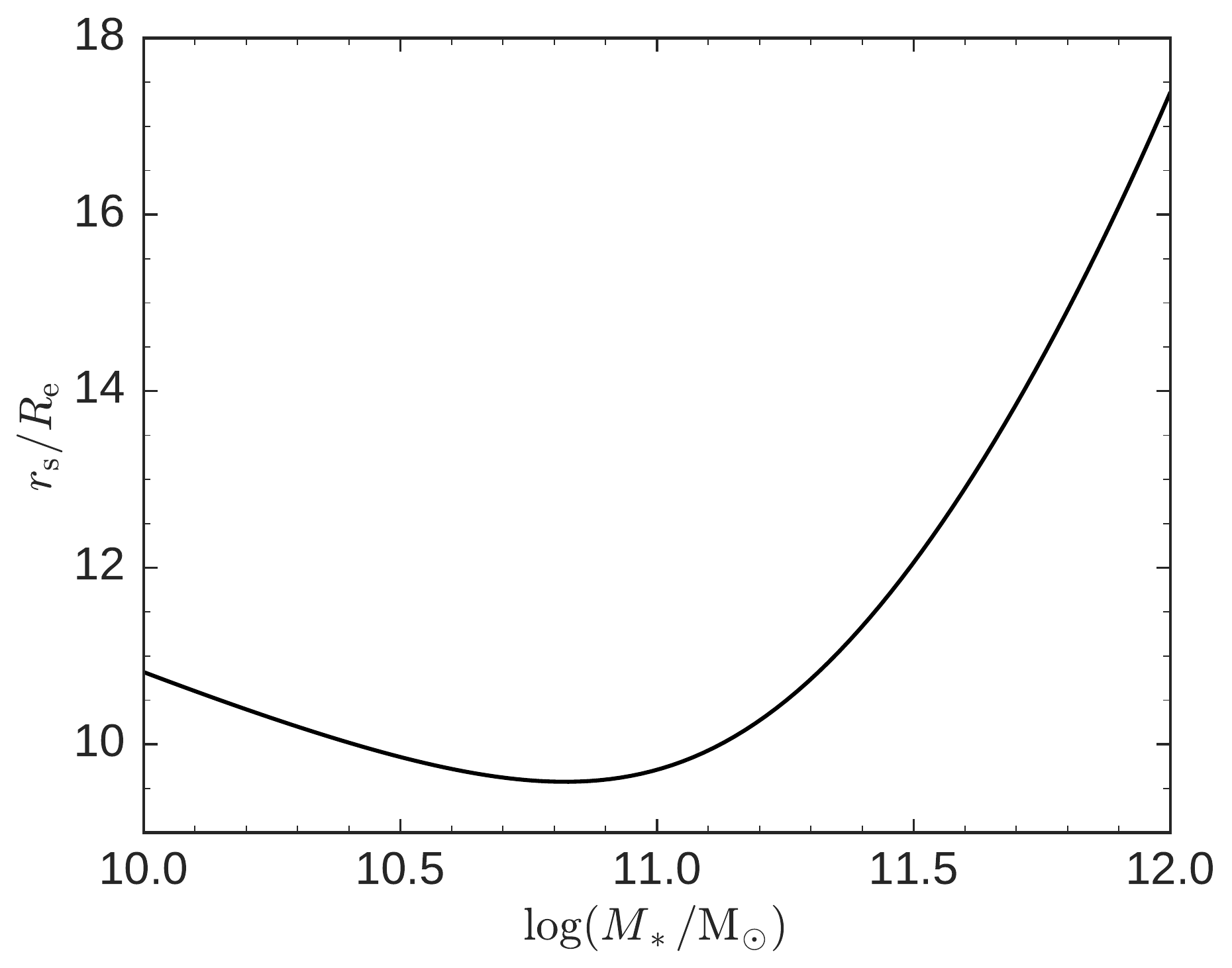}\hspace{0.01\textwidth}\\
	\caption{\label{fig:appendix_b} \textit{Top panel}: Variation of $\rs/\re$ with stellar mass in our simple galaxy model (SGM1). The ratio of the scale radius of the dark matter halo to galaxy size reaches a minimum around ${\rm log~}(M_*/\Msun) \sim 11$. }
\end{figure}

\subsection{Halo assembly epoch as a function of galaxy properties, without binning by stellar mass}
Here, we show a version of Figure \ref{fig:halo_gal}, without binning our galaxies by stellar mass. We have also included galaxies with ambiguous morphological classification. Note that the late halo assembly epoch for galaxies in the field is not obvious  without the correction we have applied to account for the strong dependence of $z_{\rm form}$ on galaxy stellar mass.

\begin{figure*}
    \includegraphics[width=1.0\textwidth]{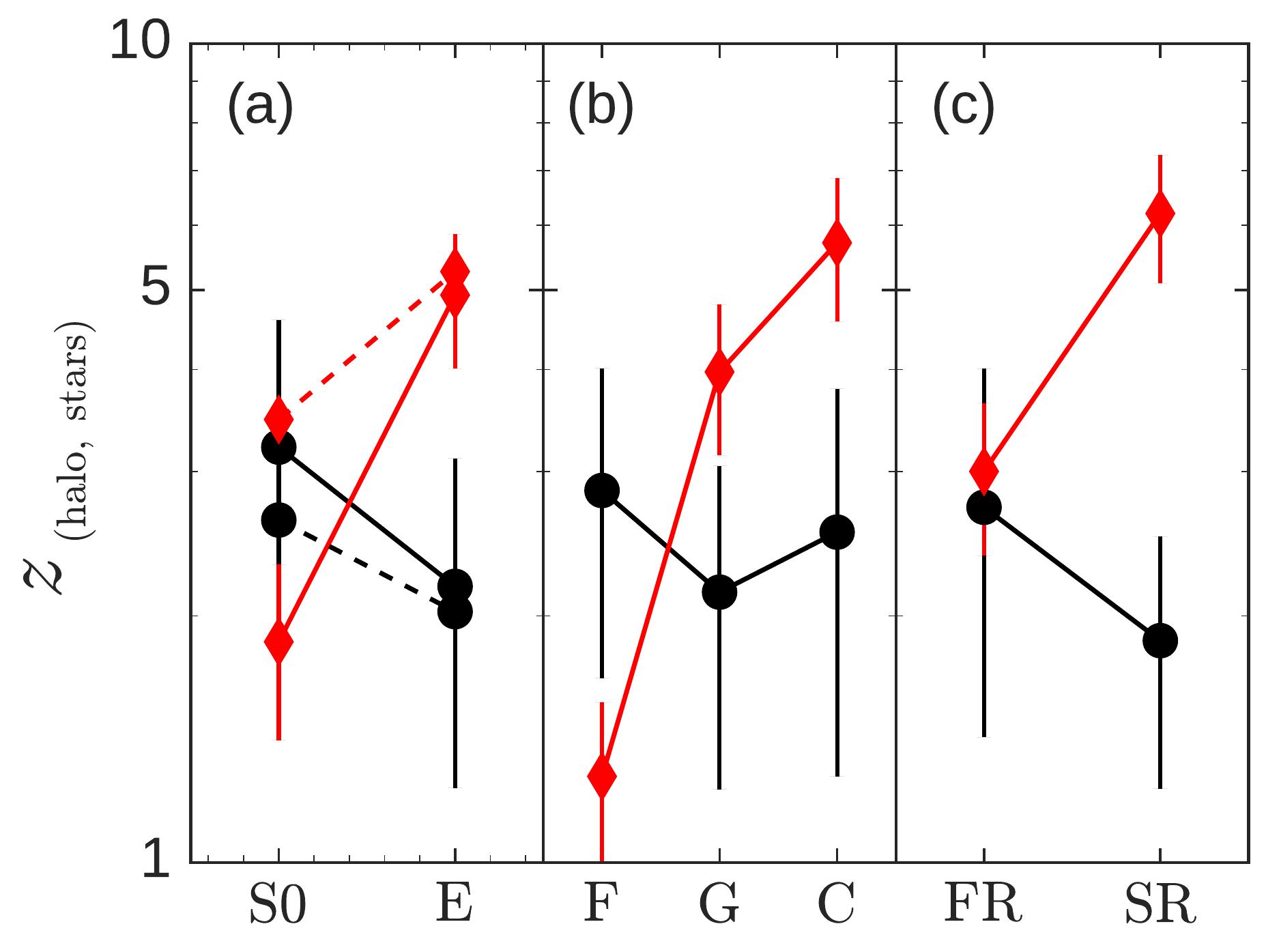}\hspace{0.01\textwidth}\\
	\caption{\label{fig:appendix_c} Summary plot showing the mean halo assembly epoch for our galaxy sample (black circles) without binning by stellar mass. The plot also shows the mean formation epoch that corresponds the luminosity--weighted ages of the central stars in our sample (red diamonds). Panel \textit{a} shows the mean assembly epoch according to galaxy morphology (E=elliptical, S0=lenticular), with the dashed lines joining the mean epochs when galaxies with ambiguous classifications are added to either morphologies. Panel \textit{b} shows mean assembly epoch as a function of galaxy environment (F=field, G=group, C=cluster) and panel \textit{c} shows the  mean assembly epoch as a function of central galaxy kinematics (FR=fast central rotator, SR=slow central rotator). Comparing the panels with those in Figure \ref{fig:halo_gal}, highlights the need to account for the strong dependence of halo assembly epoch with galaxy stellar mass. The trend we earlier observed in the field, where haloes of massive galaxies assemble at later epochs is not obvious.}
\end{figure*}

\bsp

\end{document}